\documentclass[10pt, a4paper, english, onehalfspacing, headsepline]{DoctoralThesis} 
\usepackage{tabularx}
\usepackage{graphicx}  % standard LaTeX graphics tool;
\usepackage{mathpazo}
\usepackage{amsmath}   % For getting proper math eqns
\usepackage[compress]{cite}
\usepackage{amssymb}   % Used for getting various symbols eg. gtrsim, lesssim etc.
\usepackage{bm} % For bold math (esp of lower greek letters)
\usepackage{dcolumn}% Align table columns on decimal point
\usepackage{pagecolor}
\usepackage{afterpage}
\usepackage{color}
\usepackage{multirow}
\usepackage{mathrsfs}
\usepackage[titletoc]{appendix}
\usepackage{amsfonts}
\usepackage{varioref}
\usepackage{color}
\usepackage{float}
\usepackage{subfig}
\usepackage[normalem]{ulem}
\usepackage{comment}
\usepackage{placeins}
\usepackage{caption}
\usepackage{xcolor}
\RequirePackage[colorlinks=true,citecolor=red,urlcolor=cyan,linkcolor=red]{hyperref}

\usepackage[Lenny]{fncychap}

\def\gr{general relativity }

\def\ni{\noindent}

\def\bea{\begin{equation}}
\def\eea{\end{equation}}
\labelformat{chapter}{Chapter #1} 
\labelformat{section}{Section #1} 
\labelformat{subsection}{Section #1} 
\labelformat{subsubsection}{Section #1}
\labelformat{subsubsubsection}{Section #1}
\labelformat{equation}{Eq.~(#1)} \labelformat{figure}{Fig.~#1} 
\labelformat{subfigure}{Fig.~\thefigure#1} 
\labelformat{table}{Tab.~#1} 
\labelformat{appendix}{Appendix #1}
\numberwithin{equation}{section}

\usepackage[normalem]{ulem} % To get strikethrough (\sout)

\newcommand{\ee}{\mathrm{e}}
\newcommand{\ii}{\mathrm{i}}
\newcommand{\dd}{\mathrm{d}}
\newcommand{\cf}{\varphi}

\newcommand{\om}{\omega}
\newcommand{\Om}{\Omega}
\newcommand{\Ord}{\mathcal{O}}

\newcommand{\de}{\delta}
\newcommand{\De}{\Delta}
\newcommand{\ep}{\epsilon}
\newcommand{\sst}{\sin^2\!\th}
\newcommand{\cct}{\cos^2\!\th}
\renewcommand\th{\theta}
\newcommand{\la}{\lambda}
\newcommand{\dl}{\partial}
\thesistitle{Theoretical and Observational Constraints\\ on Theories Beyond General Relativity} % Your thesis title, this is used in the title and abstract, print it elsewhere with \ttitle
\supervisor{\href{https://physics.iitgn.ac.in/tgp/members.html}{Dr. Sudipta \textsc{Sarkar}}} % Your supervisor's name, this is used in the title page, print it elsewhere with \supname
\examiner{} % Your examiner's name, this is not currently used anywhere in the template, print it elsewhere with \examname
\degree{Doctor of Philosophy} % Your degree name, this is used in the title page and abstract, print it elsewhere with \degreename
\author{\href{https://physics.iitgn.ac.in/tgp/members.html}{Rajes \textsc{Ghosh}}} % Your name, this is used in the title page and abstract, print it elsewhere with \authorname
\addresses{} % Your address, this is not currently used anywhere in the template, print it elsewhere with \addressname

\subject{Physical Science} % Your subject area, this is not currently used anywhere in the template, print it elsewhere with \subjectname
 % Keywords for your thesis, this is not currently used anywhere in the template, print it elsewhere with \keywordnames
\university{\href{https://iitgn.ac.in/}{Indian Institute of Technology Gandhinagar}} % Your university's name and URL, this is used in the title page and abstract, print it elsewhere with \univname
\department{\href{https://iitgn.ac.in/}{IIT Gandhinagar}} % Your department's name and URL, this is used in the title page and abstract, print it elsewhere with \deptname
\group{\href{https://physics.iitgn.ac.in/}{Department of Physics}} % Your research group's name and URL, this is used in the title page, print it elsewhere with \groupname
\faculty{\href{http://faculty.university.com}{Faculty Name}} % Your faculty's name and URL, this is used in the title page and abstract, print it elsewhere with 

\AtBeginDocument{
\hypersetup{pdftitle=\ttitle} % Set the PDF's title to your title
\hypersetup{pdfauthor=\authorname} % Set the PDF's author to your name
\hypersetup{pdfkeywords=\keywordnames} % Set the PDF's keywords to your keywords
}

\begin{document}

\frontmatter % Use roman page numbering style (i, ii, iii, iv...) for the pre-content pages

\pagestyle{plain} % Default to the plain heading style until the thesis style is called for the body content

%----------------------------------------------------------------------------------------
%	TITLE PAGE
%----------------------------------------------------------------------------------------

\normalsize
\begin{titlepage}
\begin{center}

\vspace*{.06\textheight}
{\scshape\LARGE \univname\par}\vspace{0.7 cm} % University name
\textsc{\Large Doctoral Thesis}\\[0.5cm] % Thesis type

\HRule \\[0.4cm] % Horizontal line
{\huge \bfseries \ttitle\par}\vspace{0.1cm} % Thesis title
\HRule \\[1.5cm] % Horizontal line
 
\begin{minipage}[t]{0.4\textwidth}
\begin{flushleft} \large
\emph{Author:}\\
\href{http://www.johnsmith.com}{\authorname} % Author name - remove the \href bracket to remove the link
\end{flushleft}
\end{minipage}
\begin{minipage}[t]{0.4\textwidth}
\begin{flushright} \large
\emph{Supervisor:} \\
\href{http://www.jamessmith.com}{\supname} % Supervisor name - remove the \href bracket to remove the link  
\end{flushright}
\end{minipage}\\[1.2cm]
\includegraphics[scale=0.1]{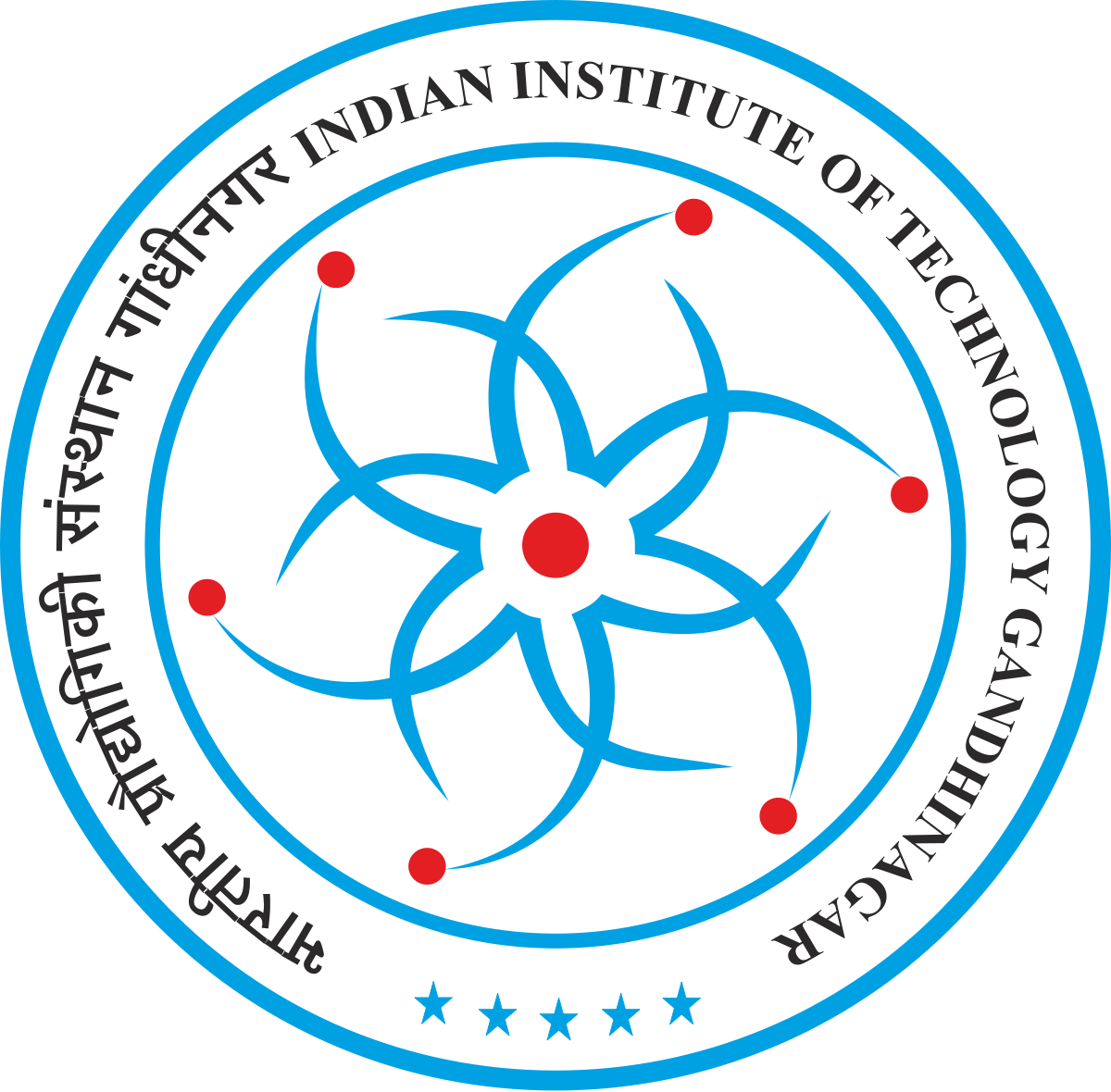}

\vfill

\large \textit{A thesis submitted in fulfillment of the requirements\\ for the degree of \degreename}\\[0.1cm] % University requirement text
\textit{in the}\\[0.1cm]
\groupname\\\deptname\\[1cm] % Research group name and department name
 
\vfill

%{\large \today}\\[4cm] % Date
%\includegraphics[scale=0.2]{IITGN_logo.png} % University/department logo - uncomment to place it
 
\vfill
\end{center}
\end{titlepage}
\cleardoublepage
%----------------------------------------------------------------------------------------
%	DECLARATION PAGE
%----------------------------------------------------------------------------------------
\begin{figure}
    \centering
    \includegraphics[scale=0.15]{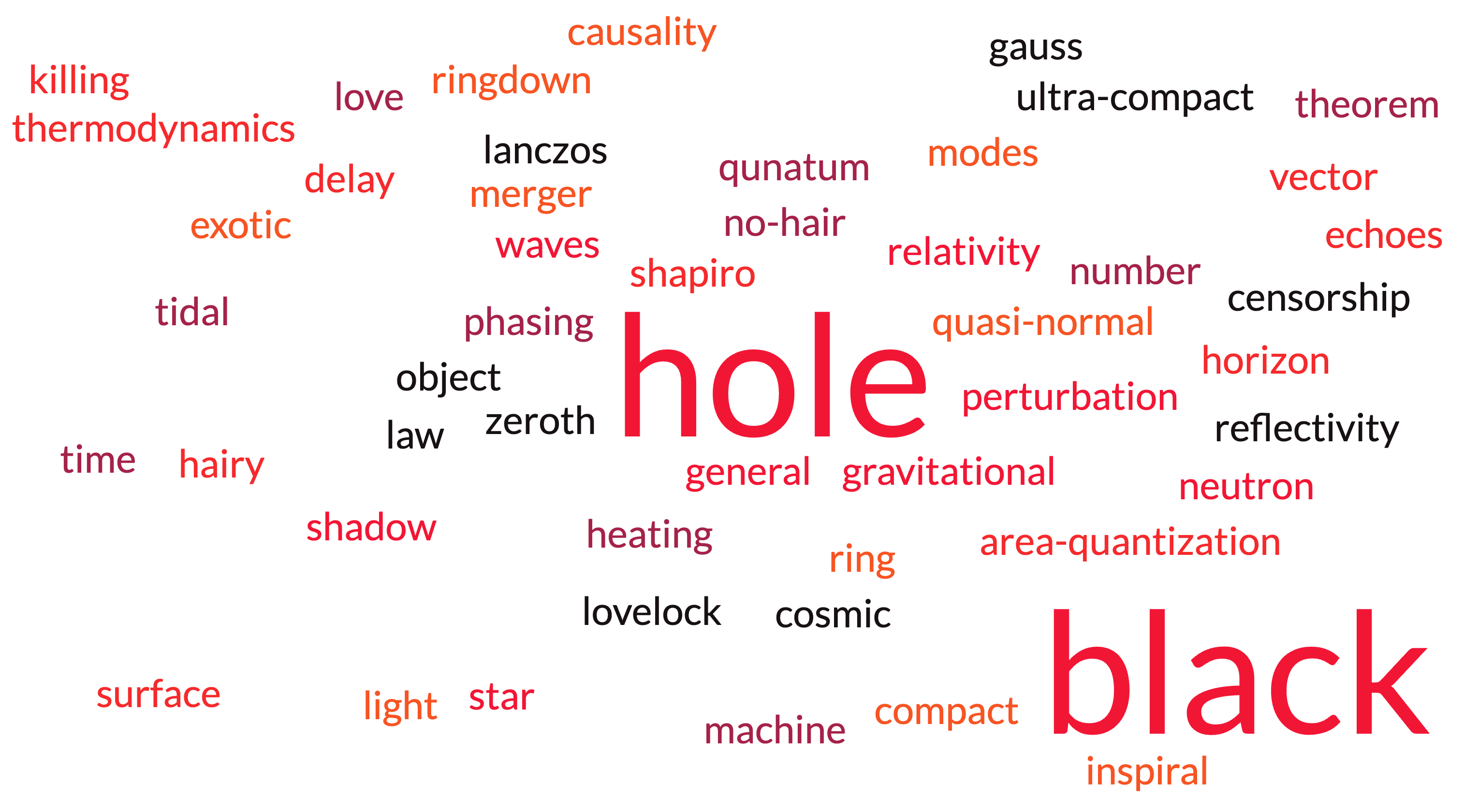}
\end{figure}
\vspace{5 cm}
\cleardoublepage
\begin{center}
\Huge \textbf{Declaration}
\end{center}
\vspace{1 cm}
%\addchaptertocentry{\authorshipname} % Add the declaration to the table of contents
\large \noindent I, \authorname, declare that this thesis titled, \textit{Theoretical and Observational Constraints on Theories Beyond General Relativity} represents my ideas in my own words and where other's ideas or words incorporated,
I have adequately cited and referenced the original sources. Also, I declare
that I have adhered to all principles of academic honesty and integrity and have not falsified or
fabricated any idea/fact/source in this submission. I understand that any violation
of the above can cause disciplinary action by the Institute and can also summon penal action from the
sources which have thus not been properly cited or from whom proper permission has not been taken.
\vspace{1 in}
 
\noindent Signed: Rajes Ghosh\\
\rule[0.5em]{25em}{0.5pt} % This prints a line for the signature
 
\noindent Date: 28/09/2023\\
\rule[0.5em]{25em}{0.5pt} % This prints a line to write the date

\cleardoublepage
%----------------------------------------------------------------------------------------
%	CERTIFICATE PAGE
%----------------------------------------------------------------------------------------
\begin{center}
\textbf{\Huge{CERTIFICATE}}
\end{center}

%\addchaptertocentry{\authorshipname} % Add the certificate to the table of contents
\vspace{1 cm}

\large \noindent It is certified that the work contained in the thesis titled Theoretical and Observational Constraints on Theories Beyond General Relativity by Mr. \authorname ~(Roll No. 17510060), has been carried out under
my supervision and that this work has not been submitted elsewhere for a degree.
\vspace{10 cm}
 
\hspace{11 cm}\noindent \supname

\hspace{11 cm}\noindent(Thesis Supervisor)

\hspace{11 cm}\noindent Department of Physics

\hspace{11 cm}\noindent IIT Gandhinagar 

\hspace{11 cm}\noindent Date: 28/09/2023\\

%----------------------------------------------------------------------------------------
%	QUOTATION PAGE
%----------------------------------------------------------------------------------------

%\vspace*{0.2\textheight}

%\noindent{\itshape It doesn't matter how beautiful your theory is, it doesn't matter how smart you are. If it doesn't agree with experiment, it's wrong..}\bigbreak

%\hfill Richard P. Feynman

\begin{center}
\dedicatory{\Large{\textmd{Dedicated To}}\ldots} 

\Large{\textbf{My Beloved Parents \& Brother,}}
\vspace{0.2 cm}

\Large{\textit{For their endless love, support and encouragement}}
\end{center}

%----------------------------------------------------------------------------------------
%	ACKNOWLEDGEMENTS
%----------------------------------------------------------------------------------------

\begin{acknowledgements}
\addchaptertocentry{\acknowledgementname} % Add the acknowledgements to the table of contents
\vspace{0.5 cm}
As my doctoral studies at IIT Gandhinagar draw to a close, I take a moment to contemplate how eventful and memorable the journey these years have been. I wish to express my heartfelt thanks to the many individuals whose help and support have made this journey thoroughly enjoyable.\\

\ni
First and foremost, it gives me immense pleasure to express my sincere gratitude towards my doctoral supervisor, Prof. Sudipta Sarkar for suggesting interesting and challenging problems that helped me better understand the subject. I really appreciate his inspiring mentorship, patient guidance and unwavering encouragement throughout this thesis work. Moreover, numerous discussions with him on various topics of physics, mathematics, philosophy and political affairs have always been a learning experience for me. I will also miss his generous invitations for delicious food and good tea/coffee.\\

\ni
I wish to thank Prof. Alok Laddha (CMI, Chennai), Prof. Jose D. Edelstein (USC, Spain), Prof. Enrico Barausse (SISSA/IFPU,  Italy), Prof. Sumanta Chakraborty (IACS, Kolkata) and Prof. P. Ajith (ICTS, Bengaluru) for their wonderful collaboration and invaluable guidance through several projects. Their insightful suggestions and advice have consistently played a pivotal role in advancing my research. I extend a special appreciation to Prof. Laddha for hosting me at CMI and Prof. Barausse at SISSA/IFPU during my visits. Their pleasant hospitality and friendly behavior always made me feel right at home.\\

\ni
I am also indebted to my Doctoral Study Committee members Prof. Arpan Bhattacharyya, Prof. Baradhwaj Coleppa and Prof. Uddipta Ghosh for their critical evaluation and valuable comments on my works during many talks that I presented at IIT Gandhinagar. I am obligated for their unconditional support and guidance throughout my PhD.\\

\ni
I want to thank previous members of our research group C. Fairoos and Akash K. Mishra for delightful collaboration in many projects. I thoroughly enjoyed numerous engaging discussions on diverse physics topics with them. I also acknowledge the help and support from my collaborators Kabir Chakravarti, Mostafizur Rahman, Sebastian H. V\"{o}lkel, Nicola Franchini, Sreejith Nair, Lalit Pathak, Selim Sk, and N. V. Krishnendu in several projects. I have learned a lot from them.\\

\ni
I must take this opportunity to thank the entire physics community at IIT Gandhinagar for fostering a welcoming, supportive, and research-driven environment. I am really proud to be a part of this amazing family. I feel fortunate to learn physics from Profs. Ravinder R. Puri, Anand Sengupta, Rupak Banerjee and Vinod Chandra. I am also grateful to Prof. Krishna Kanti Dey for many academic advises, and Prof. Atul Dixit for patiently explaining various interesting concepts of mathematics. I thank Saswati Roy, Raju Beerasent, Akshay Nagpal for providing useful help and support on numerous occasions. A special thanks also goes to Prof. B. Prasanna Venkatesh, Prof. Gopinadhan Kalon and Prof. Chandan K. Mishra for engaging in many interesting discussions and sharing delectable foods.\\

\ni
I am forever grateful to my friends Kamalesh, Abhishek(s), Gokul, Gowthama, Sunny, Rahul, Lalit, Manoj, Rik, Arnab, Sreejith, Akshat, Saptaswa, Suvigya, Lalita, Nisha, Nividha, Ipsita, Aparna, Dipika, Biswa, Divya, Vishakha, Soham, Chanchal and Setab for always standing by me and sharing some of the most cherished memories of my life. I am also thankful to my elder colleagues Akash, Amruta, Kousik, Agnivo, Abhirup, Manu, Utsav, Amit, Soumen, and Chakresh for their affection and support. I wish all of you a happy and prosperous life ahead.\\

\ni
I will never forget the fond moments spent with my lifelong friends Sudip, Neelabha, Kapil, Pankaj,  Samridhi and Kanhaiya. Thank you for making my days at IIT Gandhinagar eventful. I am also grateful to everybody at \textit{Shanti Restaurant} for serving us with delicious food, without which life would have been difficult! I also extend a warm gratitude to Andrea for his hospitality over my stay at his airbnb and invitation in Easter. A big thanks goes to all members of SISSA/IFPU for their pleasant company during my stay at Italy.\\

\ni
Lastly, I would like to thank my family, especially my parents and brother, for their endless love, support and encouragement through the ups and downs of my life. I owe everything to their struggle and sacrifice.\\

\ni
I am grateful to acknowledge the financial support from the Prime Minister Research Fellowship and special grants from IIT Gandhinagar, which have made possible several national and international visits during my Ph.D.

\vspace{3 cm}
~~~~~~~~~~~~~~~~~~~~~~~~~~~~~~~~~~~~~~~~~~~~~~~~~~~~~~~~~~~~~~~~~~~~~~~~~~~~~~~~~~~~~~~~~~~~~~~~~~~~~~~~~~~~~~~~~~~~~~~~~~~\textbf{\texttt{\Large{Rajes Ghosh}}}

\end{acknowledgements}

\vspace*{\fill}
\ni
\Large{\textbf{\emph{“Nature uses only the longest threads to weave her patterns, so each small piece of her fabric reveals the organization of the entire tapestry.”}}}\\
\\\large{-- \texttt{Richard P. Feynman, The Character of Physical Law\\ (1964, Cornell University)}}
\vspace*{\fill}

%----------------------------------------------------------------------------------------
%	ABSTRACT PAGE
\cleardoublepage
\newpage
\phantomsection
\thispagestyle{empty}
\addcontentsline{toc}{chapter}{Abstract}
\markboth{\MakeUppercase{\Huge{Abstract}}}{\MakeUppercase{\Huge{Abstract}}}
\begin{center}  {\LARGE \textbf{Abstract}}   \end{center}

\ni
This thesis embarks on a comprehensive investigation of modified gravity theories and their implications on the properties of compact objects. Our primary objective is to shed light on the fundamental nature of gravity by exploring potential departures from General Relativity (GR) through a combination of theoretical analyses and observational techniques.\\

\ni
One of our focal points is to understand black hole (BH) thermodynamics, with particular attention given to the zeroth law assuring the constancy of surface gravity on a stationary Killing horizon. Despite significant challenges posed by the complicated nature of field equations, we have successfully extended the proof of the zeroth law in Lanczos-Lovelock gravity. It marks an important step forward in the study of BH thermodynamics in modified gravity. We have also provided a general structure of higher-curvature field equations that may support zeroth law.\\

\ni
In the pursuit of a fully consistent theory of gravity, it is imperative to eliminate potential alternatives systematically. In the same spirit, we present a detailed analysis of the so-called causality constraints, which restrict the structure of higher-curvature terms based on the sign of Shapiro time shift. In contrast to Einstein-Gauss-Bonnet (EGB) theory, we show that quadratic gravity (QG) is free from such causality issues and hence, it can be considered a healthy effective theory at low energies. The findings obtained here could offer invaluable insights towards classifying consistent classical theories of gravity.\\

\ni
Compact objects exhibit a rich array of gravitational phenomena that unveil the intricate facets of strong gravity. Nevertheless, any inquiry into their nature necessitates a sound comprehension of their stability, which is intimately connected to the underlying light ring (LR) structure. In this context, we perform an in-depth analysis to ensure the presence of at least one LR outside the ergoregion of a compact object. Then, following a chain of reasoning motivated by other recent works, one may argue against the stability of horizonless objects with ergoregion and lend strong support in favor of the BH hypothesis.\\

\ni
It is well known that the no-hair properties of vacuum BHs in GR fail to hold in non-vacuum scenarios and modified gravity. However, could these hairs be so confined in the near-horizon regime that they have eluded detection through observations? Our theorem provides a negative answer to this question by showing that all existing hairs of any static, spherically symmetric, and asymptotically flat $D$-dimensional BHs must extend at least to the innermost LR, regardless of the specific theory of gravity being considered. In addition to its apparent observational relevance, our analysis bears interesting implications on the hairs of horizonless compact objects and the size of LRs as well.\\

\ni
We also explore the potential quantum-gravity deviations from GR using a model known as BH area-quantization. We utilize various gravitational wave (GW) observables, such as tidal heating in the inspiral phase and late-time echoes in the ringdown phase of a BH binary, to glean insights into possible detections of various quantum signatures. In future, as advancements in GW detectors bring about enhanced accuracy, sensitivity, and signal-to-noise ratios, we anticipate the ability to impose stringent constraints on various parameters related to area-quantization. Eventually, such explorations may provide valuable insights towards developing a consistent quantum theory of gravity.\\

\ni
Moreover, we employ one of the most important tools to probe near-horizon physics: the quasi-normal modes (QNMs) associated with perturbed BHs. In GR, finding these modes is particularly simple as the governing perturbation equation in the Schwarzschild/Kerr background decouples into radial and angular parts. However, in general, one may not have such luxury for BH solutions of a modified theory. To tackle this difficulty, we propose an efficient method of computing scalar QNMs of BHs perturbatively close to Schwarzschild/Kerr BHs. One may extend our method to incorporate gravitational perturbation, which has crucial applicability in determining the properties of the remnant BH formed as an end state of binary mergers. Notably, such techniques have recently been used for the event GW150914 to test the BH area theorem. We present a method that uses the result above and the validity of the BH second law to put stringent constraints on the topological Gauss-Bonnet coupling. Future GW observations might help to make our bound even stronger.\\ 

\ni
To provide a structured overview of the thesis, we have organized it into chapters that progressively delve deeper into these diverse aspects of modified gravity and compact objects. Each chapter is dedicated to a specific facet of our investigation, building a coherent narrative that spans both theoretical and observational explorations. We aspire to achieve nothing less than imparting valuable insights and novel perspectives that may significantly enhance our understanding of the fundamental nature of gravitation.
\cleardoublepage
\newpage

%----------------------------------------------------------------------------------------

%----------------------------------------------------------------------------------------
%	LIST OF PUBLICATIONS
%--------------------------------------------------------------------------

\clearpage
\Huge
\begin{center}
\textbf{List of Publications}
\end{center}
\large
This thesis is comprised of results from the following publications,
\begin{itemize}
\item \textbf{Rajes Ghosh}, S. Sarkar, "Black Hole Zeroth Law in Higher Curvature Gravity," \href{https://journals.aps.org/prd/abstract/10.1103/PhysRevD.102.101503}{Phys.Rev.D 102 (2020) 10, 101503 (Rapid Communication)}, \href{https://arxiv.org/abs/2009.01543}{arXiv: 2009.01543 [gr-qc]}.

\item J. D. Edelstein, \textbf{Rajes Ghosh}, A. Laddha, S. Sarkar, "Causality constraints in Quadratic Gravity," \href{https://link.springer.com/article/10.1007/JHEP09(2021)150}{JHEP 09 (2021) 150}, \href{https://arxiv.org/abs/2107.07424}{	arXiv: 2107.07424 [hep-th]}.

\item \textbf{Rajes Ghosh}, S. Sarkar, "Light rings of stationary spacetimes," \href{https://journals.aps.org/prd/abstract/10.1103/PhysRevD.104.044019}{Phys.Rev.D 104 (2021) 4, 044019}, \href{https://arxiv.org/abs/2107.07370}{arXiv: 2107.07370 [gr-qc]}.

\item K. Chakravarti, \textbf{Rajes Ghosh}, S. Sarkar, "Signature of nonuniform area quantization on gravitational waves," \href{https://journals.aps.org/prd/abstract/10.1103/PhysRevD.104.084049}{Phys.Rev.D 104 (2021) 8, 084049}, \href{https://arxiv.org/abs/2108.02444}{arXiv: 2108.02444 [gr-qc]}.

\item K. Chakravarti, \textbf{Rajes Ghosh}, S. Sarkar, "Signature of nonuniform area quantization on black hole echoes," \href{https://journals.aps.org/prd/abstract/10.1103/PhysRevD.105.044046}{Phys.Rev.D 105 (2022) 4, 044046}, \href{https://arxiv.org/abs/2112.10109}{arXiv: 2112.10109 [gr-qc]}.

\item K. Chakravarti, \textbf{Rajes Ghosh}, S. Sarkar, "Constraining the topological Gauss-Bonnet coupling from GW150914," \href{https://journals.aps.org/prd/abstract/10.1103/PhysRevD.106.L041503}{Phys.Rev.D 106 (2022) 4, L041503 (Letter)}, \href{https://arxiv.org/abs/2201.08700}{arXiv: 2201.08700 [gr-qc]}.

\item \textbf{Rajes Ghosh}, N. Franchini, S. V\"{o}lkel, E. Barausse, "Quasinormal modes of nonseparable perturbation equations: The scalar non-Kerr case," \href{https://journals.aps.org/prd/abstract/10.1103/PhysRevD.108.024038}{Phys.Rev.D 108 (2023) 2, 024038}, \href{https://arxiv.org/abs/2303.00088}{arXiv: 2303.00088 [gr-qc]}.

\item \textbf{Rajes Ghosh}, Selim Sk, S. Sarkar, "Hairy Black Holes: Non-existence of Short Hairs and Bound on Light Ring Size," \href{https://journals.aps.org/prd/abstract/10.1103/PhysRevD.108.L041501}{Phys.Rev.D 108 (2023) 4, L041501 (Letter)}, \href{https://arxiv.org/abs/2306.14193} {arXiv: 2306.14193 [gr-qc]}.

\end{itemize}

%----------------------------------------------------------------------------------------
%	LIST OF CONTENTS/FIGURES/TABLES PAGES
%----------------------------------------------------------------------------------------

\tableofcontents % Prints the main table of contents

\listoffigures%\addcontentsline{toc1}{chapter}{List of Figures} % Prints the list of figures

%\listoftables%\addcontentsline{toc2}{chapter}{List of Tables} % Prints the list of tables

%----------------------------------------------------------------------------------------
%	ABBREVIATIONS
%----------------------------------------------------------------------------------------

\begin{abbreviations}{ll} % Include a list of abbreviations (a table of two columns)

\textbf{LR} & \textbf{L}ight  \textbf{R}ing\\
\textbf{BH} & \textbf{B}lack \textbf{H}ole\\
\textbf{GR} & \textbf{G}eneral \textbf{R}elativity \\
\textbf{QG} & \textbf{Q}uadratic \textbf{G}ravity \\
\textbf{LL} & \textbf{L}ovelock - \textbf{L}anczos\\
\textbf{GW} & \textbf{G}ravitational  \textbf{W}ave\\
\textbf{QBH} & \textbf{Q}uantum \textbf{B}lack \textbf{H}ole\\
\textbf{QNM} & \textbf{Q}uasi - \textbf{N}ormal  \textbf{M}ode\\
\textbf{EGB} & \textbf{E}instein - \textbf{G}auss - \textbf{B}onnet\\
\textbf{$\Lambda$CDM} & \textbf{L}ambda  \textbf{C}old  \textbf{D}ark  \textbf{M}atter\\
\textbf{CEMZ} & \textbf{C}amanho - \textbf{M}aldacena - \textbf{E}delstein -  \textbf{Z}hiboedov\\
\end{abbreviations}

\cleardoublepage

%----------------------------------------------------------------------------------------
%	PHYSICAL CONSTANTS/OTHER DEFINITIONS
%----------------------------------------------------------------------------------------

%\begin{constants}{lr@{${}={}$}l} % The list of physical constants is a three column table

% The \SI{}{} command is provided by the siunitx package, see its documentation for instructions on how to use it

%Speed of Light & $c_{0}$ & \SI{2.99792458e8}{\meter\per\second} (exact)\\
%Constant Name & $Symbol$ & $Constant Value$ with units\\

%\end{constants}

%----------------------------------------------------------------------------------------
%	SYMBOLS
%----------------------------------------------------------------------------------------

%\begin{symbols}{lll} % Include a list of Symbols (a three column table)

%$a$ & distance & \si{\meter} \\
%$P$ & power & \si{\watt} (\si{\joule\per\second}) \\
%Symbol & Name & Unit \\

%\addlinespace % Gap to separate the Roman symbols from the Greek

%$\omega$ & angular frequency & \si{\radian} \\

%\end{symbols}

\clearpage

\newpage
\Huge {Notations and Conventions}
\vspace{1 cm}
\large

\noindent
We have adopted the following notations and conventions throughout this thesis. If used otherwise, it will be specified explicitly.
\vspace{1 cm}
\normalsize
\begin{itemize}
\item The spacetime metric is denoted as $g_{\mu\nu}$, where the Greek indices runs over the spacetime dimensions. Also, the "mostly-positive" metric signature $(-,+,+,+, \cdots)$ will be used. Einstein's summation convention is assumed to sum over repeated indices.

\item We shall use the natural units and set the speed of light ($c$), Newton's gravitational constant ($G$), Boltzmann constant ($k_B$), and Planck's constant ($\hbar$) to unity.
\end{itemize}

%----------------------------------------------------------------------------------------
%	DEDICATION
%----------------------------------------------------------------------------------------

%----------------------------------------------------------------------------------------
%	THESIS CONTENT - CHAPTERS
%----------------------------------------------------------------------------------------

\mainmatter % Begin numeric (1,2,3...) page numbering

\pagestyle{thesis} % Return the page headers back to the "thesis" style

% Include the chapters of the thesis as separate files from the Chapters folder
% Uncomment the lines as you write the chapters
%\clearpage

\chapter{{\color{red!60!black}Introduction} }\label{Chapter 1}
\large
We live in a magnificent time when unprecedented technological development is reshaping our knowledge about the universe, and the field of gravitational physics is no exception. The recent observations of gravitational waves (GWs) by the LIGO-Virgo collaboration~\cite{Abbott:2016blz, TheLIGOScientific:2016wfe, TheLIGOScientific:2016src, Abbott:2016nmj, Abbott:2017vtc, TheLIGOScientific:2017qsa} and the black hole (BH) shadow imaging by the Event Horizon Telescope (EHT)~\cite{Akiyama:2019cqa, Akiyama:2019bqs, Akiyama:2019fyp, Akiyama:2019eap} are constantly pushing the boundaries of our understanding of gravity ever further. In this chapter, we reflect on a glimpse of this scientific excitement, which will be incomplete if we do not start from the beginning.
%%%%%%%%%%%%%%%%%%%%%%%%%
\section{{\color{blue!70!brown} Galileo, Newton, Einstein and Beyond $\ldots$}}
%%%%%%%%%%%%%%%%%%%%%%%%%
The story of science is full of profound ideas and great ideals. This fact is aptly exemplified by the development of scientific concepts regarding one of the four fundamental forces of nature, namely \textit{gravity}. Galileo was perhaps the first to perform systematic studies to comprehend the true character of this then-mysterious force~\cite{drake2003galileo}. His experiments challenged the prevailing Aristotelian view by showing that all objects fall at the same rate irrespective of their masses and material compositions, a significant step towards the modern understanding of gravity.\\

\ni
Galileo's observations and experiments laid the foundation for what would later become Newton's law of universal gravitation, drawing "heaven" and "earth" in the same footing that marks the \textit{first great unification}~\cite{weinberg1994dreams, mainzer1996symmetries}. Newton built upon Galileo's work, formulating the laws of motion and developing a mathematical framework to describe the gravitational attraction between masses. According to Newton's law of universal gravitation~\cite{newton1999principia}, any two point masses in the universe attract each other by a force ($\vec{F}_{\text{g}}$) acting along the line ($\hat{r}$) joining them, and the amount of force is proportional to the product of their masses ($m_{1}\, \textrm{and}\, m_{2}$) and inversely proportional to the square of the distance ($r$) between them:
%%%%%%%%%%%%%%%%%%%%%%%%%%%%%%%
\begin{equation}
\vec{F}_{\text{g}} = - \frac{G\, m_1\, m_2}{r^2}\, \hat{r}\, .
\end{equation}
%%%%%%%%%%%%%%%%%%%%%%%%%%%%%%%
This equation is utterly sublime in its content, capturing the true essence of gravity fairly accurately. It is highly successful in explaining all earth-bound motion under gravity, the motion of the planets around the sun, tides on Earth and many more gravitational phenomena~\cite{newton1999principia}.\\

\ni
Owing to these magnificent successes, Newton's theory reigned supreme for more than two centuries. However, the rapid scientific and technological developments in the early and mid-twentieth century started to expand the horizon of science through precise observations, and soon, several limitations of the Newtonian viewpoint came to notice. The increasing list of discrepancies included the perihelion precession of Mercury~\cite{born1924einstein}, to which Newtonian gravity failed to provide a complete answer. Moreover, it was also apparent that Newton's law of gravitation is inconsistent with the newly developed relativistic description of motion.\\

\ni
The stage was all set for new groundbreaking ideas, and Einstein (1915) ingeniously accomplished this through the formulation of \textit{\gr}(GR)~\cite{Einstein:1916vd}. Profound in its consequences and instrumental in its applications, GR is, without a doubt, one of the most successful theories ever developed by humanity. It throws away the Newtonian idea of gravity being a force and elegantly replaces it with the geometry of $4$-dimensional pseudo-Riemannian differential manifold, known as the spacetime metric. In this framework, gravity emerges as a manifestation of spacetime curvature produced by matter/energy (re-introducing $G$ and $c$ for the moment)~\cite{hawking_ellis_1973, Wald:1984rg, Chandrasekhar:1985kt},
%%%%%%%%%%%%%%%%%%%%%%%%%%%%%%%
\begin{equation} \label{efe}
R_{\mu\nu}-\frac{1}{2}R\, g_{\mu\nu} + \Lambda\, g_{\mu \nu}= \frac{8\pi\,G}{c^4}\, T_{\mu\nu}\, .
\end{equation}
%%%%%%%%%%%%%%%%%%%%%%%%%%%%%%%
The presence of $c$ showcases the in-built relativistic and causal nature of the theory. Additionally, the tensorial representations and the local differential character of the above field equations reflect the two crucial guiding principles that Einstein used during the formulation of GR: the diffeomorphism invariance and the equivalence principle.\\

\ni
In the compact notations, the above field equations represent a set of ten non-linear coupled partial differential equations of both hyperbolic and elliptic nature. Without any further symmetry assumptions, these equations are formidable to tackle analytically, and one usually needs the extravaganza of sophisticated numerical techniques to solve them consistently.
%%%%%%%%%%%%%%%%%%%%%%%%%%
\subsection{{\color{red!70!blue}Two Exact Black Hole Solutions of GR}}
%%%%%%%%%%%%%%%%%%%%%%%%%%
Though a few vacuum and non-vacuum analytical solutions to Einstein's field equations are known~\cite{Stephani:2003tm, griffiths2009exact}, we shall mainly focus on two important vacuum BH solutions of GR in this thesis, namely the Schwarzschild~\cite{Schwarzschild:1916uq} and Kerr BHs~\cite{PhysRevLett.11.237}. Apart from their immense observational importance, they represent the unique astrophysically relevant (with no electric/magnetic charge) asymptotically flat BH solutions, a result known as the uniqueness theorem~\cite{heusler1996black, Mazur:2000pn, Robinson:2004zz, Chrusciel:2012jk, PhysRev.164.1776}. These BH configurations are also the simplest among all celestial objects in the sense that a complete specification of their characteristics only requires two parameters - mass and spin (neglecting electric/magnetic charge), as guaranteed by the celebrated no-hair theorems~\cite{ PhysRevLett.26.331, Bekenstein:1971hc, Bekenstein:1972ky}.\\

\ni
Among these two solutions of GR, the Schwarzschild metric was discovered by K. Schwarzschild in 1916~\cite{Schwarzschild:1916uq}. It represents the asymptotically flat spacetime outside a spherically symmetric matter distribution of mass $M$,
%%%%%%%%%%%%%%%%%%
\begin{equation}\label{Sch}
ds^2 = -\left(1-\frac{2M}{r}\right)dt^2 + \left(1-\frac{2M}{r}\right)^{-1}dr^2 + r^2 \left( d\theta^2 + \sin^2\theta\, d\phi^2 \right)\, .
\end{equation}
%%%%%%%%%%%%%%%%%%%
Interestingly, due to Birkhoff's theorem~\cite{1923rmpbookB, 2005GReGr372253J}, 
staticity and asymptotic flatness of the vacuum metric follow as a consequence of the spherical symmetry. The Schwarzschild solution has been considered extensively in various classical tests of GR~\cite{Dyson:1920cwa, born1924einstein, Will:2014kxa}, such as the perihelion precession of Mercury and light deflection around the sun. These tests show a brilliant match with observations.\\

\ni
For the case of a spherically symmetric BH, however, the Schwarzschild solution represents the exterior spacetime of an isolated static BH with an event horizon at $r_h=2M$, where $M$ represents the Arnowitt-Deser-Misner (ADM) mass of the BH. Its rotating generalization is known as the Kerr BH, which was discovered by R. Kerr in 1963~\cite{PhysRevLett.11.237}, half a century after K. Schwarzschild proposed his solution. Unlike the Schwarzschild metric, the Kerr metric is stationary and has two parameters - mass $M$ and spin $a \leq M$:
%%%%%%%%%%%%%%%%%%%%%%%%%%%
\begin{equation}
\begin{split}\label{kerr}
    ds^2=-\left(1-\frac{2\, M\, r}{\rho^2}\right) dt^2 + \frac{\rho^2}{\Delta}\, dr^2 +\rho^2\, d\theta^2 +\sin^2 \theta &\left[\frac{(r^2+a^2)^2-\Delta\, a^2 \sin^2\theta }{\rho^2}\right] d\phi^2 \\
    & -\frac{4\, M\, r\, a\, \sin^2 \theta}{\rho^2}\, dt\, d\phi\, ,
\end{split}
\end{equation}
%%%%%%%%%%%%%%%%%%%%%%%%%%%
where $\rho^2(r, \theta) = r^2+a^2\, \cos^2 \theta$, and $\Delta(r) = r^2+a^2-2\, M\, r$. The Schwarzschild solution is recovered in the limit $a \to 0$. The event horizon of Kerr BH is located at $r_h=M+\sqrt{M^2-a^2}$, the largest root of the function $\Delta(r)$. Another surface of interest is the ergosphere at $r_e = M+\sqrt{M^2-a^2\, \cos^2 \theta}$, where the time translation Killing vector $(\partial_t)^\mu$ becomes null. While the event horizon represents the limit of all stationary observers, the ergosphere denotes the limit of static observers. In other words, inside the region (called the ergoregion) between the event horizon and the ergosphere, all observers must co-rotate with the BH to remain stationary at a fixed radial distance.\\

\ni
Besides their extensive use in various classical tests of GR, the Schwarzschild and Kerr BHs are also ubiquitous in modern astrophysical observations. Especially, GW observations of the LIGO-Virgo collaboration~\cite{Abbott:2016blz, TheLIGOScientific:2016wfe, TheLIGOScientific:2016src, Abbott:2016nmj, Abbott:2017vtc, TheLIGOScientific:2017qsa} and the BH shadow observations by the EHT~\cite{Akiyama:2019cqa, Akiyama:2019bqs, Akiyama:2019fyp, Akiyama:2019eap} are in excellent agreement with the so-called "Kerr paradigm", in which the underlying BHs are assumed to be well described by Kerr solutions. Moreover, these observations provide some of the most direct evidence of the existence of BHs in our universe.

%%%%%%%%%%%%%%%%%%%%%%%%%%
\subsection{{\color{red!70!blue} Why Venturing Beyond GR?}}
%%%%%%%%%%%%%%%%%%%%%%%%%%
Despite its tremendous success and striking observational consistency, GR is continuously being put to the test that probes different energy scales, spanning local to cosmological distances. This incessant scrutiny is not only to check the validity of GR at all length scales but also to explore its limitations, some of which are discussed below. Though the proper cure for these limitations is still unknown, various theoretical and observational efforts are always on-try to obtain a systematic understanding that might help us towards the formulation of a complete theory of gravity in the future.\\

\ni
(i) {{\color{red!70!blue} Testing GR at all length scales:}} As discussed earlier, GR is very well tested in solar and stellar systems~\cite{Will:2014kxa, Stairs:2003eg, Will:2018bme}, beyond which it still lacks enough observational support~\cite{Uzan:2010, Ishak:2018his}. Being a classical theory, GR breaks down at small enough length scales (corresponding to high energies), and a complete quantum theory of gravity must replace it. These quantum modifications might play non-ignorable roles near the vicinity of BH horizons, a proposal thoroughly studied in the literature~\cite{Bekenstein:1974jk, Bekenstein:1995ju, Skenderis:2008qn, Almheiri:2012rt, Saravani:2012is, Cardoso:2016oxy, Cardoso:2016rao, Burgess:2018pmm, Oshita:2019sat}. In addition to high energy corrections, GR may also receive modifications at large length scales~\cite{Dvali:2002vf}. For example, the accelerated expansion~\cite{SupernovaSearchTeam:1998fmf, SupernovaCosmologyProject:1998vns} of our universe might be due to such a large-scale modification to GR rather than a consequence of dark energy~\cite{Deffayet:2001pu}. Such deviations from the classical GR paradigm could have distinct observational signatures that may help us build phenomenological models~\cite{Bekenstein:1974jk, Bekenstein:1995ju, Oshita:2019sat} having hints of the ultimate theory of gravity.\\

\ni
(ii) {{\color{red!70!blue} Inherent non-quantum nature:}} By construction, GR is a classical field theory that governs the gravitational interactions between bulky objects. And, like all other fundamental descriptions of nature, it is expected that gravity too should obey the laws of quantum mechanics. Unfortunately, this is not the case -- any attempt to reconcile GR with the established quantum field theory went in vain due to the appearances of non-renormalizable ultraviolet (UV) divergences. When perturbed around a fixed curved background, every loop order comes with new divergences, spoiling any hope of reabsorbing them into a finite number of counter terms of an effective action. However, this perturbative non-renormalizability can be tamed by truncating the loop expansion at a particular order. Such truncations give rise to modified theories of gravity up to a certain energy scale. For example, ignoring some non local contributions, the $1-$loop effective action has the form~\cite{Utiyama:1962sn, Birrell:1982ix},
%%%%%%%%%%%%%%%%%%
\begin{equation}\label{1loop}
\mathcal{A}_{1-\text{loop}} = \frac{1}{16\pi G} \int d^4x\, \sqrt{-g} \left[ R - 2\Lambda + \alpha\, R^2 + \beta\, R_{\alpha \beta}{R^{\alpha \beta}} \right]\, .
\end{equation}
%%%%%%%%%%%%%%%%%%%
A theory thus obtained is known as the Stelle gravity~\cite{PhysRevD.16.953}, where the UV divergences are removed till the first loop order by the renormalization of the Newton's constant ($G$), cosmological constant ($\Lambda$), and the two higher curvature couplings $(\alpha,\, \beta)$. Results from string theory~\cite{Boulware:1985wk, Zwiebach:1985uq} also suggest that GR might make sense only at low energies, which has to be supplemented by such higher curvature terms as one approaches UV scales.\\

\ni
(iii) {{\color{red!70!blue} Issue with spacetime singularities:}} All known BH solutions of GR (including Schwarzschild and Kerr discussed previously) have singular cores, where various curvature invariants blow up, and the theory loses its predictive power. In fact, as a consequence of the famous Penrose-Hawking singularity theorems~\cite{Penrose:1964wq, hawking_ellis_1973}, such spacetime singularities (together with the big bang singularity) are inevitable consequences of the spacetime dynamics governed by GR. In such a scenario, the predictions of GR can only make sense if the BH singularities are cloaked inside their event horizons, remaining causally disconnected from the domain of outer communications. This statement is known as Penrose's weak cosmic censorship conjecture~\cite{Penrose:1969pc, Wald:1984rg}, which has no proof thus far. However, as suggested by loop quantum gravity~\cite{Ashtekar:2008ay, Blanchette:2020kkk}, it is expected that these undesirable features will be automatically censored out in the yet-to-be-found quantum theory of gravity. This expectation is built upon a simple yet powerful consequence of quantum theory. According to the laws of quantum mechanics, a mass or some amount of energy can not be localized below its Compton length, a scale where quantum effects become very prominent. Thus, near a singularity, quantum effects cannot be neglected, and a consistent quantum theory of gravity must replace GR.\\

\ni
(iv) {{\color{red!70!blue} Corrections in strong gravity regime:}} The Einstein-Hilbert action of GR is invariant under the general coordinate transformations. However, one can add all sorts of higher-order curvature invariants to this action while preserving this diffeomorphism symmetry. A priori, there is no theoretical reason to discard such higher curvature terms from the action. On the contrary, following our earlier discussion, perturbative renormalizability up to a certain loop order necessarily generates these terms in the low-energy effective action~\cite{Utiyama:1962sn, Birrell:1982ix, PhysRevD.16.953}. In particular, the effects of higher curvature terms will be prominent in strong gravity regimes, such as near the vicinity of merging BHs. At a classical level, the dimensionful couplings to the higher curvature terms come with associated classical length scales at which their effects might be important. Moreover, when linearized about a solution, these modified theories may have additional propagating modes other than massless graviton, rendering a fundamental change in the gravitational dynamics from its GR counterpart~\cite{Damour:1992we, Alexeev:1996vs, Fujii:2003pa, faraoni2004cosmology}. Consequently, the GW signatures of modified gravity could differ considerably from GR's.\\

\ni
(v) {{\color{red!70!blue} Reasons from cosmology:}} Our current understanding of the universe is based on the $\Lambda$-Cold Dark Matter ($\Lambda$CDM) model, which presupposes the applicability of GR at vast cosmological length scales. A wealth of observations from cosmic microwave background radiation and galaxy dynamics agree with this model~\cite{Aghanim:2018eyx, Weinberg:2008zzc, Goldhaber_2009}. However, this triumph comes with the price of including dark matter and dark energy, whose nature is not well known, as dominant components in the overall energy content of the universe~\cite{Peebles:1982, Weinberg:2008zzc}. With this inclusion, galaxy rotation curves can be explained by the presence of dark matter~\cite{Weinberg:2008zzc}, and the late-time expansion of the universe can be explained as a consequence of dark energy sourced by a positive cosmological constant ($\Lambda$)~\cite{SupernovaSearchTeam:1998fmf, SupernovaCosmologyProject:1998vns}. However, this satisfactory match is just a curtain over a more enigmatic issue, namely the fine-tuning problem~\cite{Weinberg:1989, Carroll:2000fy, Padmanabhan:2002ji, Martin:2012bt}. Assuming its origin from vacuum energy density, the measured value ($\Lambda/8 \pi G \sim 10^{-47}$ GeV$^4$) is in severe disagreement with that of vacuum energy density ($10^{71}$ GeV$^4$) predicted by quantum field theory. In addition, the estimated value of the Hubble constant at the late and early universe shows a $5\sigma$-tension~\cite{Riess:2021jrx}. To resolve these discrepancies, we possibly have to consider alternative descriptions of gravity beyond GR~\cite{PhysRevD.75.083504, AMENDOLA2008125, Appleby:2007vb, PhysRevD.77.046009, PhysRevD.77.124024, PhysRevD.76.064004, PhysRevD.80.123528, Starobinsky_2007, Tsujikawa:2007xu}.
%%%%%%%%%%%%%%%%%%%%%%%%%%
\subsection{{\color{red!70!blue} How to  Venture Beyond GR?}}
\begin{figure}[!htp]
    \centering
    \includegraphics[scale=0.25]{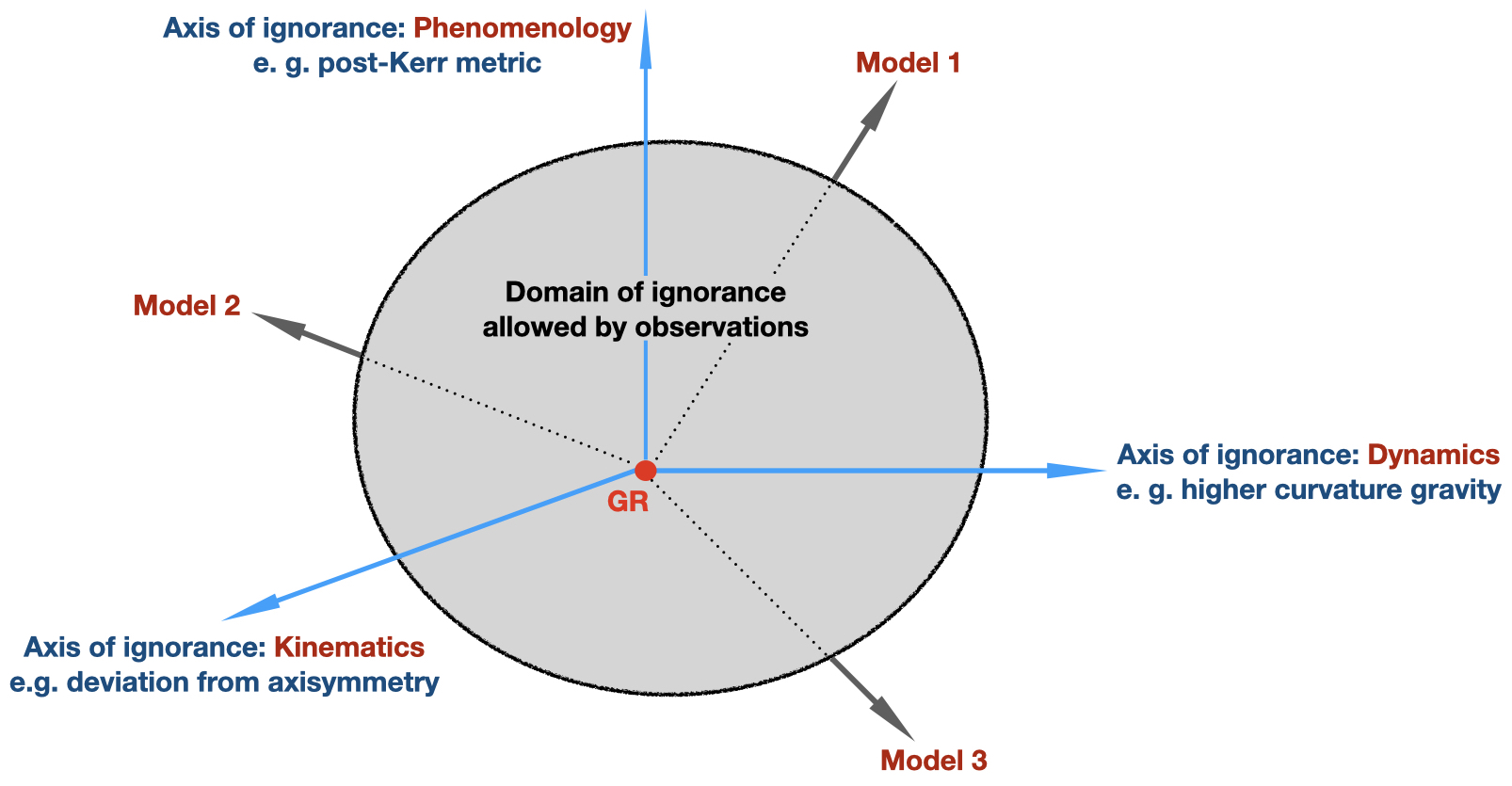}
    \caption{\textbf{Landscape of modified theories.} The three axes represent the three independent ways to modify GR denoted by the central red dot. The shaded blob signifies the observational bounds on various beyond-GR parameters introduced by different models shown by several rays emanating from the origin. It is expected that these modified models will correspond to GR as some limiting cases.}
    \label{Land}
\end{figure}

%%%%%%%%%%%%%%%%%%%%%%%%%%
The previous subsection discusses several limitations and shortcomings of GR that prompt us to study other alternatives. Although we are not in a position to rank these modified theories, we should continue exploring them, as they may provide valuable inputs towards the longed-for quantum theory of gravity. In this spirit, one can consider three distinct ways to modify GR, namely by (a) making dynamical changes to the GR action (and hence, to the field equations), (b) performing observation-oriented phenomenological changes in the solutions of GR, and (c) invoking kinematical alteration in the spacetime structure. Then, all possible modified models are nothing but some admixture of these distinct variations. A suggestive pictorial way to visualize the landscape of the modified theories of gravity is presented in \ref{Land}.\\
%%%%%%%%%%%%%%%%%%%%%%%%%%%
%%%%%%%%%%%%%%%%%%%%%%%%%%%

\ni
Let us now have a closer look at these three distinct ways to modify GR, referred to as the "axes of ignorance" in the figure above. We shall also illustrate them with known alternative models well studied in the literature.\\

\ni
(a) {{\color{red!70!blue} Dynamical modifications:}} The spacetime dynamics in GR are governed by Einstein's field equations given by \ref{efe}. Moreover, owing to Lovelock's theorem~\cite{Lovelock:1972vz}, GR is the unique second-order geometric theory of gravity in $4D$ with the metric as the sole dynamical field. Therefore, a modification of spacetime dynamics necessarily results from deviations in the GR field equations, which may be sourced either by the addition of various higher curvature terms in the action, such as quadratic gravity (QG)~\cite{PhysRevD.16.953} and Lanczos-Lovelock (LL) gravity~\cite{Lovelock:1972vz}, and/or by the inclusion of extra dynamical fields (scalar/vector/tensor) other than the metric, such as scalar-tensor theory~\cite{Damour:1992we, Alexeev:1996vs, Fujii:2003pa, faraoni2004cosmology} and Horndeski theory~\cite{Horndeski:1974wa, Kobayashi:2019hrl}. However, we shall mainly focus on the higher curvature modifications to GR. These higher curvature terms might give rise to distinguishable observational signatures, which can be used to put constraints on the corresponding coupling constants. For the purpose of this thesis, let us consider two such modified theories, namely the QG and the LL gravity.\\

\ni
Neglecting the cosmological constant, the action for the QG at arbitrary spacetime dimensions ($D$) is given by 
%%%%%%%%%%%%%%%%%%
\begin{equation}\label{QG}
\mathcal{A}_{\text{QG}} = \frac{1}{16\pi} \int d^Dx\, \sqrt{-g} \left[ R + \alpha\, R^2 + \beta\, R_{\alpha \beta}{R^{\alpha \beta}} \right]\, .
\end{equation}
%%%%%%%%%%%%%%%%%%%
In fact, for $D=4$, QG represents the most general quadratic action due to the Gauss-Bonnet theorem that expresses the Kretschmann scalar as a linear combination of the two quadratic Ricci scalars. This theory is studied extensively in the literature~\cite{PhysRevD.16.953, buchbinder1992effective, Nojiri:2001ae, Alvarez-Gaume:2015rwa, Salvio:2018crh}, and several exact BH solutions are also known~\cite{Audretsch:1993kp, Matyjasek:2004vr, Berej:2006cc, Svarc:2018coe}. It is also interesting to note that QG possesses extra dynamical modes other than massless graviton, which may become tachyonic unless the coupling constants are chosen so that $4\, (D-1)\, \alpha + D\, \beta \geq 0$, and $\beta \leq 0$~\cite{Tekin:2016vli}. Even in this constrained parameter space, the massive spin-$2$ mode is actually a ghost rendering unbounded Hamiltonian unless $\beta = 0$~\cite{Audretsch:1993kp, Tekin:2016vli}. However, at a classical level, we need not be too alarmed as long as our probe energy scale lies below the mass of the ghost, where QG makes sense as a low-energy effective theory. In \ref{Chapter_3}, we shall consider this theory to study the so-called causality constraints~\cite{Camanho:2014apa, Hollowood:2015elj, Hollowood:2016ryc, Edelstein:2016nml, deRham:2020zyh, Edelstein:2021jyu, Chen:2021bvg}.\\

\ni
As another example of a higher-curvature extension of GR, we may consider the LL theories that are unique generalizations of GR in higher dimensions ($D > 4$) having second-order field equations~\cite{Lanczos:1938sf, Lovelock:1972vz, Padmanabhan:2013xyr}. In the absence of the cosmological constant, the Lagrangian of such a theory takes the general form:
%%%%%%%%%%%%%%%%%%%%%%%%%%%
\begin{equation} \label{LL}
\mathcal{L}^{(D)}=\sum_{m=1}^{[D-1) / 2]} \alpha_{m}\, \mathcal{L}_{m}^{(D)}\ ,
\end{equation}
%%%%%%%%%%%%%%%%%%%%%%%%%%%
where $\{\alpha_m\}$ are the coupling constants with ${\alpha}_1=1$. The $m$-th order LL term, $\mathcal{L}_{m}^{(D)}$ is constructed from the $D$-dimensional curvature tensor $R^{cd}_{ab}$ and the totally anti-symmetric generalized Kronecker delta, 
%%%%%%%%%%%%%%%%%%%%%%%%%%%
\begin{equation} \label{LLm}
\mathcal{L}_{m}^{(D)}= \frac{1}{16 \pi}\, \frac{1}{2^{m}}\, \delta_{c_{1} d_{1} \ldots c_{m} d_{m}}^{a_{1} b_{1} \ldots a_{m} b_{m}}\, R_{a_{1} b_{1}}^{c_{1} d_{1}} \cdots R_{a_{m} b_{m}}^{c_{m} d_{m}}\, .
\end{equation}
%%%%%%%%%%%%%%%%%%%%%%%%%%%
The terms correspond to $m=1$ is the Einstein-Hilbert Lagrangian, whose variation gives the Einstein's field equations \ref{efe}. On the other hand, considering up to $m=2$ term leads to the Einstein-Gauss-Bonnet gravity (EGB) with Lagrangian ($\alpha_2 \equiv \lambda$)~\cite{Boulware:1985wk, Padmanabhan:2013xyr},
%%%%%%%%%%%%%%%%%%
\begin{equation}\label{EGB}
\mathcal{L}_{\text{EGB}} = \frac{1}{16\pi} \, \left[ R + \lambda\, \left(R^2 - 4\, R_{\alpha \beta} R^{\alpha \beta} + R_{\alpha \beta \gamma \delta} R^{\alpha \beta \gamma \delta} \right) \right]\, .
\end{equation}
%%%%%%%%%%%%%%%%%%%
The term inside the braces is known as the Gauss-Bonnet term, which is topological in $D=4$ and does not contribute to the field equations. We shall study the zeroth law of BH thermodynamics in LL gravity~\cite{Ghosh:2020dkk} and put stringent constraint on the EGB coupling from GW observations~\cite{Chakravarti:2022zeq} in \ref{Chapter_2} and \ref{Chapter_8}, respectively.\\

\ni
(b) {{\color{red!70!blue} Phenomenological modifications:}} In GR, several robust theoretical results like uniqueness theorems~\cite{heusler1996black, Mazur:2000pn, Robinson:2004zz, Chrusciel:2012jk, PhysRev.164.1776} and no-hair theorems~\cite{PhysRevLett.26.331, Bekenstein:1971hc, Bekenstein:1972ky} uniquely specify the structure of the metric outside any isolated BHs. However, in natural astrophysical setups, BHs are seldom isolated, and the outside environment may have nontrivial effects on the spacetime structure~\cite{Barausse:2007dy, Barausse:2014tra}. In the presence of these effects, the actual metric may show deviations from the usual Schwarzschild/Kerr BHs. And this is where an observation-oriented bottom-up approach involving phenomenological modelling of the deviated metric can be advantageous. Over the years, various such models have been proposed and studied in great detail. Among them, the Johanssen-Psaltis~\cite{Johannsen:2011dh}, Konoplya-Rezzolla-Zhidenko~\cite{Rezzolla:2014mua, Konoplya:2016jvv} and the post-Kerr~\cite{Glampedakis:2005cf} metrics are widely considered to study BH shadow and quasi-normal modes (QNMs). Unlike GR, these models come with new parameters other than the mass and spin of the BHs, which can be constrained from astrophysical observations~\cite{Ryan:1995wh, Ryan:1997hg, Gair:2007kr, Barausse:2008xv, Bambi:2011jq, EventHorizonTelescope:2021dqv, Dey:2022pmv}.\\

\ni
Intriguingly, a captivating link exists between phenomenological modeling and the previously mentioned dynamical modification. This connection arises because solutions to modified gravity theories frequently manifest in a form reminiscent of phenomenologically-motivated classes of metrics. Nevertheless, we will address them individually, as these two modifications stem from distinct perspectives, and at times, one may prove more advantageous to work with than the other. \\

\ni
There are other compelling reasons to consider phenomenological models of the spacetime metric also. Classical BHs are the only celestial objects with zero reflectivity, thanks to their all-absorbing event horizons. However, some unknown quantum gravity effects near the horizon may result in deviations from this feature~\cite{Bekenstein:1974jk, Bekenstein:1995ju, Oshita:2019sat}. For example, a model known as the "BH area-quantization" was proposed by Bekenstein and Mukhanov~\cite{Bekenstein:1974jk, Bekenstein:1995ju}, according to which the area of BH event horizons are quantized in equidistant steps resulting in selective absorption at discrete frequencies. Such a BH, dubbed as quantum BH (QBH), will have distinct GW observable signatures in both inspiral~\cite{Agullo:2020hxe, Datta:2021row, Chakravarti:2021jbv} and the late-ringdown~\cite{Cardoso:2019apo, Chakravarti:2021clm} stages of a BH-binary event. We shall consider these effects in great detail in \ref{Chapter_6} of this thesis.\\

\ni
Interestingly, phenomenological techniques can also be used to model the spacetime outside a rotating compact object without a horizon. In contrast to BHs, the spacetime outside such an object may require specifying more parameters than just mass and spin. Given this setup, one can ask many interesting questions regarding their stability under perturbations, light ring (LR) structure and the presence of hairs. We shall try to answer these questions in \ref{Chapter_4}, \ref{Chapter_5}, and \ref{Chapter_7}.\\

\ni
(c) {{\color{red!70!blue} Kinematical modifications:}} Apart from the dynamical content of field equations, spacetime also has rich kinematical structures. These structures are theory-agnostic and hence, can be studied without using any field equations. For an example, one may consider Raychaudhuri's equation~\cite{Raychaudhuri:1953yv} that encapsulate the kinematics of timelike and null congruence in a spacetime.\\

\ni
Often a great deal of kinematical structure follows directly from the underlying spacetime symmetries. Let us consider some examples to illustrate this fact. The planar nature of Schwarzschild geodesics can be traced back to the $SO(3)$ symmetry of the metric. Also, the Kerr spacetime is known to be circular~\cite{Carter:1973rla, Wald:1984rg}, having a single cross term $g_{t \phi}$ in the metric (in Boyer-Lindquist coordinates). Recently, it has been shown that the solution to any effective metric theory of gravity, which is perturbatively connected to GR, remains circular~\cite{Xie:2021bur}. Therefore, any deviation from this kinematical symmetry may signal the "non-metric" nature of the underlying theory, such as degenerate higher-order scalar-tensor (DHOST) theory, where the photons see an effective metric that depends on the scalar field~\cite{BenAchour:2016cay, Anson:2021yli}. Various observational signatures of the BH solutions of this theory have been studied recently~\cite{Takamori:2021atp}.\\

\ni
As is evident, such kinematical features have immense theoretical and observational significance, rendering the study of various modifications of spacetime symmetries and their consequences very important. For instance, the scalar wave equation in Schwarzschild/Kerr BH spacetime is known to be separable~\cite{Regge:1957td, Zerilli:1970se, Brill:1972xj, 1974ApJ193443T}, a property very useful for finding the QNM frequencies~\cite{Leaver:1985ax}. However, under a generic deviation from this kinematical structure of these metrics, the Klein-Gordon (KG) equation fails to separate in radial and angular parts, making it utterly difficult to find the QNM frequencies. In \ref{Chapter_7}, we shall discuss a systematic method of finding scalar QNMs of such BHs~\cite{Ghosh:2023etd}.
%%%%%%%%%%%%%%%%%%%%%%%%%%
\section{{\color{blue!70!brown} Some Novel Features of Modified Gravity}}
%%%%%%%%%%%%%%%%%%%%%%%%%%
In the above section, we have motivated the importance of considering modified gravity and explored several ways to perform such modifications. However, studying all possible modified theories is a Herculean task. Therefore, we must be systematic in our approach. In theory side, these modified theories of gravity and their solutions must obey certain consistency criteria, such as causality and stability under perturbations. Moreover, by demanding observational consistency, we can constrain various deviations from GR. For this purpose, we shall now turn to discuss some of the novel features of modified gravity, which will serve as a prelude to the subsequent chapters.
%%%%%%%%%%%%%%%%%%%%%%%%%%
\subsection{{\color{red!70!blue} Theoretical Explorations}}
%%%%%%%%%%%%%%%%%%%%%%%%%%
Several compact objects, both with and without horizons, are going to be the central focus of this thesis. And, we shall see that these compact objects demonstrate many general characteristics that often transcend beyond GR.\\

\ni
(A) {{\color{red!70!blue} BH thermodynamics and the Zeroth law:}} In Einstein's theory, stationary BHs possess some remarkable properties, such as rigidity and simple horizon topology~\cite{Hawking:1971vc}. However, the most interesting among these is their intriguing similarity with ordinary thermodynamic systems. In the late twentieth century, the pioneering works of Bardeen, Carter, Hawking, Bekenstein and others provide a firm mathematical ground for this analogy~\cite{Bekenstein:1972tm, Bekenstein:1973ur, Bardeen:1973gs, hawking1975}. More precisely, the study of quantum fields in BH spacetime led to Hawking's discovery of BH temperature $T$, which upheld Bekenstein's intuitive argument that a BH horizon must be attributed with an entropy $S$ proportional to its area $A$~\cite{Bekenstein:1972tm, Bekenstein:1973ur, hawking1975}:
%%%%%%%%%%%%%%%%%%
\begin{equation} \label{TS}
T = \frac{\hbar\, c^3}{2\pi G\, k_B}\, \kappa\, ,~~~~~~~\text{and}~~~~~~~ S = \frac{k_{B}\, c^3}{4 G\, \hbar}\, A\, ,
\end{equation}
%%%%%%%%%%%%%%%%%%
where we have re-introduced $\hbar$, $c$, $G$, and the Boltzmann's constant $k_B$ for the moment, and $\kappa$ is the surface gravity at the event horizon. With these definitions of temperature and entropy, it is then straightforward to show that stationary BHs of GR obey the four laws of BH thermodynamics analogous to ordinary thermodynamic systems~\cite{Bardeen:1973gs}.\\

\ni
For the purpose of this thesis, we now consider the \textit{zeroth law} in detail. It dictates that the surface gravity (and hence, the temperature) of a stationary Killing horizon is constant. Then, in GR, the rigidity theorem assures that a stationary event horizon must be a Killing horizon~\cite{Hawking:1971vc} and thus, the zeroth law also holds for BH horizons. Let us first introduce some notations to understand various proofs of this result. Let $\xi^a$ be the horizon-generating Killing vector field that is null on the Killing horizon $\mathcal{H}$. Then, the surface gravity ($\kappa$) at the horizon is given by $\xi^b_{;a}\, \xi^a = \kappa\, \xi^b$. Then, it immediately follows as a geometric fact (without using field equations) that $\kappa_{;a}\, \xi^a = 0$, i.e., surface gravity remains constant along each generator~\cite{Bardeen:1973gs, Wald:1984rg}. Therefore, the only thing that remains to be shown for completing the proof of the zeroth law is $\kappa_{;a}\, e^a_A = -R_{ab}\, \xi^a\, e^b_A = 0$, i.e., surface gravity also does not vary across (towards the transverse directions $e^a_A$) the generators.\\

\ni
There are two distinct ways to proceed from here: by assuming some symmetries of the underlying spacetime (theory-agnostic) or by using field equations along with some energy conditions on matter. For the former case, we have four known results in the literature~\cite{Racz:1995nh, heusler1996black}.\\

\ni
(i) \textbf{Assuming bifurcate Killing horizon:} If $\mathcal{H}$ consists of a bifurcation $2$-surface $\mathcal{B}$ where $\xi^a = 0$, we readily get that $\kappa_{;a}\, e^a_A = 0$ on $\mathcal{B}$. Then, the zeroth law immediately follows since $\kappa$ is Lie-dragged along the horizon generators~\cite{Racz:1995nh}. \\

\ni
(ii) \textbf{Assuming static Killing horizon:} In this case, the Killing vector field is hypersurface orthogonal and twist-free ($\omega_{a}=0$) due to Frobenius theorem. Now, we can apply $\xi_{[d}\nabla_{c]}$ (which is tangent to $\mathcal{H}$) on the defining equation for surface gravity, namely $\xi^b_{;a}\, \xi^a = \kappa\, \xi^b$.  After some algebraic manipulations, one finally obtains the variation of $\kappa$ on $\mathcal{H}$ as $4\,  \xi_{[a}\nabla_{b]}\, \kappa = - \epsilon_{abcd} \nabla^{[c} \omega^{d]}= 0$, by using $\omega_{ab}=0$. Hence, for any static spacetime, the zeroth law follows without any field equations~\cite{Racz:1995nh}.\\

\ni
(iii) \textbf{Assuming $t-\phi$ reflection isometry:} If the spacetime has $t-\phi$ reflection isometry, we must have $\xi^a \omega_a =0$ everywhere. Then, the commuting nature of the spacelike Killing vector $\psi^a$ with $\xi^a$ implies that $\kappa$ is constant along the orbits of $\psi$ (as shown earlier, this is also true along the orbits of $\xi$). Moreover, the variation of $\kappa$ orthogonal to both $(\xi^a,\, \psi^a)$ is given by $2\, \epsilon^{abcd}\, \psi_b\, \xi_c\,  \nabla_d\,  \kappa = \nabla^a\left( \xi^b \omega_b \right)$. However, this vanishes as well due to the $t-\phi$ reflection isometry, completing the proof~\cite{Racz:1995nh}.\\

\ni
(iv) \textbf{Assuming circularity:} The proof of zeroth law in this case is essentially same as the case above. Here also the commuting nature of $\psi^a$ and $\xi^a$ implies that $\kappa$ is constant along the orbits of both $\psi$ and $\xi$. Additionally, one of the integrability conditions, namely $\xi^a R_a^{\, [b} \xi^c \psi^{d]} = 0$ suggests $\nabla^a\left( \xi^b \omega_b \right)=0$. Then, following a calculation similar to case-(iii) infers that the variation of $\kappa$ orthogonal to both $(\xi^a,\, \psi^a)$ vanishes trivially on $\mathcal{H}$, which finishes the proof~\cite{Racz:1995nh, heusler1996black}.\\

\ni
However, in the absence of any extra symmetry, proof of the zeroth law requires the use of field equations. In GR, Einstein's equations help us rewrite the variation of $\kappa$ across the horizon generators in terms of the contracted stress-energy tensor, namely $\kappa_{;a}\, e^a_A = -8\, \pi\, T_{ab}\, \xi^a\, e^b_A$. The last expression vanishes if one further assumes the dominant energy condition on matter~\cite{Bardeen:1973gs}. It is evident that the same proof does not apply to modified theories of gravity since their field equations are different.\\

\ni
A generalization of this proof to arbitrary higher curvature gravity is very non-trivial. And, for a long time, the only other theory for which the zeroth law was proven is the $(R+\alpha\, R^2)$-gravity~\cite{Jacobson:1995uq}. However, we recently demonstrated its extension in the context of LL gravity~\cite{Ghosh:2020dkk}, which is the content of \ref{Chapter_2} of this thesis. Our proof supports the claim that the validity of BH thermodynamics transcends beyond GR. Such an extension gives us a helpful tool to identify a common link in the space of all possible gravity theories. Recently, a more general proof of the zeroth law to any diffeomorphism invariant theories has been proposed~\cite{Bhattacharyya:2022nqa}, stretching this link even further.\\

\ni
(B) {{\color{red!70!blue} Causality constraints in modified gravity:}}
Since the zeroth law of BH thermodynamics turns out to have a universal nature, it can not be used to constrain the structure of higher curvature terms in the gravitational action. Therefore, to achieve this task, one should use other consistency criteria, such as causality~\cite{Camanho:2014apa, Hollowood:2015elj, Hollowood:2016ryc, Edelstein:2016nml, deRham:2020zyh, Edelstein:2021jyu, Chen:2021bvg}. It has recently been used to put rather non-trivial constraints on several modified theories by Camanho, Edelstein, Maldacena, and Zhiboedov (CEMZ)~\cite{Camanho:2014apa}. They showed that the theories in which the sign of the Shapiro time shift experienced by a probe graviton depends on its polarization are acausal. Thus, a "healthy" classical theory of gravity must lead to a positive Shapiro shift for all propagations, criteria referred to by CEMZ as \textit{causality constraint}~\cite{Camanho:2014apa}.\\

\ni 
The first step of their construction is to find a shock wave solution to the underlying theory. There are two suggestive ways to produce a shock wave. Since any diffeomorphism invariant theory of gravity has Lorentz symmetry, one may boost a BH solution to a high velocity $v$, producing a shock in the limit $v \to 1$~\cite{Aichelburg:1970dh}. However, there is a more direct way to produce a shock wave solution by solving the field equations in the presence of singular stress-energy tensor $T_{uu}= - P_u\, \delta(u)\, \delta^{(D-2)}(\vec{x})$ sourced by a high energy particle with momentum $P^u$ moving along a null direction $v$ (localized at $u=0, x^i =0$). One can then solve Einstein's equations to find the following shock wave solution in GR~\cite{Camanho:2014apa}:
%%%%%%%%%%%%%%%%%%
\begin{equation} \label{grshock}
ds^2 = -du\, dv + \overline{h}_0(u, x_i)\ du^2 + \sum_{i=2}^{D-2} (dx_i)^2\, ,
\end{equation}
%%%%%%%%%%%%%%%%%%
where the profile function $\overline{h}_0(u, x_i)$ is given by
%%%%%%%%%%%%%%%%%%%
\begin{equation} \label{gr_prof}
\overline{h}_0(u, x_i) = \frac{4\ \Gamma \left(\frac{D-4}{2}\right)}{\pi^{\frac{D-4}{2}}} \ \delta(u)\ \frac{G \lvert P_u \rvert}{r^{D-4}}\, .
\end{equation}
%%%%%%%%%%%%%%%%%%%%
%%%%%%%%%%%%%%%%%%%%
\begin{figure}[!htp]
    \centering
    \includegraphics[scale=0.6]{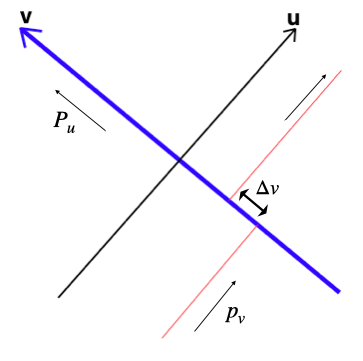}
    \caption{\textbf{A test particle crossing the shock.} The shock is confined at $(u=0, x^i=0)$. A particle crosses the shock with impact parameter $r=b$ and undergoes a Shapiro time delay $\Delta v > 0$.}
    \label{shoch}
\end{figure}
%%%%%%%%%%%%%%%%%%
Let us now consider a test particle of momentum $p_v$ crosses this shock with an impact factor $r=\sqrt{x^i\, x_i} = b$ as depicted in \ref{shoch}. In this case, a simple manipulation of the geodesic equation suggests that the particle will suffer a Shapiro time delay as it crosses the shock~\cite{Camanho:2014apa},
%%%%%%%%%%%%%%%%%%%%%
\begin{equation} \label{gr_shift}
(\Delta v)_{\textrm{GR}}\,=\, \frac{4\ \Gamma\left(\frac{D-4}{2}\right)}{\pi^{\frac{D-4}{2}}}\ \frac{G \lvert P_u \rvert}{b^{D-4}} > 0\, .
\end{equation}
%%%%%%%%%%%%%%%%%%%%%
A positive value of $(\Delta v)_{\textrm{GR}}$ for $D>4$ implies causality in the sense of Gao and Wald~\cite{Gao:2000ga}, which dictates that in a causal theory, the asymptotic structure of the spacetime fixes the maximum propagation speed which sets an upper limit on speeds for propagation in the bulk. Interestingly, for $D=4$, the Shapiro time shift is unphysical as it depends on an arbitrary IR cutoff associated with the shock~\cite{Dray:1984ha}. Similar calculations can also be performed for the propagation of scalar and graviton fields, which also shows positivity of the Shapiro time shift and hence, implies the causal nature of GR~\cite{Camanho:2014apa}.\\

\ni
However, for EGB gravity with Lagrangian given by \ref{EGB}, CEMZ demonstrated that the Shapiro time shift of graviton can always be made negative for either sign of the Gauss-Bonnet coupling constant $\lambda$, by choosing suitable polarization modes for the graviton~\cite{Camanho:2014apa}. Hence, they concluded that EGB theory is acausal since it supports superluminal propagation unless we set $\lambda=0$, reducing the theory to GR. In other words, EGB theory can not be considered a consistent classical theory of gravity as long as $\lambda \gg \ell_p^2$, where $\ell_p$ denotes the Planck length~\cite{Camanho:2014apa}.\\

\ni
An important question readily follows: Are all higher curvature theories of gravity acausal in the sense of CEMZ? We consider this question in \ref{Chapter_3} of this thesis. As we shall see, there are at least two challenges that we have to face in other modified theories. First, unlike EGB gravity which has the same shock wave solution as GR, a general modified theory will have a different shock wave solution. Moreover, solving the graviton equation of motion gets more involved due to the presence of higher-order derivatives. However, thanks to a known exact shock wave solution~\cite{Campanelli:1995ex}, we shall show that QG is free from such causality issues~\cite{Edelstein:2021jyu}. In fact, we shall propose a general class of higher curvature theories (of which QG is a member) having this property.\\

\ni
(C) {{\color{red!70!blue} Light rings of stationary spacetimes:}}
So far, our discussion was mostly centred around exploring different features of modified gravity, with minimal focus on their corresponding solutions. We now shift our attention towards examining certain intrinsic characteristics of compact objects within these modified theories, signatures that are universal and not tied to any specific theory. These compact objects are unique laboratories to probe strong gravity regimes, and understanding their features may also provide us with useful information about the underlying spacetime.\\

\ni
Various observational signatures like QNMs and shadows of such compact objects depend crucially upon an essential ingredient associated with the spacetime outside them, namely light rings (LRs). They mark the locations where the gravitational deflection of light becomes so extreme that photons can stay in circular orbits. For example, in the case of Schwarzschild BH given by \ref{Sch}, the LR is located at $r=3M$, which can be easily calculated using the null geodesic equation. A similar calculation in Kerr spacetime also reveals the presence of two LRs outside its event horizon, respectively co-rotating and counter-rotating with the BH. However, do BHs of any theory of gravity have such LRs outside them? In $4$-dimensions, a positive answer to this vital question has been provided recently~\cite{Cunha:2020azh}. The authors use some ingenious topological analysis to show that all non-extremal, stationary, axisymmetric and asymptotically flat BH spacetime with topologically spherical Killing horizon admits at least one LR outside the event horizon in each rotation sense. Later, this result was extended for static BHs with asymptotically de-Sitter (dS) and anti-de-Sitter (AdS) cases~\cite{Wei:2020rbh}.\\

\ni
However, in general, the presence of LRs does not uniquely specify the nature of the central object. It is because, apart from BHs, several known horizonless compact objects, such as boson stars and gravastars, also possess LRs around them. In this context, it has also been shown that if the central horizonless object has one LR (referred to as ultra-compact objects or UCOs in short), then an even number of them must exist~\cite{Cunha:2017qtt}. This proof does not require any field equations and is also topological. In \ref{Chapter_4} of this thesis, we shall establish a novel result, which shows that for rotating horizonless objects having ergoregion, their assumption of having one LR is redundant, and at least one light ring outside the ergoregion must exist~\cite{Ghosh:2021txu}.\\

\ni
Besides these important results, it has been observed that LRs in horizonless compact object is also connected with their stability. Unlike BHs, horizonless objects have a reflective surface that may considerably slow down the dissipation of perturbations present in spacetime. The situation becomes worse in the presence of a region outside the horizon that can potentially trap such perturbing modes. Then, perturbations can sustain inside such regions for a long time without decay, rendering the spacetime unstable. One such instability is caused if an ergoregion exists outside a horizonless object, where perturbations can grow indefinitely due to negative energy states and may ultimately destroy the central object~\cite{Friedman}. Apart from this ergoregion instability, the presence of a stable light ring (minimum of the effective potential) can also provide a trapping region for the perturbations, causing non-linear effects to inflict LR instability~\cite{Cardoso:2014sna, Cunha:2022gde, Zhong:2022jke}. Though there is no general result, a few recent numerical studies~\cite{Cunha:2022gde, Zhong:2022jke} have shown that the instability timescale can be as small as a few milliseconds for some known horizonless objects such as boson stars.\\

\ni
These results suggest that all horizonless compact objects may suffer from such instability for having stable LRs outside them. This strongly supports the so-called “BH hypothesis”, which claims that the objects with LRs are BHs. If that is the case, then EHT observations~\cite{Akiyama:2019cqa, Akiyama:2019bqs, Akiyama:2019fyp, Akiyama:2019eap} of shadows can be considered as direct evidence for the existence of BHs in our universe.\\

\ni
(D) {{\color{red!70!blue} Hairy BHs and absence of short hairs:}} Apart from causing extreme gravitational lensing discussed above, a strong gravity regime outside a BH spacetime may also give rise to exotic phenomena such as the growth of "hairs". Unlike the Schwarzschild/Kerr solutions of GR, BHs in modified gravity may have additional hairs besides mass and angular momentum. Many hairy modifications of Kerr BHs are known in the literature, and extensive studies have been performed to determine their observational signatures~\cite{Ryan:1995wh, Ryan:1997hg, Gair:2007kr, Bambi:2011jq, Isi:2019aib}. However, the presence of extra hairs can be suitably captured via observations probing only the far-away regions of spacetime if the hairs extend sufficiently outside the horizon. Therefore, it is crucial to investigate whether BHs of modified gravity can grow short hairs confined solely to the near-horizon regions.\\

\ni
In fact, even in GR, BH solutions with exotic matter content can produce hairs, of which BHs with Yang-Mills~\cite{Bizon:1990sr}, dilatonic~\cite{Kanti:1995vq}, skyrmionic~\cite{Luckock:1986tr} and axionic~\cite{Campbell} hairs are worth mentioning. However, it has been shown that any static BH solutions of GR can not have short hairs if the matter obeys the weak energy condition (WEC) and the non-positive trace condition~\cite{Nunez:1996xv, Hod:2011aa}. More explicitly, the hairs must extend to the innermost LR of the spacetime~\cite{Hod:2011aa}. The proof of this theorem also requires other non-trivial assumptions such as spherical symmetry, asymptotic flatness and dimensionality of the spacetime being four. This result has immense observational importance since it implies that for detecting the presence of hair around BHs, it is sufficient to probe till the near-LR region alone. In other words, if BHs in GR have hairs, their signatures will be captured in observations such as BH shadow and gravitational lensing.\\

\ni
In \ref{Chapter_5} of this thesis, we shall discuss a novel generalization of this important theorem~\cite{Ghosh:2023kge}. Most notably, our generalization does not use the field equations and is valid in arbitrary spacetime dimensions ($D \geq 4$). This shows the absence of BH short hairs is a theory-agnostic feature that transcends beyond GR.
%%%%%%%%%%%%%%%%%%%%%%%%%%
\subsection{{\color{red!70!blue} Confrontation with Observations}}
%%%%%%%%%%%%%%%%%%%%%%%%%%
Besides theoretical consistency, modified theories of gravity should also be confronted with numerous observations providing a compelling way to constrain any possible deviation from GR. For example, the weak-field approximation of several modified theories shows Yukawa-type correction over the usual Newtonian potential~\cite{Capozziello:2004sm, Capozziello:2009ss, Berry:2011pb}. Such small-scale correction in Newton's law has been tested in the laboratory, resulting in stringent bounds on the higher curvature couplings~\cite{PhysRevD.70.042004,PhysRevLett.98.021101}. Apart from these local experiments, stellar and cosmological observations can also lead to strong bounds on similar higher curvature couplings~\cite{Berry:2011pb, PhysRevD.86.081504, Berti:2015itd, Maselli_2017, Ayzenberg:2017ufk}. Whereas most of these experiments/observations are concerned with testing gravity in weak field limits, it is also desirable to test the validity of modified theories in strong field regimes. The recent GW observations of merging binary BHs are most suitable for this task. In such extreme gravity, higher-order post-Newtonian (PN) corrections capturing the effects of various non-GR terms can also have significant consequences over various observables, making it a natural test-bed for modified theories~\cite{Carson:2020rea}.\\

\ni
A binary coalescence event has three distinct phases -- inspiral, merger and ringdown. Among them, the merger phase gives rise to highly non-linear gravitational effects, and there is no analytical treatment available. One needs the full machinery of numerical relativity to understand its features. In contrast, there are well-known analytical/semi-analytical tools to study various observables, such as GW phasing in the inspiral phase and emission of QNMs in the ringdown phase. Moreover, GWs emitted in these stages also carry valuable information about the merging objects and the final remnants. For example, any non-Kerr modifications and/or deviations in near-horizon physics can have observable effects on the inspiral and ringdown GW signals. These effects can be captured via several well-known tests, such as the inspiral-merger-ringdown (IMR) consistency test (tests GR by finding consistency between the inspiral and ringdown waveform)~\cite{Abbott_2016, PhysRevD.94.021101, Ghosh_2017, Carson:2020cqb, PhysRevD.101.084050}, parameterized PN tests (for constraining non-GR deviations)~\cite{Chamberlain_2017, Yunes:2009ke, PhysRevD.98.084042, Carson:2019yxq, Tahura:2019dgr}, and the propagation test (tests for modified dispersion relations and variation in GW speed)~\cite{Will:1997bb, Mirshekari:2011yq, LIGOScientific:2020tif, LIGOScientific:2017zic}.\\

\ni
(A) {{\color{red!70!blue} BH area quantization and its observational signatures:}} Let us consider a binary of two BHs with total mass $m$. In the inspiral phase, the binary orbit evolution and the emitted GWs can be expressed in terms of the following energy balance and GW phasing equations~\cite{Tichy:1999pv}:
%%%%%%%%%%%%%%%%%%%%%
\begin{equation} \label{flux}
\frac{dE(v)}{dt}=-F(v),\, \, \frac{d\Phi(v)}{dt}=\pi\, f\, ,
\end{equation}
%%%%%%%%%%%%%%%%%%%%%
where $E(v)$ is orbital energy, $F(v)$ is the GW flux, $\Phi(v)$ is the orbital phasing, and $v = (\pi\, m\, f)^{1/3}$ is the reduced velocity with $f$ being GW frequency. Since classical BHs absorb at all frequencies, $F(v)$ contains an extra part in addition to the usual flux loss at infinity. This additional flux term is responsible for "tidal heating", a phenomenon well-understood within the analytical framework of PN expansion~\cite{Tichy:1999pv}. However, in the presence of some unknown quantum effects near BHs, its horizon may acquire a feature akin to atoms and only absorbs at specific characteristic frequencies~\cite{Bekenstein:1974jk, Bekenstein:1995ju, Oshita:2019sat}. For instance, BHs may have quantized horizon area, a proposal first put forward by Bekenstein and Mukhanov~\cite{Bekenstein:1974jk, Bekenstein:1995ju}. Using this model, it has been recently argued that area quantization might have detectable imprints on classical observables such as GW phasing~\cite{Agullo:2020hxe, Datta:2021row}. In the second part of this thesis, we shall elaborate on various models of area quantized BHs and study their associated observational signatures~\cite{Chakravarti:2021jbv}.\\

\ni
Similar to the inspiral phase, the ringdown GW signal is also susceptive to modifications in near-horizon physics. In particular, due to quantized area, BH horizons would absorb selectively, and the reflected modes at the horizon will show up in the late time signal as repeated echoes~\cite{Cardoso:2019apo}. Any future detection of such echoes in GW observations can provide a powerful tool for probing quantum effects in strong gravity regimes. Such effects of area quantization will be thoroughly discussed in \ref{Chapter_6}~\cite{Chakravarti:2021clm}.\\

\ni
(B) {{\color{red!70!blue} BH perturbations and associated QNMs:}} Apart from BH coalescence events, near-horizon structure can also be probed by studying QNMs associated with the decay of perturbations in BH spacetimes. In GR, finding out QNM frequencies are particularly easy due to the separable nature of perturbation equations~\cite{Regge:1957td, Zerilli:1970se, Brill:1972xj, 1974ApJ193443T, Leaver:1985ax}. However, this is not true for all BH solutions of general modified theories of gravity. In such a scenario, finding the QNM modes become utterly difficult, and one needs a unified way to handle such non-separable perturbations equations.\\

\ni
In \ref{Chapter_7}, we shall demonstrate a general method of finding scalar QNMs in a BH spacetime perturbatively "close" to Schwarzschild/Kerr metric~\cite{Ghosh:2023etd}. Using this method, we will calculate QNM frequencies and study the stability of various phenomenological BHs, such as Schwarzschild and Kerr BHs with an anomalous quadrupole moment~\cite{Glampedakis:2005cf}. Our method also shows a universal structure of the eikonal QNMs for such deformed Schwarzschild/Kerr BHs.\\

\ni
(C) {{\color{red!70!blue} Constraining the Topological EGB coupling using GW Observation:}}  Besides its crucial application in BH stability analysis, BH perturbation theory is especially useful to determine the mass and spin of the remnant BH formed as an end state of binary mergers. Recently, this technique has been used~\cite{Isi:2020tac} for GW150914 to test one of the most important properties of GR BHs, namely Hawking’s area theorem~\cite{Hawking:1971vc, Bardeen:1973gs}. In the context of BH mergers, the global version of this theorem dictates that the final BH's area must be larger than the total initial area of the component Kerr BHs in the inspiral phase. Moreover, in GR, due to entropy-area proportionality, an increase of BH area also implies an increase of BH entropy during merger. Therefore, the aforesaid test of the area law can also be viewed as a test of the global version of the BH second law.\\

\ni
However, the inclusion of a Gauss-Bonnet term in the gravitational action will modify the entropy-area relationship~\cite{Jacobson:1993xs, Liko:2007vi}. As a result, irrespective of its topological nature (that do not alter the gravitational dynamics and Kerr BH remains a solution), the validity of area theorem does not necessarily imply the validity of the second law unless the value of the coupling is constrained. In \ref{Chapter_8} of this thesis, we shall discuss how this concept can be used to put a stringent bound on the topological Gauss-Bonnet coupling~\cite{Chakravarti:2022zeq}.
%%%%%%%%%%%%%%%%%%%%%%%%%%
\section{{\color{blue!70!brown} A Brief Overview of the Thesis}}
%%%%%%%%%%%%%%%%%%%%%%%%%%
This thesis discusses various novel studies on modified gravity and the properties of compact objects. We aim to use both theoretical and observational tools to probe possible departures from GR. On the theoretical side, we consider BH thermodynamics, stability of compact objects, presence of BH hairs, and the issue of causality that may provide valuable input towards the ultimate quantum theory of gravity. Moreover, on the observational side, we employ GW observations and BH perturbation theory to explore new aspects of gravitational physics and put stringent bounds on the beyond-GR parameters. A chapter-wise overview of the thesis is given below.\\

\begin{center}
\textbf{Part I. Theoretical Studies on Modified
Gravity}
\end{center}

\begin{itemize}
\item In \ref{Chapter_2}, we generalize the zeroth law of BH thermodynamics in the context of EGB gravity. An extension to general LL theory is also discussed. 

\item \ref{Chapter_3} discusses the causality constraint in QG. In particular, we show that, unlike EGB theory, QG is free from any causality issue.

\item \ref{Chapter_4} deals with LR structure of stationary spacetimes. We prove a novel theorem regarding the existence of LRs outside the ergoregion of such a spacetime.

\item In \ref{Chapter_5}, we present a theory-agnostic generalization of the no-short hair theorem of GR. Various consequences of this novel theorem are also discussed. \\

\end{itemize}

\begin{center}
\textbf{Part II. Observational Signatures of Modified Gravity}
\end{center}

\begin{itemize}

\item \ref{Chapter_6} discusses various models of area-quantized BHs and their GW signatures. In particular, we consider its effects on tidal heating in the inspiral phase, and the possible echo signals in the ringdown phase of a binary.

\item In \ref{Chapter_7}, we present a general method of calculating scalar QNMs in a non-Kerr spacetime. A few such toy BH models are also used to illustrate our method.

\item \ref{Chapter_8} aims to find a stringent bound on the topological Gauss-Bonnet coupling using GW observations. For this purpose, the recent verification of the area law for the event GW170817 will be used. 

\end{itemize}

% Main chapter title

\cleardoublepage
\begin{center}
\vspace*{\stretch{1}}
\Huge{\textbf{Part I.}}\\
\Huge{\textbf{Theoretical Studies on Modified Gravity}}
\vspace*{\stretch{1}}
\end{center}

 \clearpage

\chapter{{\color{red!60!black}Black Hole Zeroth Law in Higher Curvature Gravity} }\label{Chapter_2}
\large
\textbf{This Chapter is based on the work: Phys. Rev. D 102 (2020) 10, 101503 (Rapid Communication) by R. Ghosh and S. Sarkar}~\cite{Ghosh:2020dkk}.\\

\ni
As discussed in the previous chapter, stationary BHs of GR can be attributed with properties, such as temperature and entropy, akin to ordinary thermodynamic systems in equilibrium. This intriguing similarity has been mathematically formulated in terms of four fundamental laws governing the BH mechanics~\cite{Bekenstein:1972tm, Bekenstein:1973ur, Bardeen:1973gs, hawking1975}. Among them, the zeroth law asserts that the surface gravity of a stationary Killing horizon is a constant, which once associated with the Hawking temperature given by~\ref{TS} becomes the zeroth law of BH thermodynamics. The validity of this law is necessary for the formulation of the remaining three laws and, thereby, for interpreting stationary BHs as thermodynamic systems.\\

\ni
Apart from its well-known proof in GR~\cite{Bardeen:1973gs}, there are several theory-agnostic proofs of the zeroth law without assuming any field equations. We have sketched some of these proofs in the previous chapter, so we shall not discuss them here. In contrast, this chapter aims to extend the proof of the zeroth law in LL gravity represented by the Lagrangian in \ref{LL}. It is a unique class of ghost-free higher curvature theories with second-order field equations in time. Interestingly, both the first law~\cite{Wald:1993nt, Iyer:1994ys} and the quasi-stationary second law~\cite{Chatterjee:2011wj, Kolekar:2012tq, Sarkar:2019xfd} have been generalized in this theory, making it a very natural candidate for studying the zeroth law. Such an extension also provides a unique tool to probe the unified nature of BH thermodynamics beyond GR, helping us understand the characteristics of stationary BHs in some well-motivated modified theories.\\

\ni
For a long time, $(R+\alpha\, R^2)$-gravity was the only theory beyond GR for which a proof of zeroth law existed~\cite{Jacobson:1995uq}. Also, it is worth mentioning that a few years ago an attempt was made to generalize the zeroth law in LL gravity, which let the
authors claimed that the constancy of surface gravity does not hold in general~\cite{Sarkar:2012wy}. However, we shall show that their claim is not completely correct and such a generalization is indeed possible. In particular, we prove that \textit{the surface gravity of a stationary Killing horizon in LL gravity is constant, provided the matter obeys the dominant energy condition and all geometric quantities have Taylor-expandable structures in series of coupling constants (requiring smooth limits to the corresponding GR quantities) at the horizon}. If one further assumes the rigidity theorem~\cite{Hawking:1971vc}, our result suggests the validity of the zeroth law for stationary BHs in LL gravity.\\

\ni
We shall end this chapter by sketching a possible way to generalize our result in other modified theories of gravity. Interestingly, it will turn out that the validity of the zeroth law for stationary Killing horizons puts severe restrictions on the form of the field equations at the horizon. Recently, such form is shown to be valid in the scalar-tensor~\cite{Dey:2021rke}, Horndeski~\cite{Sang:2021rla}, and scalar-hairy Lovelock gravity~\cite{Fang:2022nfa}, and our proof of the zeroth law has been extended in these theories. In fact, it has also been proven that, under a further assumption that the underlying metric has a smooth Taylor-expandable structure in terms of the spacetime coordinates, the aforesaid form of the field equations follows directly as a theory-agnostic fact~\cite{Bhattacharyya:2022nqa}.
%%%%%%%%%%%%%%%%%%%%%%%%%
\section{{\color{blue!70!brown} Geometry of Stationary Killing Horizons}}
%%%%%%%%%%%%%%%%%%%%%%%%%
For the purpose of discussing our result, it is useful to setup the notations first and study some properties of a stationary Killing horizon $\mathcal{H}$. It is a null hypersurface generated by a timelike Killing vector field $\xi^a$, which is null on $\mathcal{H}$. In this chapter, instead of Greek letters, we shall denote the $D$-dimensional spacetime indices by lowercase English letters. Then, a quantity of interest, namely the surface gravity ($\kappa$) at the Killing horizon can be defined by the geodesic equation, $\xi^b_{;a}\, \xi^a = \kappa\, \xi^b$. As a result, one can readily show (without using any field equations) that $\kappa_{;a}\, \xi^a = 0$, i.e., surface gravity remains constant along each generator~\cite{Bardeen:1973gs, Wald:1984rg}. However, $\kappa$ may still vary from one generator to the other.\\

\ni
To calculate the variation of surface gravity across the generators, we can construct a basis $\{\xi^a, N^a, e^a_A \}$ on $\mathcal{H}$. Here, $N^a$ is a null vector satisfying, $\xi^a N_a = -1$; $ N^a N_a = 0$. And, $\{ e^a_A \}$ are $(D-2)$ spacelike vectors along the transverse directions satisfying, $e^b_A \xi_b = e^b_A N_b = 0.$ One can decompose the spacetime metric in this basis as follows:
%%%%%%%%%%%%%%%%%%%%%%%%%%%%%%%%%%%%
\begin{equation} \label{ZLmetric}
g^{a b}=-\xi^{a} N^{b}-N^{a} \xi^{b}+\sigma^{A B} e_{A}^{a} e_{B}^{b}\ ,
\end{equation}
%%%%%%%%%%%%%%%%%%%%%%%%%%%%%%%%%%%%
where $\sigma^{AB}$ is the $(D-2)$-dimensional transverse metric on any spacelike slice of the Killing horizon $\mathcal{H}$.\\

\ni
Note that due to the stationary nature of the spacetime, the horizon generators of $\mathcal{H}$ are endowed with zero shear and expansion parameters. The congruence is also rotation-free due to its hypersurface orthogonal character. Thus, Raychaudhuri's equation implies that on the horizon, $R_{a b} \xi^{a} \xi^{b}=0$, and $R_{a b c d}\ e_{A}^{a} e_{B}^{b} e_{C}^{c} \xi^{d}=0$~\cite{Wald:1993nt, Vega:2011ue}. Using these results, one can now compute the variation of $\kappa$ across the generators as,
%%%%%%%%%%%%%%%%%%%%%%%%%%%%%%%%%%%%
\bea \label{ZLacross}
\kappa_{; a}\, e_{A}^{a}=-R_{a r p q}\ \xi^{r} N^{p} \xi^{q} e_{A}^{a}=-R_{a b}\ \xi^{a} e_{A}^{b}\ .
\eea
%%%%%%%%%%%%%%%%%%%%%%%%%%%%%%%%%%%%
Therefore, the only remaining task to establish the zeroth law is to show that the R.H.S of \ref{ZLacross} vanishes identically. This is achieved in GR, by using Einstein's field equations and the dominant energy condition on matter,
%%%%%%%%%%%%%%%%%%%%%%%%%%%%%%%%%%%%
\bea \label{ZLdom}
\kappa_{; a}\, e_{A}^{a}= 8\, \pi\, j_{a}\, 
 e^{a}_{A }\ ,\ \ \textrm{with}\ \ j_{a}=-T_{ab}\, \xi^b \propto \xi_a\ .
\eea
%%%%%%%%%%%%%%%%%%%%%%%%%%%%%%%%%%%%
Then, the zeroth law $\kappa_{; a}\, e_{A}^{a} = 0$ follows directly from the orthogonality property, $e^b_A\, \xi_b =0$. It is evident that the same proof does not apply to modified theories with different field equations. Moreover, the complicated nature of modified field equations may make such generalization to arbitrary higher curvature gravity very non-trivial.
%%%%%%%%%%%%%%%%%%%%%%%%%
\section{{\color{blue!70!brown} Zeroth Law in LL gravity: A Previous Attempt}}
%%%%%%%%%%%%%%%%%%%%%%%%%
In \ref{Chapter 1}, we gave a short review of LL gravity, see \ref{LL} and \ref{LLm} for reference. We now extensively use these equations to discuss a previous attempt to generalize the zeroth law in LL gravity as presented in Ref.~\cite{Sarkar:2012wy}, where the authors claimed that such an extension might not be possible in general. However, we will show by presenting such a generalization that their claim is not entirely correct.\\

\ni
As shown earlier, the zeroth law in GR follows as a direct consequence of the field equations and the dominant energy condition on matter. However, this is not the case in LL gravity. Even using the associated field equations and the dominant energy condition, the authors of Ref.~\cite{Bhattacharyya:2022nqa} could only arrive at the equation,
%%%%%%%%%%%%%%%%%%%%%%%%%
\bea \label{ZLacrossf}
\kappa_{; a}\, e_{B}^{a}= - \sum_{m \geq 2}\ 2^m\, \alpha_{m}\, {}^{(D-2)} E_{(m-1) B}^{A}\ e_{A}^{a}\, R_{a r p q}\, \xi^{r}\, N^{p}\, \xi^{q}\ .
\eea
%%%%%%%%%%%%%%%%%%%%%%%%%
Here, ${}^{(D-2)}E_{(m-1)B}^{A}$ represents the equation of motion corresponding to $(m-1)$-th order LL theory constructed from the intrinsic curvatures ${}^{(D-2)}R^{ABCD}$ and the metric $\sigma_{AB}$ of the horizon cross-section. Also, the zeroth law in GR forces the $m=1$ term to vanish automatically. Now, we want to show that the same is true for any general theory in the LL class. However, for simplicity, let us first consider the case for EGB gravity, keeping only the first term ($m=2$) in the R.H.S of~\ref{ZLacrossf}.
%%%%%%%%%%%%%%%%%%%%%%%%%
\section{{\color{blue!70!brown} Zeroth Law in EGB Theory}}
%%%%%%%%%%%%%%%%%%%%%%%%%
In this special case, one can considerably simplify \ref{ZLacrossf} by noting that ${}^{(D-2)}E_{(1)B}^{A}$ is nothing but the intrinsic Einstein tensor $G_{B}^{A}$ of the cross-section. Then, using \ref{ZLacross}, we finally get the following result,
%%%%%%%%%%%%%%%%%%%%%%%%%
\bea \label{ZLm2}
\left(\delta_{B}^{A}-4\, \lambda\, G_{B}^{A}\right) T_A=0\ ,
\eea
%%%%%%%%%%%%%%%%%%%%%%%%%
where we have used the notation $\kappa_{; a}\, e_{A}^{a} \equiv T_A$ to save some writing and renamed the EGB coupling constant $\alpha_2$ as $\lambda$. Recall that our goal is to prove $T_A = 0$ everywhere on $\mathcal{H}$.\\

\ni
Before we move on the general proof, it is interesting to consider the case when $\lambda$ is small compare to all associated length scale of the problem. In such a scenario, the zeroth law follows directly from the fact the $(D-2)$-dimensional matrix $M^B_A \equiv \left(\delta_{B}^{A}-4\, \lambda\, G_{B}^{A}\right)$ is invertible, making $T_A=0$ as the only solution of \ref{ZLm2}. To see this, we first write this determinant as, $\operatorname{det}\left(M_{B}^{A}\right) = 1+2\, \lambda\, (D-4)\ {}^{(D-2)}R + \mathcal{O}(\lambda^2)$. At $D=5$, its value matches exactly with the entropy density of the Gauss-Bonnet theory, which is certainly non-zero irrespective of the sign of $\lambda$~\cite{Padmanabhan:2013xyr}. However, at $D>5$, this argument can only be extended for positive values of $\lambda$. It is because of the following inequality satisfies by the determinant, $\operatorname{det}\left(M_{B}^{A}\right) = 1+2\, \lambda\, {}^{(D-2)}R + 2\, \lambda\, (D-5)\ {}^{(D-2)}R > 0$, for all $D \geq 5$ and $\lambda \geq 0$.\\

\ni
Now, we are in a position to discuss our general strategy to prove the zeroth law in EGB theory for arbitrary values of the coupling constant $\Lambda$, which will be valid at arbitrary spacetime dimensions $D \geq 5$. For this purpose, we will rewrite \ref{ZLacrossf} in a suggestive way,
%%%%%%%%%%%%%%%%%%%%%%%%%%%
\bea \label{ZLre}
4\, \lambda\, G_{B}^{A}\, T_A = T_B\ .
\eea
%%%%%%%%%%%%%%%%%%%%%%%%%%%
If we now assume that the EGB theory has a "smooth" limit to GR (the so-called Einstein branch), then various geometric quantities such as $G_{B}^{A}$ and $T_A$ must be Taylor expandable in series of $\lambda$. That is, we should be able to write, $G^A_B= (G_0)^A_B+\lambda (G_1)^A_B+\lambda^2 (G_2)^A_B+ \ldots$, and similarly, $T_A= \lambda (T_1)_A + \lambda^2 (T_2)_A+\ldots$, for all $\lambda$. Here, we have used the fact that zeroth law in GR implies $(T_0)_A$ vanishes identically. Then, substituting these expressions in \ref{ZLre} and comparing like powers of $\lambda$, it is easy to check that $(T_1)_A = (T_2)_A= \ldots = 0$. It completes our proof of the zeroth law in EGB gravity.
%%%%%%%%%%%%%%%%%%%%%%%%%
\section{{\color{blue!70!brown} Generalization for General LL Theories}}
%%%%%%%%%%%%%%%%%%%%%%%%%
In this section, we shall discuss a generalization of our proof of zeroth law in the context a general LL theory. This proof will run parallel to the EGB case presented above. For an arbitrary LL theory, \ref{ZLre} takes the general form,
%%%%%%%%%%%%%%%%%%%%%%%%%%
\bea \label{ZLmg2f}
2^{j}\, \alpha_j\, L_{B}^{A}\, T_A = T_B\, .
\eea
%%%%%%%%%%%%%%%%%%%%%%%%%%
The spatial tensor $L^{A}_{B}$ is given by 
%%%%%%%%%%%%%%%%%%%%%%%%%%
\bea \label{ZLL}
L^{A}_{B}= \sum_{m \geq 2}\, 2^{m-j}\, \beta_{m j}\, {}^{(D-2)}E_{(m-1) B}^{A}\, ,
\eea
%%%%%%%%%%%%%%%%%%%%%%%%%%
where $\beta_{m j}$ denotes the ratio $\alpha_{m}/\alpha_{j}$, with $\alpha_{j}$ being the first non-zero coupling constant in the set $\{\alpha_m \mid {m \geq 2}\}$. We can now proceed in the same way as before and extend the zeroth law for the full LL class of theories.
%%%%%%%%%%%%%%%%%%%%%%%%%
\section{{\color{blue!70!brown} Summary}}
%%%%%%%%%%%%%%%%%%%%%%%%%
In this chapter, we have discussed a generalization of the zeroth law for LL class of theories by using the corresponding field equations and dominant energy condition on matter. In addition, we need to assume an extra reasonable but non-trivial condition that geometric quantities at the Killing horizon are Taylor expandable in series of the coupling constants ($\alpha_i$). This condition restricts our result to the so-called Einstein branch of LL gravity, which has a smooth $\alpha_i \to 0$ limit to GR. Moreover, the applicability of our result for BH event horizon is hinged upon the validity of Hawking's strong rigidity theorem~\cite{Hawking:1971vc} in LL gravity, whose proof is not known yet. Though, given stark similarities with GR, we expect that an analogous theorem could be extended in LL theories as well.\\

\ni
Apart from LL gravity, our method has also been used to prove zeroth law in other modified theories, such as Horndeski gravity~\cite{Sang:2021rla}, and scalar-hairy Lovelock theory~\cite{Fang:2022nfa}. As an explicit example, let us consider the former case~\cite{Sang:2021rla}, where the authors assumed various geometric quantities on the horizon are Taylor expandable in series of the Horndeski coupling constant. Then, proceeding in parallel to our proof, the zeroth law is shown to hold for stationary Killing horizons. For other generalizations of zeroth law, another recent result of Ref.~\cite{Bhattacharyya:2022nqa} is worth mentioning. Here, besides considering smooth structures in terms of coupling constants, the authors assumed the near-horizon regimes admit a Taylor-expandable structure in terms of spacetime coordinates. Then, as a result of boost invariance, several components of the near-horizon field equations vanish and the zeroth law follows immediately~\cite{Bhattacharyya:2022nqa}.\\

\ni
We emphasis that our proof is not, in general, valid for non-Einstein branches (in which $\alpha_i \to 0$ limit is ill-defined) of LL gravity. For example, there is a known spherically symmetric solution to EGB gravity that does not have a smooth $\lambda \to 0$ limit~\cite{Boulware:1985wk}. Nevertheless, it is still possible to extend our result in a particular scenario where geometric quantities on the horizon have Taylor-expandable form in series of $1/\alpha_i$ and GR is recovered in the limit $\alpha_i \to \infty$~\cite{Boulware:1985wk}. For example, let us consider such a branch in EGB gravity. Then, we can write, $G^A_B= (G_0)^A_B+\lambda^{-1} (G_1)^A_B+\lambda^{-2} (G_2)^A_B+ \ldots$, and similarly, $T_A= \lambda^{-1} (T_1)_A + \lambda^{-2} (T_2)_A+\ldots$, for all $\lambda \neq 0$. Substituting these expressions in \ref{ZLre} and comparing the like powers of $1/\lambda$, we again achieve the zeroth law since $(T_1)_A = (T_2)_A= \ldots = 0$.\\

\ni
Our proof clearly reflects the special structure of LL gravity that allows us to write the equations regarding the variation of surface gravity across the horizon generators in a suggestively simple way, see for reference \ref{ZLm2} and \ref{ZLmg2f}. This simplicity can be traced back to the fact that LL field equations does not contain any higher order terms involving derivatives of curvature tensors. For obvious reasons, such structure will fail to hold in other modified gravity, making it extremely difficult to generalize our proof for those theories without any further assumptions. However, one may still ask an interesting question: What should be structure of the modified gravity field equations to support zeroth law? To answer this question, we may start with the general form of higher curvature gravity field equations, namely $G_{ab} + \alpha \, H_{ab} = 8\pi\,T_{ab}$. The term $H_{ab}$ contains all the contributions from higher curvature terms in the action. Then, the validity of the zeroth law requires $H_{ab} \xi^a e^{b}_{A}=0$ everywhere on the Killing horizon. Our aim is to restrict the form of $H_{ab}$ so that zeroth law holds. To achieve this task, we first note that, for a stationary Killing horizon, we must have $ H_{ab} \xi^a \xi^b = 0$. It implies the following general form of $H_{ab}$ on the horizon:
%%%%%%%%%%%%%%%%%%%%%%%%%%%%%%%
\bea
H^{ab} = C\, \xi^a\, \xi^b +D\, g_{\perp}^{ab} + E^{A}\, N^{(a} e^{b)}_{A} + F^{A B}\, e_{A}^{a}\, e_{B}^{b}\nonumber \ .
\eea
%%%%%%%%%%%%%%%%%%%%%%%%%%%%%%%
Here, $ g_{\perp}^{ab} $ is the contravariant metric on the $2$-plane perpendicular to a horizon slice. Also, the coefficients $C$, $D$, $E^A$ and $F^{A B}$ are local geometric quantities constructed from the metric and curvatures at the horizon. The above equation suggests that one needs $E^A = 0$ for the validity of the zeroth law. However, unlike GR and LL gravity, this may not be true in general without any further assumptions. It would be interesting if there is any such generalization in arbitrary gravity theory. As a closing remark, it is worth mentioning that our result shows the universal character of stationary BHs, strengthening the idea that applicability of BH thermodynamics extends beyond GR.

\cleardoublepage

\chapter{{\color{red!60!black}Causality Constraints in Quadratic Gravity} }\label{Chapter_3}
\large
\textbf{This Chapter is based on the work: JHEP 09 (2021) 150 by J. D. Edelstein, R. Ghosh, A. Laddha, and S. Sarkar}~\cite{Edelstein:2021jyu}.\\

\ni
Einstein's theory of GR is not the unique description of gravity having diffeomorphism invariance. Obeying this symmetry, one can add a plethora of higher curvature terms to the Einstein-Hilbert action, generating various modified theories. Though they possess several universal properties, such as BH thermodynamics discussed in the previous chapter, some of these theories might give rise to pathological phenomena. Therefore, it is crucial to constrain the structure of higher curvature terms in the gravitational action using various consistency criteria, such as non-negativity of energy and stability of vacuum solutions. Moreover, these criteria may also provide us with a way to classify all consistent classical gravitational theories in arbitrary spacetime dimensions.\\

\ni
Recently, an additional consistency criterion, known as the \textit{causality constraint}, has been put froward by Camanho, Edelstein, Maldacena and Zhiboedov (CEMZ)~\cite{Camanho:2014apa}. Motivated by a theorem by Gao and Wald~\cite{Gao:2000ga} which identifies causal propagations in a spacetime in terms of its asymptotic properties, they argued that theories in which the sign (positive corresponds to delay, and negative corresponds to advancement) of the Shapiro time shift can depend on the polarizations of the probe graviton are acausal. And hence, such sick theories should be ruled out on the ground of causality. They further showed that a classical observable like Shapiro shift experienced by a probe in a shock wave background has a deep connection with the three-point scattering amplitudes of the theory in the eikonal limit~\cite{Camanho:2014apa}. One can use this idea to put rather non-trivial constraints on the types of higher order curvature scalars (quadratic, cubic etc) that may be present in the gravitational action. For example, in \ref{Chapter 1}, we have discussed how CEMZ criterion rule out the presence of Gauss-Bonnet coupling in $D \geq 5$ dimensions~\cite{Camanho:2014apa}.\\

\ni
Therefore, a question of utmost importance is that whether all modified gravity theories suffer from this causality issue. Since there is an infinite number of such higher curvature theories, we have to be systematic in our approach. For this purpose, it is suggestive to start with the QG given by \ref{QG}, which include quadratic-curvature modifications to Einstein-Hilbert action. This theory is well-studied in the literature~\cite{PhysRevD.16.953, buchbinder1992effective, Nojiri:2001ae, Alvarez-Gaume:2015rwa, Salvio:2018crh}, and several exact BH solutions are known~\cite{Audretsch:1993kp, Matyjasek:2004vr, Berej:2006cc, Svarc:2018coe}. Also, unlike GR, QG possesses extra dynamical modes other than massless graviton, which are tachyonic unless the coupling constants are chosen so that $4\, (D-1)\, \alpha + D\, \beta \geq 0$, and $\beta \leq 0$~\cite{Tekin:2016vli}. Even in this constrained parameter space, the massive spin-$2$ mode is actually a ghost rendering unbounded Hamiltonian unless $\beta = 0$~\cite{Audretsch:1993kp, Tekin:2016vli}. Even in the presence of ghost mode, there are two reasons why QG could be treated as a well-defined classical theory of gravity. First, a careful analysis of the full ADM Hamiltonian shows that QG does not violate the positive energy theorem as the higher derivative terms arising from quadratic curvature scalars fall off very rapidly at asymptotic distances and fail to overwhelm the contribution coming from the Einstein-Hilbert part of the action. Moreover, at a classical level, QG makes sense as a low-energy effective theory as long as our probe energy scale lies below the mass of the ghost.\\

\ni
Although it is known that the solutions of QG can be mapped to that of GR via field redefinition~\cite{MohammadiMozaffar:2016vpf}, one should not treat our analysis of causality in the context of QG as a duplication of effort. It is because, when perturbed around an exact solution, QG contains both massless and massive modes distinct from GR. Therefore, even under a field redefinition, three-point graviton couplings of QG are not mapped to those of GR. Hence, CEMZ construction is not sufficient to dictate causality of QG. In fact, we need to calculate the Shapiro time shift experienced by a probe graviton in the background of a known exact shock wave solution of QG and check for causality separately. The QG shock has a structure~\cite{Campanelli:1995ex}, 
%%%%%%%%%%%%%%%%%%%%%%%%%%%%%%%
\begin{equation} \label{CCshock}
ds^2 = -du\, dv + h_0(u, x_i)\ du^2 + \sum_{i=2}^{D-2} (dx_i)^2\, ,
\end{equation}
%%%%%%%%%%%%%%%%%%%%%%%%%%%%%%%
with the profile $h_0(u,x_i) = f(r)\, \delta(u)$, where
%%%%%%%%%%%%%%%%%%%%%%%%%%%%%%%
\begin{equation} \label{CCprofile}
\begin{array}{ccl}
f(r) &=& -\, \displaystyle\frac{8 \pi G\, \lvert P_u \rvert\, \Gamma\left(\frac{D}{2}-1\right)}{\pi^{\frac{D}{2}-1}}\, \left[\frac{(-2 \beta)^{2 - \frac{D}{2}}}{\Gamma\left(\frac{D}{2}-1\right)} \left(\frac{r}{\sqrt{-\beta}}\right)^{2 - \frac{D}{2}}\, K_{2 -\frac{D}{2}} \left(\frac{r}{\sqrt{-\beta}}\right) \right. \\ [1em]
& & \left. \quad\qquad\qquad\qquad\qquad\qquad\qquad -\, \displaystyle\frac{1}{D-4}\left(\frac{1}{r}\right)^{D-4}\right] ~.
\end{array}
\end{equation}
%%%%%%%%%%%%%%%%%%%%%%%%%%%%%%%
Here, we have considered the case when $\beta\, \leq\, 0$ and $D > 4$. The distance from the shock is given by the transverse coordinate $r=\sqrt{x^i x_i}$, and $K_n(x)$ is the modified Bessel function of second kind. The case $D=4$ involves a logarithmic profile with an arbitrary infra-red cut-off, and hence the corresponding Shapiro shift is unphysical similar to GR.\\

\ni
In this chapter, we aim to compute the Shapiro shift experienced by a probe graviton as it "crosses" the shock given by \ref{CCprofile} and study the issue of causality constraint in QG. We shall show that QG is free from any such issue as the time shift is polarization independent and always positive.  In any case, due to the presence of the ghost modes, it is widely accepted that the QG vacuum is unstable. Thus, it should be emphasized that we are not upholding QG as a viable theory of gravity but considering it only as a toy model which may offer interesting perspective on causality constraints in modified theories. From this perspective, our result may contribute towards a finer classification of consistent gravitational theories, where additional restrictions~\cite{Chowdhury:2019kaq, Chandorkar:2021viw} other than CEMZ criterion have to be imposed.
%%%%%%%%%%%%%%%%%%%%%%%%%
\section{{\color{blue!70!brown} Scalar Probe in the Shock Wave Background of QG}}
%%%%%%%%%%%%%%%%%%%%%%%%%
Compared to the GR solution in \ref{gr_prof}, the structure of QG shock profile as given by \ref{CCprofile} is more involved due to presence of the higher derivative terms in the action. However, for a minimally-coupled massless scalar probe $\phi(u,v,x_i)$, the governing Klein-Gordon equation takes a simpler form in the shock wave background:
%%%%%%%%%%%%%%%%%%%%%%%%%%%%
\bea \label{CCKG}
\partial_u\, \partial_v\, \phi + h_0(u,x_i)\, \partial_{v}^2\, \phi = 0\, .
\eea
%%%%%%%%%%%%%%%%%%%%%%%%%%%%
Here, we have neglected the the transverse variations of $\phi$, since they are highly suppressed by the $u$-derivatives of the profile $h_0(u, x_i)$ as the probe moves in ($u = 0^-$) and out ($u = 0^+$) of the shock with impact parameter $r = b$, see \ref{shoch}. In order to solve the above differential equation, we can make use of Fourier transformation in the $v$-direction. Finally, the variation in $\phi$ as the probe crosses the shock is given by,
%%%%%%%%%%%%%%%%%%%%%%%%%%%%
\bea \label{CCphi_var}
\phi \left(u=0^+,\ v,\ x_i \right) = e^{-i\, p_v\, \Delta v}\, .\, \phi \left(u=0^-,\ v,\ x_i \right)\, .
\eea
%%%%%%%%%%%%%%%%%%%%%%%%%%%%
It is obvious that the probe undergoes a Shapiro shift, $\Delta v_{\textrm{scalar}} = \int_{0^-}^{0^+} du\, h_0(u, b) = f(b)$. The same result can be obtained by computing the eikonal scattering amplitude in the deflectionless limit ($t/s \to 0$) as prescribed in Ref.~\cite{Camanho:2014apa}. For massless scalar and $\beta \leq 0$, the $t$-channel amplitude is as follows,
%%%%%%%%%%%%%%%%%%%%%%%%%%%%%
\begin{flalign}
{\cal A}_{4}(s, t)\, =\, \frac{4\, \pi\, G}{t}\,\left [\, \frac{1}{-\beta\, t - 1}\, \left(\, 2\, s\, u -\, \frac{1}{3}\, t^{2}\, \right)\, -\, \frac{1}{3}\, \frac{t^2}{-\, 2\, \beta\, t + 1}\,\right ]\ ,
\end{flalign}
%%%%%%%%%%%%%%%%%%%%%%%%%%%%%%
where $t = -\, \vert \vec{q} \vert^{2}$. In the eikonal limit ($t/s \to 0$), the above expression simplifies to,
%%%%%%%%%%%%%%%%%%%%%%%%%%%%%%
\begin{flalign} \label{CCeikonal}
{\cal A}_{4}^{\textrm{eik}}(s, t)\, =\, -\, \frac{8\, \pi\, G}{\beta}\, \frac{s^{2}}{t (t + \frac{1}{\beta})}\ .
\end{flalign}
%%%%%%%%%%%%%%%%%%%%%%%%%%%%%%
It is important to note that $m_g = 1/\sqrt{-\beta}$, represents the mass of the ghost mode. Then, in the impact parameter space, the phase is given by,

\begin{flalign} \label{CCscalar}
\delta(\vec{b}, s) =\, \frac{1}{2s}\, \int \frac{d^{D-2}\, \vec{q}}{(2 \pi)^{D-2}}\, \, \, e^{i \vec{b}.\, \vec{q}}\, \, {\cal A}_{4}^{\textrm{eik}}(s, t) := - p_{v}\, \triangle v_{\textrm{scalar}}\, .
\end{flalign}
Following Appendix A~[\ref{app1}], we get $\Delta v_{\textrm{scalar}} = f(b)$. As we shall see in the subsequent sections that the the amount of Shapiro shift experienced by a probe graviton is also the same and $\Delta v$ is always positive.
%%%%%%%%%%%%%%%%%%%%%%%%%
\section{{\color{blue!70!brown} Computation of Graviton Time Shift}}
%%%%%%%%%%%%%%%%%%%%%%%%%
Let us now consider the motion of a probe graviton in the shock wave background $\bar{g}_{\mu \nu}$ of QG given by \ref{CCshock}. We denote the gravitational perturbation in this background as $g_{\mu \nu} = \bar{g}_{\mu \nu} + h_{\mu \nu}$, with $\lvert h_{\mu \nu} \rvert \ll 1$. For our purpose, we only consider the transverse and traceless (TT) part of the perturbations denoted by $h_{i j}$. It is apparent that in QG, $h_{i j}$ will satisfy a fourth order equation of motion (EoM) due to the presence of quadratic curvature terms in the action. The following table lists down the contributions coming from various such terms~\cite{Benakli:2015qlh}.
%%%%%%%%%%%%%%%%%%%%%%%%%%
\renewcommand{\arraystretch}{2.5} 
\begin{table}[h!]
\begin{center}
\begin{tabular}{|c|c|}
\hline
~\bf{Terms in the action}~ & ~\bf{Contribution at first order in $h_{ij}$}~ \\
\hline
$R$ &	$-\frac{1}{2}\ \Delta h_{ij}$ \\
\hline
$R^2$ & $0$ \\
\hline
$R_{ab}R^{ab}$ & $-\frac{1}{2}\ \Delta^2 h_{ij}$ \\
\hline
\end{tabular}
\caption{Contribution to graviton EoM in TT-sector. Notation: $\Delta := - 4\left(\partial_{u} \partial_{v} + h_{0} \partial_{v}^{2}\right)$.}
\label{CCtable}
\end{center}
\end{table}
%%%%%%%%%%%%%%%%%%%%%%%%%%
Therefore, the EoM satisfied by the perturbation $h_{i j}$ is given by,
%%%%%%%%%%%%%%%%%%%%%%%%%%
\bea \label{EoM}
\left(\Delta+\beta\, \Delta^{2}\right) h_{i j} = 0 ~.
\eea
%%%%%%%%%%%%%%%%%%%%%%%%%%
Note that the graviton EoM in GR is recovered by letting $\beta \to 0^-$. Unlike GR, the solution space of \ref{EoM} contains both massless graviton satisfying $\Delta h_{i j} = 0$, and massive graviton satisfying $(\Delta - m_g^2)\, h_{i j} = 0$. However, in sharp contrast to EGB case~\cite{Camanho:2014apa}, the graviton EoM in QG does not contain any transverse derivatives ($\partial_i$) due to the absence of Riemann-squared term in the action. Therefore, we expect that the Shapiro time shift will be independent of graviton polarizations in QG. This observation will play a major role in our later discussion.\\

\ni
Now, to calculate the Shapiro time shift experienced by a probe graviton, we shall proceed in parallel to the previous scalar case. The first step is to solve \label{CCEoM} for $h_{i j}$, which can be achieved by performing a Fourier transform in v-direction. The details of the derivation is tedious and will be presented in Appendix A~[\ref{app2}]. Here, to compute the time shift, we shall only consider the general solution:
%%%%%%%%%%%%%%%%%%%%%%%%%%%
\bea \label{CCsol}
h_{i j}\left(u, v, x_{i}\right) = \int \frac{d p_v}{i p_v}\ e^{i p_v v}\ \widetilde{h}_{i j}(u, p_v, x_i) + K_{i j}^{(2)}\left(u, x_{i}\right) + C_{ij} ~,
\eea
%%%%%%%%%%%%%%%%%%%%%%%%%%%%
where a tilde denotes Fourier-space variables which are functions of $v$-momentum $p_v$ and the other coordinates $(u, x_i)$, 
%%%%%%%%%%%%%%%%%%%%%%%%%%%%
\bea \label{CCht}
\begin{array}{ccl}
\widetilde{h}_{i j}\left(u, p_v, x_{i}\right) &=& \displaystyle\bigg[\ 4 i \beta p_v\ \widetilde{K}_{ij}^{(0)} \left(p_v, x_{i}\right)\ \text{exp}\left( \frac{u}{4 i \beta p_v}\right) \\ [.3em]
& & \qquad +\ \displaystyle\widetilde{K}_{i j}^{(1)} \left(p_v, x_{i}\right) \bigg]\ \text{exp}\left( -i p_v \int^{u}\!\! du\ h_{0}\left(u, x_{i}\right)\right) ~.
\end{array}
\eea
%%%%%%%%%%%%%%%%%%%%%%%%%%%%
The presence of three undetermined profile functions denoted by $\widetilde{K}_{ij}^{(m)}\left(p_v, x_{i}\right)$ and an additive constant $C_{ij}$ in the solution affirms the fact that we are solving a fourth order differential equation. Then, using the general solution \ref{CCsol}, it is trivial to calculate the Shapiro time shift experienced by a probe graviton as it crosses the shock with impact parameter $r=b$,
%%%%%%%%%%%%%%%%%%%%%%%%%%%%
\bea \label{CCts}
\Delta v = \int_{0^{-}}^{0^{+}}\!\! du\ h_{0}(u, b) = \int_{0^{-}}^{0^{+}}\!\! du\ f(b)\ \delta(u) = f(b) ~.
\eea
%%%%%%%%%%%%%%%%%%%%%%%%%%%%
Interestingly, the value of $\Delta v$ is exactly same as the scalar case. Also, as per our expectation, $\Delta v$ is independent of graviton polarizations.\\

\ni
Now, we shall proceed to the remaining task of determining the sign of the Shapiro shift $\Delta v$, which is connected to causality. And, we shall show that $\Delta v$ is always positive for arbitrary values of the coupling $\beta \leq 0$, and dimensions $D \geq 5$. For this purpose, it is suggestive to write \ref{CCts} in the following form,
%%%%%%%%%%%%%%%%%%%%%%%%%%%%
\bea \label{CCShap}
\Delta v = (\Delta v)_{\mathrm{GR}} \times \left( 1 - \frac{1}{2^{n-1} \Gamma(n)} x^{n} K_{-n}(x) \right) ~,
\eea
%%%%%%%%%%%%%%%%%%%%%%%%%%%%
where $x = b/\sqrt{-\beta}$ is the dimensionless impact parameter, $n = (D-4)/2$ is the reduced dimension, and $(\Delta v)_{\mathrm{GR}}$ is the Shapiro delay in GR [see \ref{gr_shift}],
%%%%%%%%%%%%%%%%%%%%%%%%%%%%
\bea \label{CCShapGR} 
(\Delta v)_{\mathrm{GR}} = \frac{4 \Gamma(n)}{\pi^{n}} \frac{G |P_u|}{b^{2 n}} > 0\, .
\eea
%%%%%%%%%%%%%%%%%%%%%%%%%%%%
Thus, the positivity of $\Delta v$ is equivalent to the fact that the quantity $y(x)/[2^{n-1} \Gamma(n)]$, with $y(x) = x^n\, K_{-n}(x)$, is bounded above by unity in the entire allowed range of $x \in (0^+, \infty)$. The function $y(x)$ has the following properties:\\

\ni
(i) In the limit $x \to 0^+$, one can show that $y(x)$ approaches the value $2^{n-1} \Gamma(n)$ from the positive side, for all real values of $n$. \\

\ni
(ii) Moreover, one has $y'(x) = -x^n\, K_{-n+1}(x) < 0$, since $K_{-n+1}(x)$ has no root in $x \in (0^+, \infty)$. Therefore, the function $y(x)$ is monotonically decreasing in the above range of $x$.\\

\ni
(iii) And, in the limit $x \to \infty$, we have $y(x) \to 0$.\\

\ni
These properties imply that $y(x)$ starts with the value $2^{n-1} \Gamma(n)$ at $x=0$ and then, it decreases smoothly to zero as $x$ approaches infinity. In other words, the quantity $y(x)/[2^{n-1} \Gamma(n)]$ is bounded above by unity in the range $x \in (0^+, \infty)$, as required for showing $\Delta v > 0$. In addition, the behavior of $y(x)$ suggests that the time delay matches with the corresponding GR value in the limit $\beta \to 0^-$ and it vanishes as $\beta \to -\infty$. Some additional comments on this result are in order.\\

\ni
In the small $\beta$ regime, the difference between the Shapiro time delay in GR and QG is exponentially suppressed as $y(\beta \to 0^-) \sim (b\, m_g)^{n-1/2}\, \textrm{exp}[-b\, m_g]$, where $m_g = 1/\sqrt{-\beta}$ is the ghost mass. Then, any detector that measures the time delay having a resolution $\Delta T > \Lambda^{-1} \gg \textrm{exp}[-b\, m_g]$ can hardly distinguish GR and QG as far as low-$\beta$ Shapiro time shift is concerned, where $\Lambda < m_g$ is the energy scale up to which QG can be treated as a low energy effective description of gravitational dynamics.\\

\ni
Also, for a fixed value of $\beta$, if we let the impact parameter $b \to 0$ (contact interaction), the Shapiro time delay in QG diverges like that of GR -- though the form of divergences are different. In GR, the time delay diverges as $\Delta v \sim b^{-2n}$ in the limit $b \to 0$ for all $D \geq 5$. Whereas in QG, the divergences are as $\Delta v \sim -\textrm{ln}(b)$ for $D=6$, and $\Delta v \sim b^{-(D-6)}$ for $D>6$. However, the most interesting case is for $D=5$, where QG time delay remains finite as $b \to 0$ with a fixed value of $\beta \neq 0$, i.e., $\Delta v \to 4\, m_g\, G\, |P_u| > 0$.
%%%%%%%%%%%%%%%%%%%%%%%%%
\section{{\color{blue!70!brown} Subtleties with Field Redefinition}}
%%%%%%%%%%%%%%%%%%%%%%%%%
This section discusses the effect of field redefinition on the shock wave solution of QG and on the amount of Shapiro time delay in this shock background. It is well known that there exists a redefinition of the metric so that any QG solution can be mapped to a corresponding GR solution with some exotic matter degrees of freedom~\cite{MohammadiMozaffar:2016vpf}:
%%%%%%%%%%%%%%%%%%%%%%%%%
\begin{flalign} \label{CCmap}
\overline{g}^{\mu \nu}\, = \, g^{\mu \nu} + 2\, \beta\, R^{\mu \nu} \ ,
\end{flalign}
%%%%%%%%%%%%%%%%%%%%%%%%%
where we have used an over-bar to denote quantities in GR. With this field redefinition, one can readily check that QG shock given by \ref{CCshock} indeed maps to the GR shock given in \ref{grshock}. Thus, it seems that our calculation of Shapiro time shift presented above is unnecessary, as one can directly get it from GR using the same field redefinition. However, we shall show by a careful argument that this is not the case and our calculation of time delay in QG is not redundant.\\

\ni
The field redefinition which maps $g_{\mu \nu}\, \rightarrow\, \overline{g}_{\mu \nu}$, will also map perturbations $h_{\mu\nu}$ in the QG shock background to perturbation $\overline{h}_{\mu \nu}$ over the GR shock. Then, using the mapping given by \ref{CCmap}, we can easily show that the TT-perturbations $\overline{h}_{i j}$ satisfy the following equation: 
%%%%%%%%%%%%%%%%%%%%%%%%%%
\bea \label{CChbar}
\overline{\Delta}\ \overline{h}_{ij}\, =\, 4\, \left[h_0 - \overline{h}_0\right]\, \partial_{v}^{2}\, (1\, +\, \beta\, \Delta)\, h_{ij}\, ~,
\eea
%%%%%%%%%%%%%%%%%%%%%%%%%%
where $\overline{\Delta}$ and $\overline{h}_0$ are the Laplacian operator and the shock-wave profile in GR given by \ref{gr_prof}. Therefore, the mapped equation is, in general, different from the corresponding graviton EoM in the shock-wave background of GR, namely $\overline{\Delta}\ \overline{h}_{ij}=0$. Since $h_0 \neq \overline{h}_0$, the only way such a mapping is possible when the QG perturbations represent a massive graviton satisfying $(1\, +\, \beta\, \Delta)\, h_{ij}\, =\, 0$.\\

\ni
Thus, the time delay of QG can not be generically obtained from that of GR by using any field redefinition. Intuitively, this failure in mapping can be traced back to the additional spin-$2$ ghost mode of propagation present in QG, which has no counterpart in GR.
%%%%%%%%%%%%%%%%%%%%%%%%%
\section{{\color{blue!70!brown} Generalization to Other Modified Theories}}
%%%%%%%%%%%%%%%%%%%%%%%%%
We devote this section to discuss possible generalization of our analysis in other modified theories. For a general higher curvature theory, the metric perturbation equation will be more difficult to solve than \ref{EoM} due to the presence of complicated higher derivative terms in $(\partial_u,\partial_v, \partial_i)$. In such a scenario, Shapiro time shift might not possess any common feature and has to be obtained case by case. However, there is indeed a way to extend our result for theories whose graviton EoM is as follows,
%%%%%%%%%%%%%%%%%%%%%%%%%%
\bea \label{CCgen_eom}
\left( 1 + \gamma \Delta^n \right) \Delta h_{ij} = 0 ~,
\eea
%%%%%%%%%%%%%%%%%%%%%%%%%%
where $n$ is a non-negative integer. Note that GR and QG are two special cases with $n=0$ and $n=1$, respectively. For another example, one can readily check that a Lagrangian of the form $\mathcal{L} = \sqrt{-g} (R + \alpha R^2 + \gamma R_{ab} \nabla_c \nabla^c R^{ab})$ gives rise to such an EoM with $n=2$. The above differential equation is special for two reasons:\\

\ni
(i) The perturbation equation is factored into a GR part and a part coming from higher curvature terms with coupling $\gamma$. Therefore, its spectrum contains both massless and massive graviton similar to the QG case.\\

\ni
(ii) The EoM does not contain any transverse derivatives of the profile function $h_0(u, x_i)$.\\

\ni
Following our method, one can now integrate \ref{CCgen_eom} to obtain the time shift as $\Delta v = f(b)$ in the shock wave background given by \ref{CCshock}. However, the form of $f(r)$ will be different for different theories (labelled by the integer $n$) in this class. Hence, the sign of the Shapiro shift could also vary depending on the choice of such theories.
%%%%%%%%%%%%%%%%%%%%%%%%%
\section{{\color{blue!70!brown} Summary}}
%%%%%%%%%%%%%%%%%%%%%%%%%
Several recent studies of modified gravity have suggested that a viable alternative of GR cannot be solely characterized by diffeomorphism invariance~\cite{Camanho:2014apa, Chowdhury:2019kaq, Chandorkar:2021viw}. Such a theory must be "healthy" in the sense that it must satisfy other consistency criteria, such as causality constraint as formulated by CEMZ~\cite{Camanho:2014apa}. With these constraints at hand, the nature of recent studies in classical gravity has marked a shift from any previous analysis subscribing to the "anything goes" notion~\cite{Adams:2006sv}.\\

\ni
Using CEMZ criteria, we have demonstrated that QG is probably more well-behaved than naively expected due to the presence of ghost. Unlike EGB gravity, QG is shown to be free from such causality issue similar to GR. For this purpose, we calculated the Shapiro time shifts experienced by a massless scalar and a probe graviton as it crosses the shock wave. Interestingly, for the both cases time shift turns out to be the same and positive for all values of the coupling constant, inferring causality. Moreover, a careful analysis of field redefinition shows that this time shift is not merely the GR result in disguise. Given the importance of this analysis, we have also discussed possible generalization of our result in other modified theories of gravity. A general class of theories, of which GR and QG are two members, that might be free from such causality constraint has been proposed.\\

\ni
Besides the aforesaid generalization, it will be interesting to study the causality constraints in theories involving cubic or higher order curvature terms. The first step in this direction should be to find shock wave solutions in some modified theories. As a starting point, it might be useful to find such a solution in the most general quadratic theory of gravity, namely QG$+$Riemann-squared term, and study the causality problem in this background.\\

\ni
Let us now summarize the goals of various recent efforts towards finding a classification scheme for consistent classical gravity theories. The results from Refs.~\cite{Camanho:2014apa, Chowdhury:2019kaq, Chandorkar:2021viw, Edelstein:2021jyu} suggest that such consistent theories must obey the following necessary conditions: (i) the existence of a stable vacuum, (ii) the
validity of a positive energy theorem, and (iii) the causality properties as advocated
and quantified by CEMZ. All these conditions imply that QG can be treated as a low-energy effective theory of gravity. Also, it is worth mentioning that apart from the above list, there could be other necessary conditions one has to impose on consistent gravitational theories. One such condition involving the classical Regge growth is formulated in Ref.~\cite{Chowdhury:2019kaq}, which restrict (bound) the form of the tree-level scattering amplitudes. Thus, it will be interesting to perform a detailed analysis of the Regge growth in QG and verify whether it satisfies the bound. We leave this question for a future attempt.

\cleardoublepage

\chapter{{\color{red!60!black}Light Rings of Stationary Spacetimes} }\label{Chapter_4}
\large
\textbf{This Chapter is based on the work: Phys. Rev. D 104 (2021) 4, 044019 by R. Ghosh, and S. Sarkar}~\cite{Ghosh:2021txu}.\\

\ni
The main focus of the previous two chapters was to investigate several properties of higher curvature theories and test their consistency. Now, we shall shift our concentration to studying the universal characteristics of various compact objects, both with and without horizons. These objects showcase a multitude of gravitational phenomena, which may reveal hitherto unknown aspects of strong gravity and provide us with invaluable inputs towards the ultimate theory of gravity. Numerous observational probes, such as GWs, QNMs and shadows, hinge crucially upon a vital ingredient, namely the light ring (LR), situated outside a compact object. These are the locations where the gravitational deflection of light becomes so extreme that photons can stay put in circular orbits around the central objects. For example, the study of null geodesics outside a Schwarzschild BH given by \ref{Sch} suggests the existence of a single unstable LR at $r=3M$. In contrast, there are two LRs for each sense of rotation (co-rotating and counter-rotating) outside a Kerr BH.\\

\ni
The above examples of Schwarzschild and Kerr BHs naturally raise the question: Do all BHs have LR(s) outside their event horizons? Interestingly, a recent work~\cite{Cunha:2020azh} has answered this question affirmatively by showing that any stationary, axisymmetric, asymptotically flat,
$1 + 3$ dimensional BH spacetime with a non-extremal and topologically spherical Killing horizon admits at least one LR outside the horizon for each rotation
sense. This remarkable result follows from a novel topological argument without referring to any field equations. Later, a similar proof has been generalized to the asymptotically de-Sitter (dS) and anti-de-Sitter (AdS) non-rotating cases~\cite{Wei:2020rbh}.\\

\ni
Despite the aforesaid results in BH spacetime, the existence of LRs does not uniquely fix the nature of the central compact object. Several known horizonless objects, such as boson stars and gravastars, also possess LRs. In fact, a similar theorem~\cite{Cunha:2017qtt, Guo:2020qwk} as in the case of BHs dictates that under the assumption of an initial LR, the spacetime of a horizonless compact object always supports an even number of them. And, among these LRs, at least one must be stable. Interestingly, the proof of this theorem does not require assuming any field equations or energy conditions on matter.\\

\ni
In the presence of LRs, horizonless compact objects can behave as BH mimickers by demonstrating the same observable signatures. For example, the shadow observed by EHT can also be produced by a non-BH object with LRs, creating an alarming ambiguity in detecting BHs. Thus, the most relevant question of importance is whether such horizonless compact objects are stable under perturbations. Because if they prove to be unstable, we can exclude such objects on mere physical grounds. In particular, 
it was argued in Refs.~\cite{Keir:2014oka, Cardoso:2014sna} that stable LRs outside a horizonless compact object severely slow down the decay of perturbations by trapping them. As a result, these trapped long-lived modes can back react on the spacetime, causing nonlinear instability that could destroy the central object. However, it might be too hasty to rule out the existence of horizonless compact objects at the first sign of instability, since the LR instability timescale could be huge (say, of the order of the universe's age). Then, despite their instability, horizonless objects will be astrophysically and observationally relevant. Nevertheless, for some specific models of static horizonless compact objects like boson and Proca stars, a recent work~\cite{Cunha:2022gde} has demonstrated that the LR instability timescale is relatively short ($\sim 10^3$ light-crossing time), unless the stable LR potential well is very shallow.\\

\ni
In addition to the issue of instability timescale, there is a second loophole of using the result of Ref.~\cite{Cunha:2017qtt} to conclude that the horizonless spacetime suffers from the LR instability. Note that a stable LR is only guaranteed if the spacetime has an initial LR to begin with~\cite{Cunha:2017qtt}. However, in general, a horizonless compact object may not contain any such LR, and the instability argument fails. However, we shall now show that any stationary, axisymmetric, and asymptotically flat spacetime
in $1+3$ dimensions with an ergoregion must have at least
one light ring outside. Therefore, at least for the particular case of rotating horizonless objects with ergoregion, the presence of one LR is assured. Then, Ref.~\cite{Cunha:2017qtt} implies that there must be a stable LR, and the spacetime will be prone to both LR and ergoregion instability~\cite{Zhong:2022jke}. Therefore, our work provides a strong support in favor of the \textit{BH hypothesis}, which claims that the objects with LRs are BHs.
%%%%%%%%%%%%%%%%%%%%%%%%%
\section{{\color{blue!70!brown} LRs of Stationary Compact Objects}}
%%%%%%%%%%%%%%%%%%%%%%%%%
The first step towards proving our result is to construct the metric outside a stationary compact object. Such a metric will be independent of time ($t$), when expressed in terms of some suitable coordinates. Moreover, if we were interested in only BH solutions of any effective field theory, the underlying stationary spacetime would have been axisymmetric too (i.e., there exists a $\partial_\phi$ Killing vector), as a consequence of the rigidity theorem~\cite{Hawking:1971vc, Hollands:2022ajj}. However, since our aim is to consider compact objects both with and without horizons, the axisymmetry is not generally guaranteed. Nevertheless, to have a well-defined notion of LRs, we shall assume the underlying spacetime has this additional Killing vector $\partial_\phi$, and write down the metric in such coordinates that make the symmetry apparent. Furthermore, we consider the metric is invariant under the simultaneous inversions $(t, \phi) \to (-t, -\phi)$. Then, the most general $4$-dimensional metric for an asymptotically flat stationary, axisymmetric spacetime is as follows,
%%%%%%%%%%%%%%%%%%%%%%%%%
\bea \label{LR_statm}
ds^2=g_{tt}dt^2+g_{rr}dr^2+g_{\theta \theta}d\theta^2+g_{\phi \phi} d\phi^2+2g_{t \phi} dt d\phi\ .
\eea
%%%%%%%%%%%%%%%%%%%%%%%%%
In terms of this symmetry-adapted coordinate, all metric components are functions of $(r, \theta)$ alone, and there is no other cross components except $g_{t \phi}$. Note also that the asymptotic flatness requires the following behaviour of various metric components at large r: $ g_{t t} \to -1+C/r + {\cal O}(1/r^2)\ ,\  g_{t \phi} \sim \pm\ r^{-1}\, \sin^2\theta\ \text{and},\ g_{\phi \phi} \sim \ r^{2}\, \sin^2\theta$.\\

\ni
The ergoregion is a Killing horizon $\mathcal{H}$, where the asymptotic time-translation Killing vector $\partial_t$ becomes null. Its location is given by the largest positive root ($r_e$) of the equation $g_{tt} = 0$. In case of the non-extremal Killing horizon, $-g_{tt}'(r_e) > 0$. In addition, other metric components, such as $g_{rr}$, $g_{\theta \theta}$, and $g_{\phi \phi}$ are always positive in the region of our interest, i.e., $r \geq r_e$ (away from the axis)~\cite{Cunha:2017qtt}. Then, the study of circular null geodesics suggests that the location of LRs are given by the equations (see Appendix B~[\ref{app3}]), $\partial_\mu H_{\pm} = 0$, where we have
%%%%%%%%%%%%%%%%%%%%%%%%%
\bea \label{LR_statf}
H_{\pm}(r,\theta) = \frac{-g_{t \phi} \pm \sqrt{\Delta}}{g_{\phi  \phi}},
\eea
%%%%%%%%%%%%%%%%%%%%%%%%%
with $\Delta=g_{t \phi}^2-g_{tt}\, g_{\phi \phi} > 0$ outside the ergoregion. It is easy to check that, for a Schwarzschild/Kerr metric, $H_\pm$ boils down to the familiar circular null geodesic potentials. The two signs represent two opposite sense of rotations with respect to the central object~\cite{Cunha:2017qtt}. If the object is rotating in the "negative" sense, meaning $g_{t \phi} > 0$, the critical points of the potential $H_+$ ($H_-$) represent counter-rotating (co-rotating) LRs. Whereas for $g_{t \phi} < 0$, the roles of $H_{\pm}$ are swapped.\\

\ni
Now, to prove our theorem, we must show that there exists at least one root $\{r=r_l,\, \theta=\theta_l\}$ of the equation $\partial_{\mu} H_{\pm}=0$ in the region $r_e \leq r < \infty$ and $0 < \theta < \pi$. For this purpose, let us consider the $\theta$-equation first. In terms of a local coordinate $\rho = \sqrt{g_{\phi \phi}}$, it is easy to see that $H_{\pm} \sim \pm \rho^{-1}$ near the axis~\cite{Cunha:2020azh}. Moreover, using the fact that $\partial_\theta \rho$ is positive (negative) as $\theta \to 0$ ($\theta \to \pi$), we obtain~\cite{Cunha:2020azh}
%%%%%%%%%%%%%%%%%%%%%%%%%
\bea \label{LR_beh}
\partial_{\theta} H_{\pm} \sim \mp \frac{\partial_{\theta} \rho}{\rho^{2}} \sim\left\{\begin{array}{ll}
\mp \infty & \text{as}\ \theta \rightarrow 0, \\
\pm \infty & \text{as}\ \theta \rightarrow \pi\ .
\end{array}\right.
\eea
%%%%%%%%%%%%%%%%%%%%%%%%%
Thus, $\partial_{\theta} H_{\pm}$ must have at least one root in $(0,\pi)$ at all values of $r > r_e$. In other words, as r varies, we get a trajectory of solutions $\theta_0(r)$ of the equation $\partial_{\theta} H_{\pm}=0$. On the other hand, using \ref{LR_statf}, we can express $\partial_r H_{\pm}(r,\theta)$ in the following suggestive form:
%%%%%%%%%%%%%%%%%%%%%%%%%
\bea \label{LR_statlr}
 \partial_r H_{\pm}(r,\theta)=\pm \frac{1}{2 \sqrt{\Delta}\ g_{\phi \phi} }\left[L(r,\theta)-R(r,\theta)\right],
\eea
%%%%%%%%%%%%%%%%%%%%%%%%%
with $R(r, \theta)= -g_{tt} g_{\phi \phi}'\ \pm\ \left(2/g_{\phi \phi}\right) \left(\sqrt{\Delta} \mp g_{t \phi} \right)\left(g_{t \phi}'g_{\phi \phi} - g_{\phi \phi}' g_{t \phi}\right)$, and $L(r,\theta)= -g_{tt}' g_{\phi \phi}$. Unlike the topological proof given in Refs.~\cite{Cunha:2017qtt, Cunha:2020azh}, we now use an algebraic method to show that there must exist at least one root of $\partial_r H_{\pm} = 0$, where two functions $L$ and $R$ become equal. To achieve this, let us study their properties near ergoregion and asymptotic infinity: \\
\\(i) At the ergoregion $r=r_e$, we have $g_{t t}(r_e)=0$ and $\sqrt{\Delta(r_e)} = \pm g_{t \phi}(r_e)$. Here, $+$ve ($-$ve) sign is understood for the case $g_{t \phi} > 0$ ($g_{t \phi} < 0$). In other words, we are always considering the counter rotating LRs. And, hence it follows that $R(r=r_e)=0$. Furthermore, the function $R(r) \sim r$ as $r \to \infty$.\\
\\(ii) In contrast, the function $L(r,\theta)>0$ at $r=r_e$ and approaches to unity asymptotically, i.e., $L(r) \to 1$ at large r.\\
\\Therefore, $L(r)$ and $R(r)$ must have at least one intersection $r=r_0(\theta)$ in the region $r_e < r < \infty$ for all values of $\theta \in (0, \pi)$. Then, for all $\theta_0 \in (0,\pi)$ satisfying $\partial_\theta H_\pm =0$, there exists a solution $r=r_0(\theta_0)$ of $\partial_r H_\pm=0$, and vice versa. Hence, it is obvious that the two solution curves $r=r_0(\theta)$ and $\theta=\theta_0(r)$ must intersect at least at a point $(r_l,\theta_l)$ where both equations are satisfied. This ensures the existence of at least one (counter-rotating) LR outside the ergoregion irrespective of the fact whether the central object is BH or horizonless, completing our proof.\\

\ni
Some comments on our result are in order. Let us first recall that for a Kerr BH, a counter-rotating LR always remains outside the ergoregion for all values of spin $0<a<M$, verifying our result. Whereas the co-rotating LR migrates inside the ergoregion for spin values $M/\sqrt{2} \leq a < M$. Interestingly, for a general rotating BH which by definition possess an ergoregion outside its horizon, our theorem boils down to the result of Ref.~\cite{Cunha:2020azh}. We should also emphasis that our proof crucially depend on the asymptotic flatness condition. Thus, for spacetimes that are asymptotically dS or AdS, our result is not applicable in general. Also, during the proof, we have assumed the spacetime dimension to be four. However, the technique used here seems to be extendable for $D>4$ cases as well. We shall illustrate this fact with a simple example, namely for the static and spherically symmetric BH spacetimes.
%%%%%%%%%%%%%%%%%%%%%%%%%
\section{{\color{blue!70!brown} LRs of spherically symmetric BHs}} \label{LR_spherical}
%%%%%%%%%%%%%%%%%%%%%%%%%
Any general static and spherically symmetric $D$-dimensional ($D \geq 4$) BH spacetime is described by the metric
%%%%%%%%%%%%%%%%%%%%%%%%%
\bea \label{LR_sphm}
ds^2 = - f(r)\, dt^2 + \frac{1}{k(r)}dr^2 + h(r) d\Omega_{(D-2)}^2\, ,
\eea
%%%%%%%%%%%%%%%%%%%%%%%%%
where $r$ is a Schwarzschild-like radial coordinate and the metric components are positive outside the outermost event horizon at $r=r_H$, which is the largest positive root of $k(r) =0$. Then, staticity implies that the norm of $\partial_t$ must vanish at $r=r_H$, implying $f(r_H) = 0$~\cite{Vishi2003}. We shall also assume the BH is non-extremal, i.e., $f'(r_H) > 0$. Moreover, in case of asymptotically flat spacetime, we also have $f(r)\rightarrow 1 - C/r^{D-3} + \mathcal{O}(r^{-(D-2)}),$ $  
k(r)\rightarrow 1 - C/r^{D-3} + \mathcal{O}(r^{-(D-2)}),$ $h(r) \sim r^2\,$ as  $ r \to \infty$. Though $C$ is related to the ADM mass of the spacetime, we will not assume any particular sign of $C$.\\

\ni
Study of circular null geodesics in this spacetime readily gives the condition: $L(r_l) = R(r_l)$ for the location of the LR, where $L(r) = h(r)\, f'(r)$ and $R(r) = f(r) \, h'(r)$. This is, in fact, a special case of the stationary case discussed in the previous section with $g_{t \phi} = 0$. Now, to show that at least one solution to this LR equation must exist outside the horizon, we proceed to investigate the behavior of $L(r)$ and $R(r)$:\\
\\(i) At the event horizon $r=r_H$, the function $R(r)$ vanishes. Whereas at large values of $r$, $R(r) \sim r$.\\
\\(ii) In contrast, the function $L(r)>0$ at $r=r_h$ and it falls-off as $L(r) \sim r^{-(D-4)} $ at asymptotic infinity.\\
\\Therefore, the functions $L(r)$ and $R(r)$ must intersect at least at one point in the region $r_h \leq r < \infty$ for any $D\geq 4$, which proves our claim. In fact, it is obvious that there will be an odd number of LRs outside the event horizon. It can also be shown by considering second derivative of the geodesic potential obtained from \ref{LR_statf} for non-rotating case, $H_{\text{static}} = \sqrt{f(r)/h(r)}$, that these LRs are alternatively stable and unstable (in Lyapunov sense). Moreover, the innermost and outermost LRs are always unstable due to asymptotic flatness.\\

\ni
Unlike the stationary case, our proof for the static case is valid for any dimensions $D \geq 4$. However, it breaks down if the asymptotic flatness condition is relaxed. However, there exists a recent topological proof for asymptotically dS and AdS cases, confirming the existence of at least one LR outside the BH horizon~\cite{Wei:2020rbh}. 
%%%%%%%%%%%%%%%%%%%%%%%%%
\section{{\color{blue!70!brown} Summary}}
%%%%%%%%%%%%%%%%%%%%%%%%%
The presence of LRs in a spacetime are linked with various observable features, which provide us with unique opportunities to probe gravity. Therefore, a question of central importance is whether all compact objects support LR(s) outside them. In the presence of horizon, this question has an affirmative answer for both rotating and non-rotating cases~\cite{Cunha:2020azh}. However, the opposite statement fails to be valid, since it has been shown that horizonless ultra-compact objects always possess an even number of LRs, at least one of which is stable~\cite{Cunha:2017qtt}. Thus, it seems the so-called BH hypothesis, that claims objects with LRs are BHs, may not be valid. Interestingly, LRs themselves come to rescue by inflicting instability in horizonless spacetimes. It has been argued that the presence of stable LRs outside such a spacetime results in nonlinear instability~\cite{Cardoso:2014sna}, which may lead to fragmentation of the central ultra-compact objects. In fact, several recent works have pointed out the corresponding instability timescale can be relatively short~\cite{Cunha:2022gde, Zhong:2022jke}. This, in turn, suggests that horizonless ultra-compact objects may not survive in astrophysical timescales, providing a strong basis to the BH hypothesis.\\

\ni
Our result aims to plug a subtle loophole in the above argument by  proving the proof to a crucial assumption of the existence of one initial LR used in Ref.~\cite{Cunha:2020azh}, for spacetimes with an ergoregion. In particular, we show that any stationary, axisymmetric, and
asymptotically flat spacetime in $1+3$ dimensions with an
ergoregion must have at least one LR outside it. Unlike the previous results, our proof is completely algebraic and is valid irrespective of whether the central object has horizon or not. In fact, the boundary
conditions on various metric coefficients at the ergoregion
and the asymptotic infinity are enough to confirm the presence
of the LR. When combined with previous results~\cite{Cunha:2017qtt, Cardoso:2014sna}, our work significantly strengthen the validity of the BH hypothesis. Though a critique may argue that the horizonless objects with ergoregion suffers from the ergoregion instability~\cite{Friedman}, so why even bother to consider them in the first place. However, such an instability may not be sufficient to rule them out, since it is well-known that a mild absorptivity (as small as $~ 0.4 \%$) at the surface of the object can completely quench the ergoregion instability~\cite{Maggio:2017ivp, Maggio:2018ivz}. In such a situation, our result is absolutely necessary to (possibly) rule them out due to the presence of LR instability.\\

\ni
An important generalization of our result could be to consider a rotating star without an ergoregion. In principle, such a non-compact star may not possess any LR outside. However, at least for the case where the size ($r_s$) of the star is close to the would-be ergosphere ($r_e \lesssim r_s$), the functions can still satisfy $L(r_s) > R(r_s)$. This will particularly happen if the spacetime outside the star is assumed to be Kerr, as considered in Ref.~\cite{Zhong:2022jke}. Then, following a similar argument presented above, such object will have at least one LR outside. It will also be interesting if our algebraic proof can be further extended for $D>4$ dimensions, and for asymptotically dS/AdS cases. We leave these problems for a future attempt.
 
\cleardoublepage

\chapter{{\color{red!60!black}Hairy Black Holes and No-short Hair Theorem} }\label{Chapter_5}
\large
\textbf{This Chapter is based on the work: Phys.Rev.D 108 (2023) 4, L041501 (Letter) by R. Ghosh, Selim Sk, and S. Sarkar}~\cite{Ghosh:2023kge}.\\

\ni
As elucidated in the previous chapters, compact objects are unique laboratory to probe various aspects of gravity. Even among these objects, BHs stands out for their inherent simplicity. In stark contrast to other relativistic configurations, the spacetime outside a stationary and asymptotically flat vacuum 
BH solution of GR obeys the so-called uniqueness property~\cite{heusler1996black, Mazur:2000pn, Robinson:2004zz, Chrusciel:2012jk, PhysRev.164.1776}. That is, such a spacetime is uniquely represented by the Kerr metric having only two parameters, namely the mass $M$ and spin $a$. Even in non-vacuum scenario, it is believed that BH solutions of GR will satisfy the no-hair hypothesis~\cite{Ruffini:1971bza}, dictating that the gravitational collapse washes away all information about any additional parameters (termed as "hairs") and the final BH can be specified only in terms of conserved quantities such as mass, angular momentum and electric/magnetic charges measured at asymptotic infinity. Heuristically  speaking~\cite{Nunez:1996xv}, any matter fields residing outside a
BH would either be emitted away or absorbed by the BH itself unless those
fields were associated with conserved charges at asymptotic infinity. The initial motivation supporting this claim came from the seminal works of Bekenstein~\cite{Bekenstein:1971hc, Bekenstein:1972ky}, which state that stationary BHs cannot support any exterior scalar, vector, or spin-$2$ meson fields. These results soon led to the belief that the no-hair conjecture is true irrespective of the matter content.\\

\ni
However, this vast generalization was invalidated by the first counterexample found in Einstein-Yang-Mills theory, which supports the so-called "colored" BH solutions that depend on an additional parameter not associated to any conserved charge. The increasing list of counterexamples also contained several other BH solutions with dilatonic~\cite{Kanti:1995vq}, skyrmionic~\cite{Luckock:1986tr} and axionic~\cite{Campbell} hairs. Moreover, beyond the framework of GR, the presence of various putative higher curvature terms can also lead to hairy BH solutions. Over the years, many extensive studies have been performed to understand their non-Kerr signatures~\cite{Ryan:1995wh, Ryan:1997hg, Gair:2007kr, Bambi:2011jq, Isi:2019aib}. However, in order to suitably capture the presence of extra hairs via only far-away observations, these hairs must extend sufficiently outside the horizon. In contrary, BHs with "short" hairs (confined solely to the near-horizon regions) might mimic Kerr-like signatures when probed in the far field regions, though their near-horizon structure could be very different from that of the Kerr BH. Given this crucial
observational relevance, we want to investigate whether BHs can support short hairs.\\

\ni
Note that there is no a priori reason to suspect such short hairs can not exist outside a BH. On contrary, the non-linear character of matter fields could lead to the growth of BH hairs, which might be short as well. Nevertheless, under the assumptions of the weak energy condition (WEC) and the non-positive trace condition on matter, it has been established that static and spherically symmetric BH solutions in GR adhere to the \textit{no-short hair theorem}. It provides a lower bound on the length of existing hairs, requiring them to extend at least three-halves of the horizon radius~\cite{Nunez:1996xv}, which interestingly correspond to the innermost LR~\cite{Hod:2011aa}. In simpler terms, if there are hairs around a BH, their effects will become detectable within the vicinity of the LR region alone. Consequently, the presence of BH hairs would be evident through various astrophysical observations, such as shadow imaging and the study of QNMs, which are sensitive to the underlying LR structure.\\

\ni
Driven by its important theoretical and observational implications, we now pose the following question: Do the underlying field equations or the dimensionality of spacetime play a fundamental role in determining the extend of hairs around a BH? More specifically, we seek to understand whether a result akin to the no-short hair theorem in GR can be derived independently of the gravitational field equations in any spacetime dimensions ($D \geq 4$). If such a proposition holds true, it would represent a novel extension of the theorem to encompass any theory of gravity that permits the existence of hairy BHs. Curiously, we shall now demonstrate that the above question bears a negative answer. In particular, we shall prove that all existing hairs of any static, spherically symmetric, and asymptotically flat $D$-dimensional BHs must extend at least up to the innermost LR, regardless of the specific theory of gravity being considered. This generalization has two significant consequence. Firstly, it aids in comprehending the unified characteristics of BH solutions in a theory-agnostic manner. Secondly, it underscores that the no-short hair property cannot be regarded as a test of GR. In addition, our analysis also leads to other salient implications. For instance, we shall investigate the existence of short hairs on horizonless compact objects, expanding upon the approach outlined in Ref.~\cite{Peng:2020hkz}. Moreover, we will also delve into several generalizations of the studies presented in Refs.~\cite{Hod:2013jhd, Chakraborty:2021dmu}, constraining the size of LRs in various higher-dimensional/modified theories of gravity.
%%%%%%%%%%%%%%%%%%%%%%%%
\section{{\color{blue!70!brown} Geometry of Hairy BHs}}
%%%%%%%%%%%%%%%%%%%%%%%%
The first step towards proving our result is to consider the general static and spherically symmetric $D$-dimensional ($D \geq 4$) BH spacetime is described by the metric
%%%%%%%%%%%%%%%%%%%%%%%%%
\bea \label{Hair_Metric}
ds^2 = - f(r)\, dt^2 + \frac{1}{k(r)}dr^2 + h(r) d\Omega_{(D-2)}^2\, ,
\eea
%%%%%%%%%%%%%%%%%%%%%%%%%
where $r$ is a Schwarzschild-like radial coordinate so that he metric components are positive outside the outermost non-extremal event horizon at $r=r_H$, which is the largest positive root of $k(r) =0$. Also, the "areal coordinate" denoted by $h(r)$ is assumed to be a strictly increasing function outside the horizon. Then, the staticity of the spacetime implies that the norm of $\partial_t$ must vanish at $r=r_H$~\cite{Vishi2003}, i.e., $f(r_H) = 0$. Moreover, in case of asymptotically flat spacetime, we also have $f(r)\rightarrow 1 - C/r^{D-3} + \mathcal{O}(r^{-(D-2)}),$ $  
k(r)\rightarrow 1 - C/r^{D-3} + \mathcal{O}(r^{-(D-2)}),$ $h(r) \sim r^2\,$ as  $ r \to \infty$. Though $C$ is related to the ADM mass of the spacetime, we will not assume any particular sign of $C$. Finally, using the asymptotic flatness and non-extremality of the horizon, one can deduce that $f'(r_H) > 0$, and $k'(r_H) > 0$.\\

\ni
The metric in \ref{Hair_Metric} represents a BH solution of a theory of gravity (which we will not explicitly assume) sourced by an energy-momentum tensor $T_{\mu \nu}$. Now, we want to investigate various properties of this non-zero energy-momentum tensor, which encodes the presence of hairs. Firstly, as a consequence of spherical symmetry, $T^\mu_\nu$ must be invariant under any rotations in the transverse $(D-2)$-plane spanned by the coordinates $\{\theta_i\}, \, i = 1, \, 2, \cdots, (D-2) $. Thus, $T^t_{\theta_i}\,=\, T^r_{\theta_i}\,=\,0$, as they single out a particular direction in space. Also, the transverse sector of $T^\mu_\nu$ should be proportional to the identity matrix, which is the only covariant matrix under all rotations. Putting all these conditions together, $T^{\mu}_{\nu}$ have just four non-zero independent components, namely $\{T^t_t :=-\rho\, ;\, T^t_r \, ;\, T^r_r := p\, ;\, T^{\theta_1}_{\theta_1 } := p_T \}$ \footnote{Note that $T^t_r$ can be set to zero by staticity ($t \to -t$).}. Here, the quantities $\rho$, $p$, and $p_T$ are functions of $r$ only and represents energy density, radial, and tangential pressure, respectively. Besides them, physical invariants such as $T^\mu_\nu\, T^\nu_\mu$ should be non-divergent at the event horizon, as demanded by its regular nature. Then, using the radial component of the energy-momentum conservation equation $\nabla_\mu\,T^\mu_\nu\,=\,0$, we obtain
%%%%%%%%%%%%%%%%%%%%%%%%
\begin{align}\label{Hair_rconserve}
      \hat{P}'(r) \,=\,\frac{h^{D/2-1}}{2\, f}\,(p+\rho)\,\Delta\,+\, \frac{h^{D/2-1}}{2}\,h'\, T\, ,
 \end{align}
%%%%%%%%%%%%%%%%%%%%%%%%
where $'$ denotes a radial derivative, $\hat{P}= h^{D/2}\, p$, $\Delta\,=\,(f\,h'\,-\,h\,f')$, and $T$ represents the trace of the $T^\mu_\nu$. The above equation (valid for any $D \geq 4$) is a theory-agnostic generalization of the corresponding 4-dimensional GR result given in Ref.~\cite{Hod:2011aa}.\\

\ni
With the help of \ref{Hair_rconserve}, we now show that $\hat{P}(r)$ cannot vanish arbitrarily close to the horizon if matter obeys the following three conditions~\cite{Nunez:1996xv, Hod:2011aa}:\\
\\(i) The WEC constrains the energy and radial pressure as $\rho \geq 0$, and $\rho + p \geq 0$.\\
\\(ii) The energy-momentum tensor is assumed to have a non-positive trace ($T \leq 0$), which implies $p + (D-2)p_T \leq \rho$. This condition ensures the presence of BH hairs by restricting the matter theories.\\
\\(iii) The energy density and pressure falls off faster than $r^{-D}$ as spatial infinity is approached. It suggests that $\hat{P} \to 0$ as $r \to \infty$. As pointed out in Refs.~\cite{Nunez:1996xv, Hod:2011aa}, we need this condition to rule out any extra conserved charges. In other words, the hairs under consideration are not "secondary", according to the terminology of Ref.~\cite{Coleman:1991ku}.\\

\ni
As we shall see, these conditions specify the behavior of $\hat{P}$ near the horizon. In this regime, we can rewrite \ref{Hair_rconserve} in term of a well-behaved coordinate, namely the proper radial distance defined as $dx = k^{-1/2}\, dr$. In fact, the equivalence principle ensures the regularity ("no drama at the horizon") of this coordinate in the vicinity of the horizon. Therefore, we obtain 
%%%%%%%%%%%%%%%%%%%%%%%%%%
\begin{align}\label{radialp}
 \frac{d \hat{P}}{dx} \,=\,\frac{h^{D/2-1}}{2\,f}\,(p+\rho) \Big(f\,\frac{dh}{dx}\,-\,h\,\frac{df}{dx}\Big) +\, \frac{h^{D/2-1}}{2}\, T\, \Big(\frac{dh}{dx}\Big)\, .
\end{align}
%%%%%%%%%%%%%%%%%%%%%%%%%%
Since the horizon is assumed to be a regular surface, both sides of this equation must be finite. Now, using non-extremality $f'(r_H) >0$ along with WEC and non-positive trace conditions discussed above, one readily gets
%%%%%%%%%%%%%%%%%%%%%%%%%%
\begin{align}\label{Hair_rpH}
    p\,(r_H) = - \rho\,(r_H) \leq 0. 
\end{align}
%%%%%%%%%%%%%%%%%%%%%%%%%%
Here, it is really important to understand the non-triviality of the non-extremality condition, in particular. Interestingly, if the BH would have been an extremal one, then the term $(1/f)\, (df/dx)$ is a $0/0$-indeterminate form at the horizon (for non-extremal case, this term diverges), which may lead to a finite limit and invalidate the requirement given by \ref{Hair_rpH}. Then, using $k(r) > 0$, one also have

\begin{align} \label{Hair_PH}
    \hat{P}(r)\,\leq \,0\, ,  \hspace{3mm}\textit{and} \hspace{3mm} \hat{P}'(r) < 0\, , 
\end{align}
in the vicinity ($r \to r_H$) of the BH horizon. With these tools at our disposal, we now proceed to state and prove the central theorem.\\

\ni
\emph{Theorem}.--- If the matter content satisfies all three conditions stated above and there exists a non-empty interval $r_H\,\leq\,r\,\leq\,r_p$ where the function $\Delta(r)\,=\,(f\,h'\,-\,h\,f')\leq \,0$, we must have $\hat{P}'(r_H\,\leq\,r\,\leq\,r_p)\leq\,0$.\\

\ni
The proof of this theorem is straightforward and follows directly from \ref{Hair_rconserve} and \ref{Hair_PH}. Notice that as a consequence of the WEC, $T \leq 0$ on matter and the condition $h'(r) > 0$, both the terms in the RHS of \ref{Hair_rconserve} are non-positive. Then, it becomes immediately obvious that $\hat{P}'(r) \leq 0$ in the region $r_H\,\leq\,r\,\leq\,r_p$, completing the proof of the theorem. Therefore, the only fact remains to be shown is the existence of $r_p > r_H$, which we shall consider later. However, let us now ponder on what consequence does this theorem have on the length of the hair. Since $\hat{P}'(r) \leq 0$ in the domain $r_H \leq r \leq r_p$, \ref{Hair_PH} implies that $\hat{P}(r)$ should remain non-positive at least up to $r=r_p$. Thus, if hair exists, $|\hat{P}(r)|$ must have a local maxima at $r_{\text{hair}} \geq r_p$, where $r_{\text{hair}}$ denotes the extent of "hairosphere" outside the horizon~\cite{Nunez:1996xv, Hod:2011aa}. In other words, under the assumptions on matter content mentioned above, the existing hairs on an asymptotically flat, static, and
spherically symmetric BH solution of any theory of gravity
cannot be solely confined to the near-horizon regime and it must extend till $r_p$.\\

\ni
Let us now turn to investigate whether an $r_p > r_H$ exists up to which $\Delta(r) \leq 0$. In this context, we show that not only such an $r_p$ exists, it actually corresponds to the location of the innermost LR of the BH spacetime. To see this, it is suggestive to consider the equatorial 
 geodesics,
%%%%%%%%%%%%%%%%%%%%
\bea \label{Hair_geod}
    \Dot{r}^2\,= \, k(r) \left[ \frac{E^2}{f(r)}\,- \, \frac{L^2}{h(r)}\,-\, \epsilon\right]\, ,
\eea
%%%%%%%%%%%%%%%%%%%%
where the dot denotes a derivative with respect to an affine parameter, and $\epsilon = 1\, (\epsilon = 0)$ characterizes the timelike (null) geodesics. For circular orbits, we must have $\Dot{r}^2\,=0\,=\,(\Dot{r}^2)'$, which implies
%%%%%%%%%%%%%%%%%%%%
\begin{equation} \label{Hair_EL}
    {\Delta(r)}\, E^2\,=\,\epsilon\,f^2(r)\,h'(r)\, , \hspace{4mm}  {\Delta(r)}\, L^2\,=\, \epsilon\,h^2(r)\,f'(r)\, .
\end{equation}
%%%%%%%%%%%%%%%%%%%%
Thus, the allowed region for such orbits to exist must satisfy $\Delta(r) \geq 0$, equality is achieved only for null geodesics. Moreover, the positive roots of the equation $\Delta(r_\gamma) = 0$ represent the locations of the LRs. Then, using the asymptotic flatness and near-horizon boundary conditions, one can easily show that an odd number of such LRs must exist outside the horizon, as elaborated in \ref{LR_spherical}. These LRs divide the interval $[r_H, \infty)$ into an even number of regions, outermost of which must satisfy $\Delta(r)\,\geq \, 0$ due to asymptotic flatness. Thus, we must have $\Delta(r)\,\leq \, 0$ in the innermost region $r_H \leq r \leq r_\gamma^1$, where $r_\gamma^1$ denotes the location of the innermost LR. This completes our proof that the hairosphere of a BH must extend till the innermost LR, $r_{\text{hair}} \geq r_\gamma^1$. The novelty of our analysis lies in its generality, i.e., it is valid irrespective of the underlying field equations and for any spacetime dimensions $D \geq 4$. Now, we shall focus on several important consequences of \ref{Hair_rconserve} in the following sections.
%%%%%%%%%%%%%%%%%%%%%%%%
\section{{\color{blue!70!brown} Size of Static Shells}}
%%%%%%%%%%%%%%%%%%%%%%%%
In this section, we try to answer the question: Can a static shell of finite thickness exist entirely inside the region $r_H < r_1 < r_2 < r_\gamma^1$, where $r_1$ and $r_2$ are respectively the inner and outer radius of the shell? For this purpose, one should note that $\hat{P}(r_1) = \hat{P}(r_2)$ must vanish, as there is no matter
present outside the shell. Then, we run into a contradiction with the earlier theorem which dictates that $\hat{P}(r)$ must remains constant or decrease till the innermost LR. Therefore, irrespective of the underlying theory of gravity, a static shell of finite thickness cannot be entirely confined within the innermost LR.
%%%%%%%%%%%%%%%%%%%%%%%%
\section{{\color{blue!70!brown} Bound on LR Sizes of BHs in EGB Gravity}}
%%%%%%%%%%%%%%%%%%%%%%%%
So far, our analysis was completely theory-agnostic. However, if one indeed uses the field equations of a particular theory, it is possible to get an upper bound on the size of the
innermost LR of static, spherically symmetric and asymptotically flat BHs. This interesting result was first demonstrated in the context of $4$-dimensional BHs in GR~\cite{Hod:2013jhd}, and later generalized to $D > 4$~\cite{Chakraborty:2021dmu}.It was also shown that a similar statement is true in $5$-dimensional EGB gravity~\cite{Chakraborty:2021dmu}. Now, using our powerful theorem, we shall now extend this result for an arbitrary dimensions $D \geq 5$.\\

\ni
The Lagrangian of EGB gravity is given in \ref{EGB}, which leads to GR in the limit of vanishing coupling constant ($\lambda \to 0$). A variation of this action gives us the necessary field equations, $G^{(1)}_{\alpha\beta}\,+\,\lambda\,G^{(2)}_{\alpha\beta}\,=\,8\pi\,T_{\alpha\beta}$.The explicit forms of the $G^{(1)}_{\alpha\beta}$ and $G^{(2)}_{\alpha\beta}$ can be found in Ref.~\cite{Zhou:2011wa}. Then, any spherically symmetric and static BH solution of these field equations can be written as
%%%%%%%%%%%%%%%%%%%%%%%%
\begin{align} \label{Hair_EGBMetric}
    ds^2\,=\, -\, e^{-2\delta(r)}\,\mu(r)\, dt^2\,+\,\frac{1}{\mu(r)}\,dr^2\,+\,r^2\,d\Omega^2_{D-2}\, ,
\end{align}
%%%%%%%%%%%%%%%%%%%%%%%%%
where the regular non-extremal event horizon is located at $r\,=\,r_H$, so that $\mu(r_H)\,=\,0$, $\mu'(r_H)\,>\,0$, and $\delta(r)$ and its radial derivative is finite there \cite{Hod:2013jhd}. Moreover, asymptotic flatness implies $\mu(r) \to 1$ and $\delta(r) \to 0$ at infinity. Now, using the above field equations of EGB theory, we get two coupled differential equations,
%%%%%%%%%%%%%%%%%%%%
\begin{equation} \label{Hair_field1}
\begin{split}
    &\delta' =\,- \frac{8\, \pi\, r^3\, (p+\rho)}{(D-2)\, \mu\, \left[r^2 + 4\, \alpha \, (1-\mu)\right]}\, , \\
    \\
    & \mu' = \frac{2\, r^3\, (D-3)}{r^2\,+\,4\, \alpha \, (1-\mu)} \Bigg[ \frac{1-\mu}{2\, r^2}+ \frac{\alpha\,(D-5)\, (1-\mu)^2}{(D-3)\, r^4} -\frac{8\, \pi\, \rho}{(D-2)(D-3)}\Bigg]\, ,
\end{split}
\end{equation}
%%%%%%%%%%%%%%%%%%%%
where $T^t_t\,=\,-\rho$, $T^r_r\,=\,p$ and $\alpha\,=\, (D-3)(D-4)\lambda/2$. Then, at the location of the innermost LR $r_\gamma^1$, we have
%%%%%%%%%%%%%%%%%%%%
\begin{align}\label{Hair_lightring2}
\nonumber (D-1)\mu_\gamma\,-\,(D-3)\,+\,\frac{8\, \alpha\, \mu_\gamma(1-\mu_\gamma)}{(r_\gamma^1)^2} & -\,\frac{2\, \alpha\, (D-5)(1-\mu_\gamma)^2}{\left(r_\gamma^1\right)^2}\\
& =\,\frac{16\, \pi\, (r_\gamma^1)^2\, p(r_\gamma^1)}{D-2} \leq 0\, .
\end{align}
%%%%%%%%%%%%%%%%%%%%
Here, $\mu_\gamma$ and  $\delta_\gamma$ are shorthands for $\mu(r_\gamma^1)$ and $\delta(r_\gamma^1)$, respectively. Also, in the last line of the above equation we have used our theorem that dictates $p(r_\gamma^1) \leq 0$. Now, we want to solve \ref{Hair_field1} for $\mu(r)$, a function needed to determine the LR size. For this purpose, let us define a useful quantity as,
%%%%%%%%%%%%%%%%%
\begin{align}
    m(r)\,=\,\frac{r_H}{2}\,+\, \Omega_{D-2} \int_{r_H}^r \rho(x)\,x^{D-2}\, dx\ .
\end{align}
%%%%%%%%%%%%%%%%%
Here, $\Omega_{D-2}\,=\,2\pi^{(D-1)/2}/\Gamma[(D-1)/2]$ is the surface element of unit $(D-2)$-sphere and we have chosen the boundary condition, $m(r_H)\,=\,r_H/2\,>\,0$. In terms of this mass function, the solution takes the following form,
%%%%%%%%%%%%%%%%%
\begin{align}\label{Hair_EGBf}
    \mu(r)\,=\,1\,+\,\frac{r^2}{4\, \alpha}\Bigg[ 1-\sqrt{1+\frac{16\, \alpha\, M(r)}{r^{D-1}}}\, \Bigg]\, ,
\end{align}
%%%%%%%%%%%%%%%%%
where $M(r)\,=\,8\, \pi\, m(r)/(D-2)\Omega_{D-2}$. Finally, \ref{Hair_lightring2} can be re-expressed as a simpler polynomial equation:
%%%%%%%%%%%%%%%%%
\begin{align}\label{Hair_poly}
    (r^1_\gamma)^{2D-6}\,+\,16\, \alpha\, M_\gamma\,  (r^1_\gamma)^{D-5} \,-\,(D-1)^2\, M^2_\gamma\,\leq\,0\, ,
\end{align}
%%%%%%%%%%%%%%%%%
with $M_\gamma = M(r_\gamma^1)$. At this stage, let us first discuss the limiting case when $\alpha \to 0$. It is easy to check that one gets back the $D$-dimensional GR result~\cite{Hod:2013jhd, Chakraborty:2021dmu}, $r_{\gamma, (GR)}^1 \leq \left[(D-1)\, M_\gamma \right]^{1/(D-3)} = r_\gamma^{ST}$, as $M_{ST}\,\geq\, M_\gamma$, as expected. Therefore, the size of the innermost LR of a static, spherically symmetric and asymptotically flat D-dimensional BHs in GR sourced by matter obeying the aforementioned conditions is bounded above by the radius ($r_\gamma^{ST}$), the LR of Schwarzschild-Tangherlini (ST) BH \cite{Tangherlini:1963bw} with mass $M_{ST} := M(r \to \infty)$.\\

\ni
Coming back to EGB gravity with $\alpha \neq 0$, we realize that one should be careful about choosing the value of the coupling $\alpha$ so that the central object is actually a BH and not a naked singularity. With this caveat in mind, we shall now prove that Boulware-Deser (BD) BH \cite{Boulware:1985wk} has the largest LR among all static, spherical symmetric and asymptotically flat BH solutions of EGB gravity. To achieve this, it is instructive to define two polynomials in $r\,\in [0,\infty)$ as
%%%%%%%%%%%%%%%%%
\begin{align}
    F_1(r)\,&=\,r^{2D-6}\,+\,16\, \alpha\, M_\gamma\, r^{D-5}\,-\,(D-1)^2\, M^2_\gamma\, , \\
    F_2(r)\,&=\,r^{2D-6}\,+\,16\, \alpha\, M_{BD}\, r^{D-5}\,-\,(D-1)^2\, M^2_{BD}\, , 
\end{align}
%%%%%%%%%%%%%%%%%
where $M_{BD}\,:=\,M(r\to \infty)\,\geq M_\gamma$. As ensured by the Descartes' rule of signs, both of these polynomials $F_1$ and $F_2$ have single positive root $r_m$ and $r_\gamma^{BD}$ respectively, for either signs of $\alpha$. Here, $r_\gamma^{BD}$ represents the LR of a BD BH with mass $M_{BD}$. In contrast, \ref{Hair_poly} suggests that the innermost LR of any BH solution of EGB gravity is bounded above by $r_m$, i.e., $r_\gamma^{1} \leq r_m$. Moreover, since $M_{BD} \geq M_\gamma$, we must have $F_2(r=0) \leq F_1(r=0)$. Then, it is suggestive to evaluate $F_2(r)$ at the location of the root ($r_m$) of $F_1(r)$. Some simple algebraic manipulation gives us, 
%%%%%%%%%%%%%%%%%%%%
\begin{equation}
    F_2(r_m) = \left[\frac{r_m^{2D-6}}{M_\gamma} + (D-1)\,M_{BD}\right]\,\left(M_\gamma-M_{BD}\right) \leq 0\, .  
\end{equation}
%%%%%%%%%%%%%%%%%%%%%
The above inequality, in turn, implies that $r_\gamma^{BD} \geq r_m \geq r_\gamma^1$, which completes the proof. In future, it would be interesting to see whether a similar result can be extended for other higher curvature theories.
%%%%%%%%%%%%%%%%%%%%%%%%
\section{{\color{blue!70!brown} Possible Extension for Horizonless Compact Objects}}
%%%%%%%%%%%%%%%%%%%%%%%%
So far, we concentrated solely on hairs in BH spacetimes. However, let us now outline a possible way in which our result can be generalized to horizonless compact objects. The associated spacetime metric is still given by \ref{Hair_Metric}, but this time in the absence of a horizon, we set the inner boundary conditions at the center of the object $r\,=\,0$. As suggested in Ref.~\cite{Hod:2014ena}, we shall choose $L(0)\,=\,0$, and $R(0)\,>\,0$ with the assumption that $h(0) = 0$, where $L(r) = h(r)\, f'(r)$ and $ R(r) = f(r) \, h'(r)$. Interestingly, due to asymptotic flatness, the boundary conditions at spatial infinity remain unchanged. Then, using regularity of various components of the stress-energy tensor, we must have $\hat{P}(r)\,=\,0$ at the center of the compact object. We shall also assume that the WEC on matter and asymptotic fall-off of $\hat{P}(r)$ is same as in the case of BHs. However, in order to extend the no-short hair result for horizonless objects, we need to assume that the trace of the energy-momentum tensor $T^{\mu}_{\nu}$ is non-negative ($T\,\geq\,0$) \cite{Peng:2020hkz}, which is exactly opposite to the BH scenario. \\

\ni
The asymptotic fall-off condition of $\hat{P}(r)$ makes sure of the existence of an even (but non-zero) number of LRs; otherwise in the absence of LRs, $\hat{P}(r)$ would increase monotonically with $r$. Then, one can also deduce that $\hat{P}(r)$ must have a local maximum at the innermost LR ($r=r_\gamma^1$) or beyond. Thus, similar to BHs, the hairs of horizonless compact objects must also extended till the innermost LR, generalizing the results of Ref.~\cite{Peng:2020hkz} in a theory-agnostic fashion. It is interesting to note that for "usual" matter content, one always have $\rho \gg (p,\, p_T)$, which suggests that the condition $T 
\geq 0$ is not satisfied. Thus, the no-short hair theorem is not applicable for ordinary celestial objects. However, $T \geq 0$ condition may hold for objects made of "exotic" matter. In that case, the no-short hair result can give us useful information about their structures.
%%%%%%%%%%%%%%%%%%%%%%%%
\section{{\color{blue!70!brown} Summary}}
%%%%%%%%%%%%%%%%%%%%%%%%
Hairy BHs may showcase potentially distinguishable signatures in both GW and shadow observations. However, to suitably capture the presence of hairs via only far-away observations, these hairs must extend sufficiently outside the horizon. In the context of $4$-dimensional static, spherically symmetric and asymptotically flat BH solutions, this no-short hair property is guaranteed provided the matter content obeys the WEC and non-positive trace condition~\cite{Nunez:1996xv, Hod:2011aa}. It adheres a lower bound on the length
of existing hairs, requiring them to extend at least to the innermost LR of the underlying BH spacetime. In this chapter, we have extended this important result in a theory-agnostic and dimension-independent way. Our result has enormous theoretical and observational relevance. On theory side, it provides a unified way to understand several features of modified gravity BHs. Moreover, it gives us an observational tool to probe the presence of BH hairs, which may give rise to intriguing new phenomena, including potential modifications in gravitational lensing effects and BH shadow observables.\\

\ni
Besides the aforesaid results, there are other important consequence of our work. One noteworthy finding is to establish an upper bound on the size of the innermost LR of BH solutions of EGB theory in dimensions $D \geq 5$. This upper limit has important implications, as it can impose constraints on both the shadow size, as discussed in \cite{Chakraborty:2021dmu}, and the real component of the eikonal QNMs of EGB black holes when subjected to perturbations, as explored in \cite{Hod:2013jhd}. Furthermore, we touch upon the potential extension of the no-short hair theorem to horizonless compact objects, offering some preliminary insights in this direction.\\

\ni
The implications of our findings are significant and warrant further consideration. For instance, an extension of our BH-result could be to consider asymptotically flat hairy wormhole spacetimes. In this case, it seems possible to prove the no-short hair theorem by setting the inner boundary condition at the throat ($r=b>0$) as long as $R(b) > L(b)$, and assuming other necessary conditions such as WEC with $T>0$ hold true. Apart from this, exploring the extension of our work to scenarios where certain assumptions, like spherical symmetry, asymptotic flatness, or the WEC on matter, are relaxed presents an intriguing avenue for future explorations. Notably, investigating the short-hair characteristics of rotating BHs, as suggested in \cite{Hod:2014sha}, holds particular observational significance. However, we defer these potential extensions to future research endeavors.

\cleardoublepage

\clearpage

\begin{center}
\vspace*{\stretch{1}}
\Huge{\textbf{Part II.}}\\
\Huge{\textbf{Observational Signatures of Modified Gravity}}
\vspace*{\stretch{1}}
\end{center}

 \clearpage 

\chapter{{\color{red!60!black}Black Hole Area-Quantization and Its Observational Signatures} }\label{Chapter_6}
\large
\textbf{This Chapter is based on the works: Phys. Rev. D 104 (2021) 8, 084049 and Phys. Rev. D 105 (2022) 4, 044046 by K. Chakravarti, R. Ghosh, and S. Sarkar}~\cite{Chakravarti:2021jbv, Chakravarti:2021clm}.\\

\ni
Apart from various theoretical consistency discussed in earlier chapters, modified theories of gravity should also be confronted with numerous observations that provide a compelling way to constrain possible deviations from GR. For this purpose nothing is more suited than the GW observations by the LIGO-Virgo collaboration~\cite{Abbott:2016blz, TheLIGOScientific:2016wfe, TheLIGOScientific:2016src, Abbott:2016nmj, Abbott:2017vtc, TheLIGOScientific:2017qsa}. A wealth of such observations has opened up a new avenue of testing GR in strong gravity regimes. For example, there has been a recent surge of works towards understanding the possible signatures of quantum gravity near the BH horizon. Such quantum effects may result in deviations from the all-absorbing nature of classical BHs~\cite{Bekenstein:1974jk, Bekenstein:1995ju, Oshita:2019sat}. One such model known as the "BH area-quantization" was proposed by Bekenstein and Mukhanov~\cite{Bekenstein:1974jk, Bekenstein:1995ju}, according to which the area of BH event horizons is quantized in equidistant steps resulting in selective absorption at discrete frequencies. Such a BH, dubbed as quantum BH (QBH), will have distinct signatures in Hawking radiation spectrum~\cite{Bekenstein:1995ju}, and more interestingly, in both inspiral~\cite{Agullo:2020hxe, Datta:2021row, Chakravarti:2021jbv} and the late-ringdown~\cite{Foit:2016uxn, Cardoso:2019apo, Chakravarti:2021clm} stages of a BH-binary event.\\

\ni
Before we investigate these effects in great detail, let us first have a closer look at the BH area-quantization and understand its consequences. According to the proposal of Bekenstein and Mukhanov~\cite{Bekenstein:1974jk, Bekenstein:1995ju}, BH horizon area is discretized as $A = \alpha\,  \ell_p^2\, N$, where $\alpha$ is a phenomenological constant, $\ell_p = \sqrt{\hbar\, G/c^3} \sim 1.6 \times 10^{-35}\, \textrm{m}$ is the Planck's length, and $N$ is positive integer. Though the original proposition was motivated by concept of adiabatic invariance, such quantization might also arise as a prediction of any quantum theory of gravity. For instance, Loop Quantum Gravity predicts a similar quantization scheme for the BH horizon area.\\ 

\ni
In order to understand various ingredient of the above model, we notice that $N$ is usually a huge number, which is evident from the smallness of the Planck's length. In particular, a solar mass Schwarzschild BH corresponds to $N \sim 10^{78}$. On the other hand, the constant $\alpha$ was originally chosen to be $8\pi$~\cite{Bekenstein:1974jk}. However, over years several other choices are also motivated. Using arguments based on Bohr's correspondence principle, a value $\alpha = 4 \ln 3$ is proposed in \cite{Hod:1998vk}. Later, the study of Schwarzschild QNMs, suggests a value of $\alpha = 8\pi$~\cite{Maggiore:2007nq}. However, for the purpose of this chapter, we shall treat $\alpha$ as a free parameter in the model to be measured from observations.\\

\ni
In contrast to the idea of BH area quantization, it has been argued that, in general, the entropy should be quantized in a uniform spectrum~\cite{Kothawala:2008in}. In the context of GR, Bekenstein's entropy formula applies and uniform quantization of BH entropy or area are essentially the same. However, in the presence of putative modification of GR, BH entropy  is no longer proportional to the horizon area, and may contain sub-leading correction terms (both power law and logarithmic)~\cite{Wald:1993nt, Bombelli:1986rw}. In fact, even in the context of GR, such sub-leading terms may appear in the entropy-area relationship as a consequence of tracing over hidden (by the horizon) degrees of freedom of a quantum field in a state different from the vacuum~\cite{Das:2005ah, Das:2007mj, Sarkar:2007uz}. In these scenario, uniform entropy quantization will lead to a non-uniform area quantization.\\

\ni
As a consequence of area discretization (both uniform and non-uniform), BHs can only absorb at certain characteristic frequencies. As an example, it was showed in Ref.~\cite{Agullo:2020hxe} for uniform area-quantization that such frequencies corresponding to transition $N \to N+n$ are given by
%%%%%%%%%%%%%%%%%%%%%%
\bea \label{AQ_uniform}
\omega_n = \frac{\alpha\, \kappa}{8 \pi}\, n + 2\, \Omega_H + \mathcal{O}\left( N^{-1}\right)\ ,
\eea 
%%%%%%%%%%%%%%%%%%%%%%
where $\kappa = \frac{1-\chi^2}{2M(1+\sqrt{1-\chi^2})}$ and $\Omega_H = \frac{\chi}{2M(1+\sqrt{1-\chi^2})}$ are respectively the surface gravity and angular velocity at the horizon of a Kerr BH with mass $M$ and spin $\chi$. Thus, a uniformly area-quantized BH will essentially behave like a reflecting star at frequencies, $f \neq f_n := \omega_n/2\pi$. Just to get a sense of the numbers, let us consider Schwarzschild BHs ($\chi = 0$). Then, it is easy to see that for typical astrophysical BHs
detected by LIGO-Virgo collaboration with mass
$M \sim (10-50) M_\odot$, the values of the low-$n$ characteristic frequencies are approximately $100$ Hz for $\alpha = 8 \pi$. Thus, it seems that astrophysical BHs magnify the Planck-scale discretization to the realm of
GW-observations. \\

\ni
In the subsequent sections, we shall discuss how the above formula for characteristic frequencies will change for a non-uniform area-quantization. Moreover, we shall discuss its effects on the GW emissions in both inspiral and ringdown stage of a binary. In particular, we shall focus on the tidal heating and the phasing of the GW waveform in the inspiral stage, and emission of late-time echo signals in the ringdown phase.
%%%%%%%%%%%%%%%%%%%%%%%%
\section{{\color{blue!70!brown} Features of Non-uniform Area Quantization}}
%%%%%%%%%%%%%%%%%%%%%%%%
As discusses earlier, the presence of sub-leading corrections in the area-entropy relationship can result in non-uniform area discretization. Since it is possible to have both power-law and logarithmic corrections to Bekenstein's entropy formula, we may consider two different models:
%%%%%%%%%%%%%%%%%%%%%%
\begin{align} \label{AQ_power}
\textrm{Power-law correction:}\kern 10pt A = \alpha\, \ell_{p}^{2}\, N\, \left ( 1 + C \, N^{\nu}\right)\, ;
\end{align}
%%%%%%%%%%%%%%%%%%%%%%
\begin{align} \label{AQ_log}
\textrm{Logarithmic correction:}\kern 10pt A = \lambda\, \ell_{p}^{2}\, W(x)\, ;
\end{align}
%%%%%%%%%%%%%%%%%%%%%%
where $x = \textrm{exp}\left(\alpha\, N/ \lambda\right)/\lambda$, the constants $(C, \nu, \lambda)$ are model parameters, and $W$ is the Lambert $W$-function arises when we invert the modified area law with logarithmic correction. It is important to note that the constant $\nu$ parameterizing the sub-leading power-law correction must be negative, which is required to recover Bekenstein's entropy formula as $N \to \infty$. Moreover, we get back the uniform area-quantization as limiting cases $C \to 0$ and $\lambda \to 0$, respectively.
%%%%%%%%%%%%%%%%%%%%%%%%
\subsection{{\color{red!70!blue} The Transition Frequencies}}
%%%%%%%%%%%%%%%%%%%%%%%%
Classically, there is no
restriction on the variations of horizon area ($A$), as BH mass $M$ and angular momentum $J$ can vary continuously:
%%%%%%%%%%%%%%%%%%%%%%
\bea \label{AQ_vary}
\frac{\kappa}{8 \pi}\, \delta A = \delta M - \Omega_H\, \delta J\, ,
\eea
%%%%%%%%%%%%%%%%%%%%%%
where $A = 8 \pi \, M^2 \left( 1 + \sqrt{ 1- \chi^2}\right)$ is the horizon area of a Kerr BH with $\chi = J/M^2$. However, the situation changes dramatically as the BH area and angular momentum are quantized. Such a BH will behave like an atom and can only undergo transitions to discrete mass/energy levels as it interacts with external perturbations, which we shall consider as the dominant mode ($l = m = 2$) of GW emission as observed by LIGO-Virgo. Thus, we will only focus on $(N, j) \to (N+n, j+2)$ transitions, where $\Delta J = 2\hbar$ is the change in angular momentum. The corresponding angular frequencies are given by,
%%%%%%%%%%%%%%%%%%%%%%
\begin{align} \label{AQ_powerom}
\textrm{Power-law correction:}\kern 10pt \omega_{N,n} = \frac{\alpha\, \kappa}{8 \pi}\, \left\{1 + C \left(1 + \nu\right) N^{\nu} \right\}\, n\, + 2\, \Omega_H +\, {\mathcal O} \left(N^{-1}\right)\, ;
\end{align}
%%%%%%%%%%%%%%%%%%%%%%
\begin{align} \label{AQ_logom}
\textrm{Logarithmic correction:}\kern 10pt \omega_{N,n} = \frac{\alpha\, \kappa}{8 \pi}\, \frac{W(x)}{1+W(x)}\, n + 2\, \Omega_H +\, {\mathcal O} \left(N^{-1}\right)\, .
\end{align}
%%%%%%%%%%%%%%%%%%%%%%
A few points are important to note. Both these formulas reduce to \ref{AQ_uniform} in the limits $C \to 0$ and $\lambda \to 0$, as expected. However, unlike uniform quantization, these characteristic frequencies generally depend on the initial state ($N$) of transitions.
%%%%%%%%%%%%%%%%%%%%%%%%
\subsection{{\color{red!70!blue} The Overlap Condition}}
%%%%%%%%%%%%%%%%%%%%%%%%
Due to the spontaneous decay via Hawking radiation, the above transition lines are not sharply defined. Following a calculation by Page~\cite{Page:1976ki}, one can evaluate the corresponding line width $\Gamma(M,\chi)$ for classical BHs, which will serve as a upper bound on the actual line width for QBH since area-quantization enhances the stability~\cite{Agullo:2020hxe}. One can find out a numerically fitted formula for $\Gamma(M,\chi)$ in Ref.~\cite{Datta:2021row}.\\

\ni
Now, the features of a area-discretized BH can be suitably distinguished from that of classical BHs only if the transition lines are all distinct~\cite{Coates:2021dlg}.However, due to line broadening, there is a possibility of overlap among the nearly lines. To get a better handle on the overlap,
we define the following quantity:
%%%%%%%%%%%%%%%%%%%%%%%%
\bea \label{AQ_overlap}
R(\chi, N) = \frac{ \Gamma( M, \chi)}{\omega_{N,n} - \omega_{N,n-1}}\, .
\eea
%%%%%%%%%%%%%%%%%%%%%%%%
As long as this ratio is less than unity, there are no overlaps. This statement can be written mathematically as,
%%%%%%%%%%%%%%%%%%%%%%
\begin{align} \label{AQ_powerov}
\textrm{Power-law correction:}\kern 10pt \Gamma_B < 1 + C\, \left( 1+ \nu \right) N^\nu\, ;
\end{align}
%%%%%%%%%%%%%%%%%%%%%%
\begin{align} \label{AQ_logov}
\textrm{Logarithmic correction:}\kern 10pt \Gamma_B < \frac{W(x)}{1+W(x)}\, ;
\end{align}
%%%%%%%%%%%%%%%%%%%%%%
which we shall refer as the \textit{no-overlap conditions}. Here, we have used the notation that $\Gamma_B = \left(8\pi  / \alpha \kappa \right) \Gamma$. Since we are interested to study stellar mass BHs with $N \sim 10^{78}$,  to have a detectable difference from the uniform area-quantization, we must choose $\nu \sim -10^{-2}$ for power-law corrected area law. However, for the logarithmic correction, the RHS of \ref{AQ_logov} is indistinguishable from unity, which is the case for uniform area-quantization.  This
suggests that we will not be able to probe any log-correction of entropy using GW observables and therefore, from now onwards, we shall solely focus on the power-law corrected area-quantization.
%%%%%%%%%%%%%%%%%%%%%%%%
\section{{\color{blue!70!brown} Observing Effects in BH Binary Inspiral Phase}}
%%%%%%%%%%%%%%%%%%%%%%%%
Let us consider the binary inspiral of two QBHs with masses and spins $(M_1, \chi_1)$ and $(M_2, \chi_2)$. As they revolve around a common center of mass, GWs are emitted according to the flux equation given by \ref{flux}. In effect, the orbital energy is lost as the emitted GWs either takes away energy to infinity or get absorbed at the horizon of the components BHs. Classically, a BH absorbs radiation of all frequencies that incident upon its horizon, which adds an additional flux term causing the so-called "tidal heating" (TH), which is well-understood within the analytical framework of PN expansion~\cite{Alvi:2001mx, Chatziioannou:2012gq, Chatziioannou:2016kem}. And, the corresponding orbital phasing can be calculated using the following relation~\cite{Tichy:1999pv},
%%%%%%%%%%%%%%%%%%%%%%%%
\begin{align} \label{AQ_phasing}
\Psi(f)= 2\left(\frac{t_0}{M_1+M_2}\right) v^3-2 \phi_0-\frac{\pi}{4} -\frac{2}{M_1+M_2} \int^v d x\left(v^3-x^3\right) \frac{E^{\prime}(x)}{F(x)}\, .
\end{align}
%%%%%%%%%%%%%%%%%%%%%%%%
Here, $F(x)$ contains the flux due to tidal heating alone. However, due to area quantization, the component BHs will absorb selectively at characteristics frequencies given by \ref{AQ_powerom}. As a result, it causes dephasing, for which the relevant formulas are given by Eqs.(10-12) of Ref.~\cite{Datta:2021row}:
%%%%%%%%%%%%%%%%%%%%%%%%%%
\begin{align} \label{AQ_TH}
\Psi_{\mathrm{THQBH}} =\, &   \frac{3}{128\, \rho} \left(\frac{1}{v}\right)^5 \left[-\frac{10}{9}\, v^5\, \Psi_5\, \{3 \log (v)+1\}-\frac{5}{168}\, v^7\, \Psi_5\, (952 \rho+995) \right. \nonumber \\ 
& \left. +\, \frac{5}{9}\, v^8\, \{3 \log (v)-1\}\left(-4\, \Psi_8+\Psi_5\,  \psi_{\mathrm{SO}}\right)\right]\, ,
\end{align}
%%%%%%%%%%%%%%%%%%%%%%%%%%
where $\rho$ is the symmetric mass ratio, $\Psi_5 = \mathcal{H}^{(1)}\, \mathcal{A}_5^{(1)} + \mathcal{H}^{(2)}\, \mathcal{A}_5^{(2)}$, $\Psi_8 = \mathcal{H}^{(1)}\, \mathcal{A}_8^{(1)} + \mathcal{H}^{(2)}\, \mathcal{A}_8^{(2)}$, and $v = \left[\pi \left(M_1 + M_2\right) f\right]^{1/3}$ is the PN velocity parameter. Therefore, the TH contribute as the $2.5$-PN leading order effect. Moreover, $\psi_{\mathrm{SO}}$ is the spin-orbit interaction term, 
%%%%%%%%%%%%%%%%%%%%%%%%%%
\begin{align}
\psi_{\mathrm{SO}} =\, & \frac{1}{6}\left[(-56 \rho-73 \sqrt{1-4 \rho}+73)\left(\hat{L} . \hat{S}_1\right) \chi_1 \right . \nonumber \\
& \left . +(-56 \rho+73 \sqrt{1-4 \rho}+73)\left(\hat{L} . \hat{S}_2\right) \chi_2\right]\, .
\end{align}
%%%%%%%%%%%%%%%%%%%%%%%%%%
Here, $\hat{L}$ and $\hat{S}_i$ are the directions of the orbital angular momentum and the $i$-th ($i = 1,2$) component’s spin, respectively. Moreover, we have
%%%%%%%%%%%%%%%%%%%%%%%%%%
\begin{align}
& \mathcal{A}_5^{(i)} = \left(\frac{M_i}{M}\right)^3\left(\hat{L} . \hat{S}_i\right) \chi_i\left(3 \chi_i^2+1\right)\, ; \\
& \mathcal{A}_8^{(i)} = 4 \pi \mathcal{A}_5^{(i)}+\left(\frac{M_i}{M}\right)^4\left(3 \chi_i^2+1\right) \times\left(\sqrt{1-\chi_i^2}+1\right) .
\end{align}
%%%%%%%%%%%%%%%%%%%%%%%%%%
Finally, the profiles $\mathcal{H}^{(i)}$'s represent the absorptivity of the horizons, which for a classical BH is unity at all frequencies. However, for an area-quantized BH it is defined as a sum (not normalized) of Gaussian (G) centered at the characteristics frequencies $f_{n} = \omega_n / 2 \pi $ with full-width-half-maximum (FWHM) denoted by $\Gamma$,
%%%%%%%%%%%%%%%%%%%%%%%%
\bea \label{AQ_absorb}
\mathcal{H}(f) = \sum_n G\left(\mu = f_{N,n}\, ,\, \sigma = \frac{\Gamma}{2\, \sqrt{2\, \mathrm{log}\, 2}} \right)\, .
\eea
%%%%%%%%%%%%%%%%%%%%%%%%
Here, in the context of inspiralling binaries, the sum over $n$ goes till the corresponding critical frequency $f_{N,n} \leq f_c$, with $f_c$ being the frequency when two BHs come into contact to initiate the merger phase.
%%%%%%%%%%%%%%%%%%%%%%%%
\begin{figure}[!htp]
    \centering
    \includegraphics[scale=0.8]{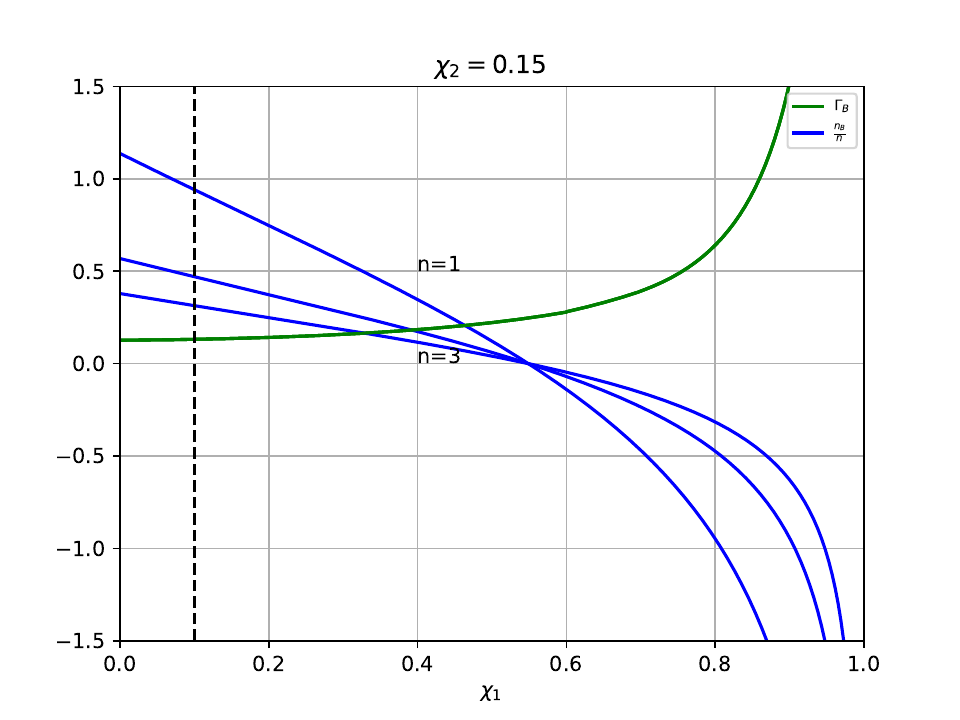}
    \caption{\textbf{$\Gamma_B$ and $n_B/n$ are plotted as a function of $\chi_1$.} Other parameters are chosen to be: $\alpha = 8\pi$, $M_1 = 20 M_\odot$, $M_2 = 30 M_\odot$, and $\chi_2 = 0.15$. The black dashed line represents a particular choice of $\chi_1 = 0.1$, for which there is no allowed transition in the case of uniform area-quantization as $n_B/n < 1$. However, it is possible to choose $C<0$ so that there are some allowed transitions for $\chi_1 = 0.1$. Also, the region where $n_B \leq 0$ is disallowed, since it implies $\omega_c \leq 2\Omega_H$.}
    \label{spider}
\end{figure}
%%%%%%%%%%%%%%%%%%%%%%%%
One can express this inequality in a more suggestive way by using \ref{AQ_absorb},
%%%%%%%%%%%%%%%%%%%%%%%%
\bea \label{AQ_nB}
n \leq n_{max} \equiv \left[ \frac{n_B}{1 + C\, N^{\nu}\, \left(1 + \nu \right)} \right] \, ,
\eea  
%%%%%%%%%%%%%%%%%%%%%%%%
where $n_B = \left(8\pi/\alpha\kappa\right)\left(\omega_c - 2\, \Omega_H\right)$ is a dimensionless quantity and $[x]$ denotes the integer part of $x$. Now, to calculate the contact frequency $f_c(M_1, M_2, \chi_1, \chi_2)$ for slowly spinning BHs, one may use Kepler's law~\cite{Datta:2021row}. On the other hand, at high values of $\chi$ where Kepler's law is not applicable, they are not of much concern because such spin values are mostly ruled out by the no-overlap condition in \ref{AQ_powerov}.
Moreover, it is instructive to combine \ref{AQ_nB} and \ref{AQ_powerov} as,
%%%%%%%%%%%%%%%%%%%%%%%%
\bea \label{AQ_Comb}
\Gamma_B\, <\, 1 + C\, N^\nu\, (1 + \nu)\, \leq\, \frac{n_B}{n}\, .
\eea
%%%%%%%%%%%%%%%%%%%%%%%%
Therefore, the $n$-th transition is allowed only if the above inequality is satisfied. Also, since $n_B$ and $\Gamma_B$ are independent of the details $(C, \nu)$ of area-quantization, \ref{AQ_Comb} gives us a very effective way to check for forbidden transitions, see \ref{spider}. For example, in the case of uniform area-quantization ($C=0$), transitions with $n > n_B$ are disallowed in the inspiral regime.
%%%%%%%%%%%%%%%%%%%%%%
\begin{figure}[!htp]
\centering
\includegraphics[scale=0.8]{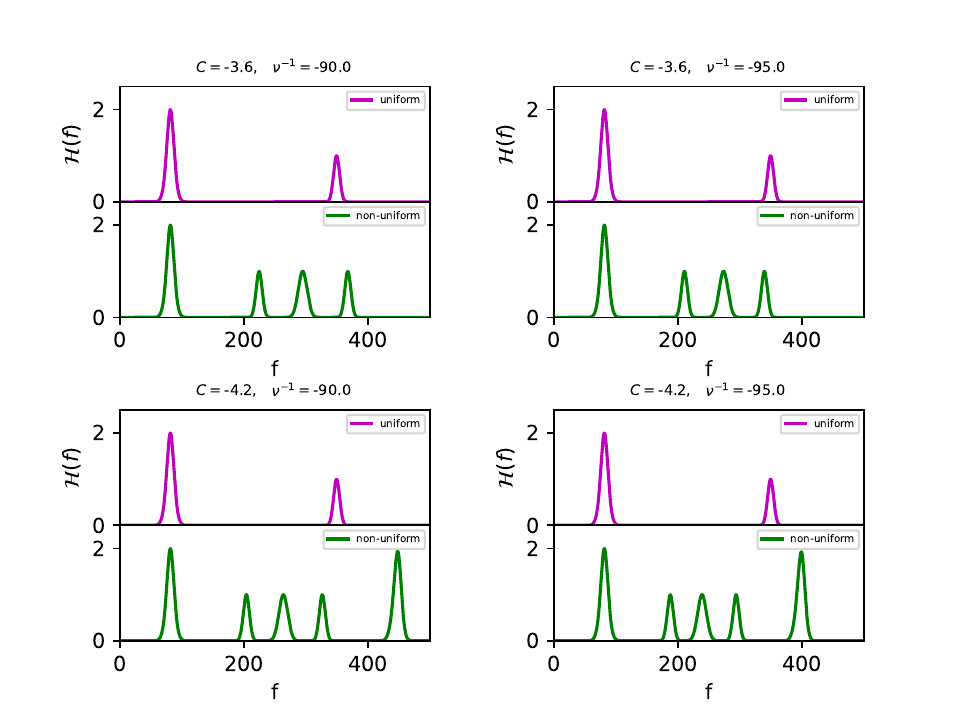}
\caption{\textbf{Enhanced horizon absorption profiles for non-uniform area quantization}. We have chosen $\alpha = 8\pi$, $\left\lbrace M_1, M_2\right\rbrace=\left\lbrace 20 M_\odot,30 M_\odot\right\rbrace$;  $\left\lbrace \chi_1, \chi_2 \right\rbrace=\left\lbrace 0.1, 0.15\right\rbrace$ and the values of $(C,\nu)$ are shown on the subplots. For cases when both BHs have nearby transition lines, the absorption profile reaches a value $\approx 2$.} 
\label{horizon}
\end{figure}
%%%%%%%%%%%%%%%%%%%%%%%
In contrast, for $C \neq 0$, we will observe quite distinct effects depending on the sign of the parameter $C$. As the value of $C$ is increased in the positive side, the phenomena of TH gets progressively quenched and the QBH starts behaving like a perfectly-reflecting compact star. On the other hand, the values $C < 0$ increase the value of $n_{max}$, which in turn enhances the TH effect compared to that of uniform area-quantization. However, one can not decrease the values of $C$ indefinitely because of the no-overlap condition.\\

\ni
We have demonstrated the effects of different choices of $(C < 0, \nu \sim -10^{-2})$ on the allowed transition lines in \ref{horizon} below. The choices of these parameters are made in such a way that the correction due to non-uniform area-quantization is subleading. Moreover, as discussed previously, the chosen negative values of $C$ are to make the spectrum richer compared to uniform area-quantization. With this enhanced horizon absorption, one can also calculate the dephasing $\Psi_{\text{THQBH}}$ due to TH using \ref{AQ_TH}.
%%%%%%%%%%%%%%%%%%%%%%
\begin{figure}[!htp]
\centering
\includegraphics[scale=0.8]{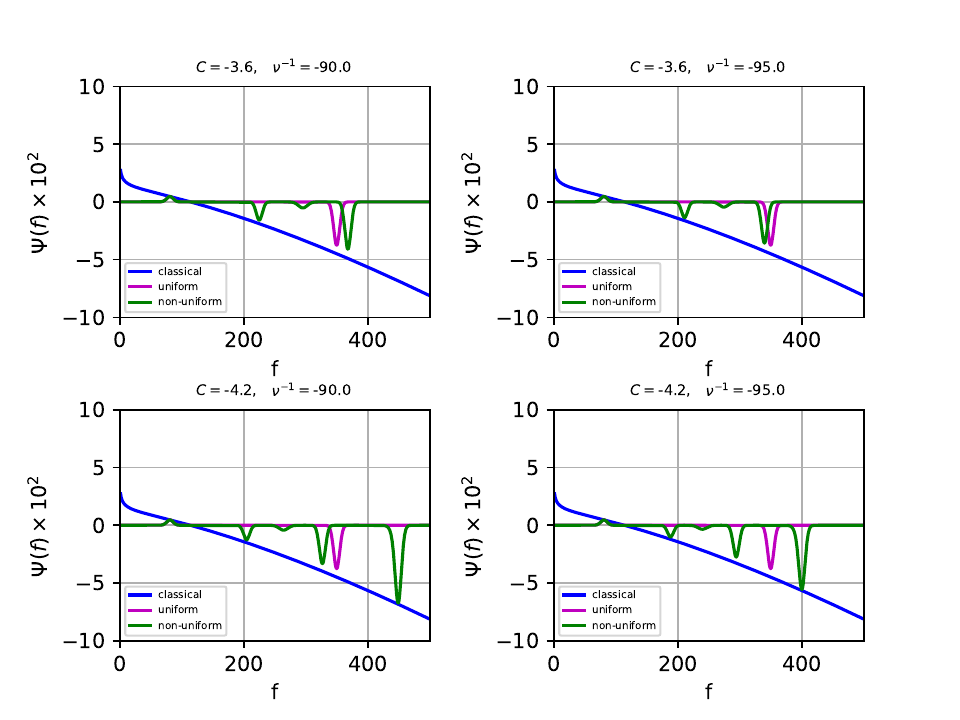}
\caption{\textbf{GW dephasing due to TH for uniform and non-uniform quantization.} Parameters are chosen to those of \ref{horizon}. The difference between classical BH and uniform/non-uniform area-quantization is notable.} 
\label{dephasing}
\end{figure}
%%%%%%%%%%%%%%%%%%%%%%%
In particular, \ref{dephasing} clearly demonstrates that classical BHs, thanks to their all-absorbing nature, results in more phase accumulation compared to the quantum BHs having uniform/non-uniform area discretization. This feature gives us a useful probe to detect any quantum structure in the vicinity of BH horizons. Moreover, as expected from earlier discussions, non-uniform area quantization with $C<0$ leads to a greater dephasing than the uniform one. However, let us emphasize that for both uniform and non-uniform area-quantization, our choice of $\alpha = 8\pi$ leads to a cumulative dephasing not more than a few radians. Therefore, the corresponding effect is considerably smaller than some of the more significant effects like eccentricity or precession. Nevertheless, our work suggests that with better accuracy in the advanced detectors, we may be able to reliably test the hypothesis of BH area-quantization in future. In fact, it would be an interesting exercise to add the TH dephasing $\Psi_{\text{THQBH}}$ for a quantum BH with $(\alpha, \, C,\, \nu)$ as free parameters to the existing binary BH coalescence templates, and carry out a simultaneous parameter estimation. Such constraints will undoubtedly provide us with valuable information to understand possible quantum nature of BHs. We leave such exercises for future attempts.
%%%%%%%%%%%%%%%%%%%%%%%%
\section{{\color{blue!70!brown} Observing Effects in BH Binary Ringdown Phase}}
%%%%%%%%%%%%%%%%%%%%%%%%
Similar to the inspiral phase, BH area-quantization can also leave its observable imprints on the postmerger ringdown phase. In the paradigm of classical BHs, the emitted GWs in this phase is well-described as a superposition of various damped sinusoidal modes known as QNMs, derived by solving a Schrodinger-type perturbation equation with perfectly ingoing boundary condition at the horizon. However, as a result of area discretization in \ref{AQ_power}, BHs can only absorb at certain characteristic frequencies given by \ref{AQ_powerom}. This feature, in turn, will lead to alteration in the near-horizon boundary conditions, which we shall elaborate in the subsequent sections. In any case, these modifications will ultimately affect the late-time behavior of the ringdown signal by inclusion of the so-called GW echoes.\\

\ni
Generation of possible echo signals in the postmerger phase of a binary has been studied extensively in literature~\cite{Cardoso:2016rao, Cardoso:2016oxy, Cardoso:2017cqb, Cardoso:2017njb, Mark:2017dnq, Correia:2018apm, Bueno:2017hyj} and their signatures in GW data are looked for~\cite{Abedi:2016hgu, Abedi:2017isz, Conklin:2017lwb, Westerweck:2017hus, Tsang:2018uie, Nielsen:2018lkf, Lo:2018sep}. However, most of these work implement the change in near-horizon boundary condition in a rather \textit{ad hoc} fashion. In contrast, BH area-quantization gives a concrete theoretical underpinning of such modifications. For example, in Ref.~\cite{Cardoso:2019apo}, the authors argued that the selective absorption of an area-discretized BH can be modelled by placing a double-barrier potential near the horizon, which essentially filters out the non-characteristic frequencies from entering into the QBH. In this chapter, we further develop this model by introducing new perspectives. We shall demonstrate that there is an ambiguity in the placement of the near-horizon double-barrier and discuss how this ambiguity leads to strikingly different features in the underlying echo signal. In particular, we shall propose a new model different from that in Ref.~\cite{Cardoso:2019apo}, which breaks the universality of echo-time (the time difference between two consecutive echoes) by making it dependent on the model of area quantization. This gives us a definitive way to distinguish between uniform/non-uniform area-quantization.
%%%%%%%%%%%%%%%%%%%%%%%%
\subsection{{\color{red!70!blue} Modelling the Quantum Filter}}
\begin{figure}[!htp]
\centering
\includegraphics[scale=0.8]{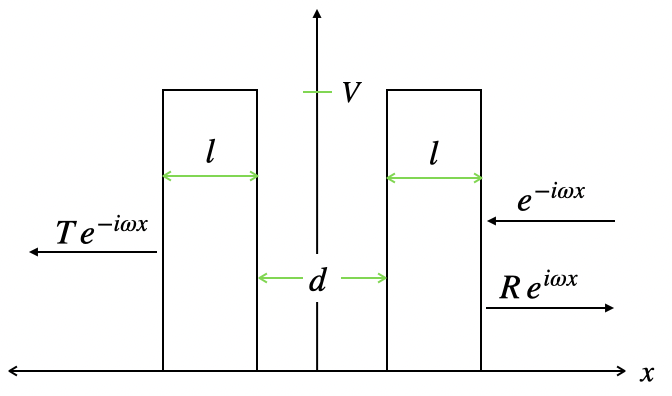}
\caption{\textbf{Symmetric double-barrier potential.} The rightmost wall, where the perturbations are reflected back, will be called as the "surface of reflection". Similarly, the leftmost wall will be referred as the "surface of absorption".} 
\label{barrier}
\end{figure}

%%%%%%%%%%%%%%%%%%%%%%%%
Due to its all-absorbing nature, the event horizon of a classical BH is endowed with zero reflectivity $R (f) = 0$ or unit transitivity $T (f) = 1$ at all frequencies. However, the behavior of a QBH horizon is strikingly different as it only absorbs at characteristic frequencies,
%%%%%%%%%%%%%%%%%%%%%%
\begin{align} \label{AQ_Schfreq}
\omega_{N,n} = \frac{\alpha\, \kappa}{8 \pi}\, \left\{1 + C \left(1 + \nu\right) N^{\nu} \right\}\, n\, ,
\end{align}
%%%%%%%%%%%%%%%%%%%%%%
where for simplicity, we set the spin $\chi = 0$ and work with Schwarzschild BHs. Therefore, QBHs must have frequency-dependent reflectivity $R(\omega)$. Neglecting the line width due to Hawking radiation for the time being, we must have $R(\omega_{N,n}) = 0$, whereas $R( \omega \neq \omega_{N,n}) = 1$. As a result, the absorption spectrum of a QBH consists of sharply peaked lines about the characteristic frequencies. Then, following Ref.~\cite{Cardoso:2019apo}, we can model this behavior by placing a near-horizon double-barrier potential, see \ref{barrier}. The only difference is that in Ref.~\cite{Cardoso:2019apo}, the authors considered uniform area
quantization, whereas we are
interested in studying the effects of the non-uniform ones.
%%%%%%%%%%%%%%%%%%%%%%%%
%%%%%%%%%%%%%%%%%%%%%%%%
Now, to mimic the absorption profile of a QBH, we need to suitably choose the barrier height $V$, width $l$, and separation $d$ such that horizon transmissivity $T(\omega)$ is very close to unity at the characteristic frequencies given by \ref{AQ_Schfreq} and vanishingly small otherwise. Using the result derived in Ref.~\cite{PhysRevB.36.4203}, we can express the transmissivity as 
%%%%%%%%%%%%%%%%%%%%%%%%
\begin{equation} \label{AQ_T}
T^{-2} = \mathcal{A}^2 + \mathcal{B}^2 +2\, \mathcal{A}\, \mathcal{B}\, \cos\left(2\, \omega\, d - \delta \right)\, ,
\end{equation}
%%%%%%%%%%%%%%%%%%%%%%%%
where $\beta^2 = V - \omega^2$, $\mathcal{A} =  \mathcal{M}^2\, \text{sinh}^2(\beta l)$, $\mathcal{B} = \text{cosh}^2(\beta l) + \mathcal{K}^2\, \text{sinh}^2(\beta l)$, and $\text{tan}(\delta) =  \mathcal{K}\, \text{sinh}(2 \beta l)\left[\text{cosh}^2 (\beta l) - \mathcal{K}^2\, \text{sinh}^2 (\beta l)\right]^{-1}$. Moreover, we have used the shorthands that $2 \mathcal{M} = \beta/\omega + \omega/\beta$, and $2 \mathcal{K} = \beta/\omega - \omega/\beta$. Thus, the transition probability oscillates between the two envelopes corresponding to the maxima and minima of the sinusoidal part:
%%%%%%%%%%%%%%%%%%%%%%%%
\begin{align} \label{AQ_env} 
\text{Upper envelope:}\, &\mid T \mid_u\, = \left(\mathcal{A} - \mathcal{B} \right)^{-1}= \, 1\, ;\nonumber \\
\text{Lower envelope:}\, &\mid T \mid_l\, = \left[ 2\, \mathcal{M}^2\, \text{sinh}^2(\beta l) + 1 \right]^{-1}\, .
\end{align}
%%%%%%%%%%%%%%%%%%%%%%%%
Now, to mimic the features of area-quantized Schwarzschild BH, we need two conditions, namely $T(\omega_{N,n}) = T_u = 1$, and $T(\omega \neq \omega_{N, n}) = T_l \approx 0$. Among them, the first condition implies the following restriction on barrier separation $d$,
%%%%%%%%%%%%%%%%%%%%%%%%%
\begin{align} \label{AQ_d3}
d = \left(\frac{3\, \pi}{2\, \omega_{N,n}}\right)\, n = \frac{d_{\text{uniform}}}{1 + C \left(1 + \nu\right) N^{\nu}}\, ,  
\end{align}
%%%%%%%%%%%%%%%%%%%%%%%%%
where  $d_{\text{uniform}} = 12\pi^2/(\alpha\, \kappa)$ is the corresponding quantity for uniform area quantization ($C = 0$). The above equation provides a one-to-one correspondence between the barrier separation and the model of area-quantization specified by $(C, \nu)$. On the other hand, the second condition can be satisfied by choosing the barrier height $V >> \omega^2$, which dictates the phase $\delta$ to be an integer multiple of $\pi$.\\

\ni
Interestingly, \ref{AQ_d3} suggests a way to distinguish between the uniform and non-uniform area-quantization. Note that for the former case, $\kappa\, d$ has a universal value for all Schwarzschild BH, independent of their masses. In contrast, for non-uniform area-quantization, $\kappa\, d$ depends on the mass of the BH through $N$. Therefore, if we fix the value of $\alpha$, multiple observations of echo from different sources can help us put constraint on the quantization parameters $(C,\nu)$. 
%%%%%%%%%%%%%%%%%%%%%%%%
\subsection{{\color{red!70!blue} Placing the Quantum Filter}}
%%%%%%%%%%%%%%%%%%%%%%%%
It is reasonable to assume that the quantum modifications due to area discretization will only be important very close to the horizon, say till a radius $r \leq r_{\epsilon} := 2M (1+\epsilon)$ for some positive values of $\epsilon << 1$. Then, the double-barrier must be placed inside this quantum extent, i.e., in between $-\infty < x \leq x_\epsilon := x(r_\epsilon)$. Here, we have introduced the so-called "tortoise coordinate" defined as $x(r) = r + 2M\, \text{log}\left(r/2M - 1\right)$. In this coordinate, the Schwarzschild horizon $r = 2M$ moves to $x \to - \infty$.\\

\begin{figure}[!htp]
\centering
\includegraphics[scale=0.5]{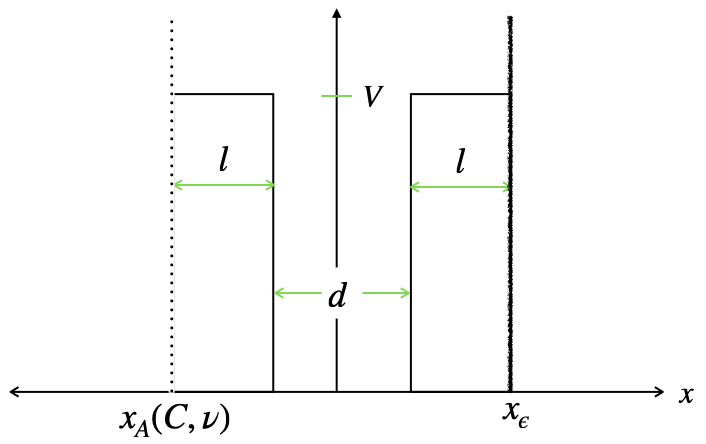}
\caption{\textbf{Depiction of Model 1.} The dotted line represents the absorption surface that varies with $(C, \nu)$, and the thick line denotes the fixed reflection surface.} 
\label{model1}
\end{figure}

\ni
However, as discussed earlier, there are two distinct ways to place this quantum barrier near the horizon. The first model is motivated by the construction in Ref.~\cite{Cardoso:2019apo}, where the surface of reflection (rightmost wall) is aligned with the quantum extent $x = x_{\epsilon}$ outside the horizon. In other words, the location of the surface of reflection of the barrier is the same for all Schwarzschild BH with different masses. Then, as a result of the $(C, \nu)$ dependence of the barrier separation $d$, the surface of absorption (leftmost wall) located at $x_A(C, \nu)$ of the double-barrier will vary for different models of the quantization. See \ref{model1} for a pictorial representation of this model.\\
%%%%%%%%%%%%%%%%%%%%%%%%
%%%%%%%%%%%%%%%%%%%%%%%%

\begin{figure}[!htp]
\centering
\includegraphics[scale=0.5]{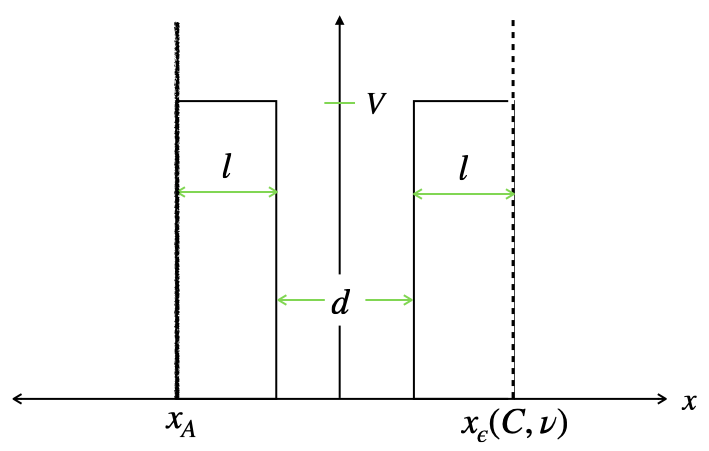}
\caption{\textbf{Depiction of Model 2.} The dotted line represents the reflection surface that varies with $(C, \nu)$, and the thick line denotes the fixed absorption surface.} 
\label{model2}
\end{figure}

\ni
Whereas, in Model 2, one fixed the location $r_A = 2M(1+a)$ (with $a < \epsilon \ll 1$) of the surface of absorption independent of $(C, \nu)$ and allows the surface of reflection to vary $x_\epsilon(C, \nu)$. For a pictorial representation of this model, see \ref{model2}. However, we need to careful in choosing the values of $x_A$ and $x_\epsilon$ so that the whole barrier lies inside the quantum regime $-\infty < x \leq x_\epsilon$. 
%%%%%%%%%%%%%%%%%%%%%%%%
%%%%%%%%%%%%%%%%%%%%%%%%
%%%%%%%%%%%%%%%%%%%%%%%%
\subsection{{\color{red!70!blue} Setting Up the Cauchy Evolution for Gravitational Perturbations}}
%%%%%%%%%%%%%%%%%%%%%%%%
Let us now consider the dynamics of a massless gravitational perturbation $\Psi(x,t)$ in the background of an area-quantized Schwarzschild BH. Then, the perturbation is governed by a Schrodinger-type second order equation,
%%%%%%%%%%%%%%%%%%%%%%%%%
\begin{equation} \label{AQ_pert}
\left[ \partial_t^2 - \partial_x^2 + V\right] \Psi(x,t) = 0\, ,
\end{equation}
%%%%%%%%%%%%%%%%%%%%%%%%%
where the effective potential is denoted by, $V = V_{\text{Sch}} + V_{\text{Barrier}}$ with $ V_{\text{Sch}}(r) = \left(6/r^2\right) (1-2M/r) (1-M/r)$ representing the usual Schwarzschild potential and $V_{\text{Barrier}}$ representing the double-barrier potential near the horizon. The maximum of $V_{\text{Sch}}$ coincides with the Schwarzschild LR at $r = 3M$, which plays a crucial role in the formation of echo. Unlike classical BH, perturbations with non-characteristic frequencies will be reflected back by the near-horizon quantum filter of a QBH. These reflected perturbation will finally reach the LR to excite the photon sphere modes and starts the initial ringdown signals. However, after some time, the ringdown modes may again reflect back from the horizon and returns to the LR, where it partially transmits through the potential maximum at $V_{Sch}$, and the remaining part is reflected back towards the horizon. As this process repeats itself, a series of late-time echoes is produced.\\

\ni
Then, to study the evolution of perturbations in the QBH background, we must put a reflecting boundary condition at $x = x_{\epsilon}$, 
%%%%%%%%%%%%%%%%%%%%%%%%%
\begin{equation} \label{AQ_bc}
\Psi(x_{\epsilon},t) \propto e^{-i \omega\left(t+x-x_{\epsilon}\right)}+R(\omega)\, e^{i \omega\left(t-x+x_{\epsilon}\right)}\ ,
\end{equation}
\noindent
%%%%%%%%%%%%%%%%%%%%%%%%%
where $R(\omega)$ is the reflectivity of the QBH horizon. Whereas the boundary condition at spatial infinity remains the same as classical BH scenario, namely perfectly outgoing $\Psi \propto e^{-i \omega (t-x)}$. Now, it is suggestive to rewrite the master equation given by~\ref{AQ_pert} in double-null coordinates defined by $u := t-x,\, \text{and}\, \, v := t+x$ as $
\left[ 4\, \partial_u\, \partial_v + V \right] \Psi(u,v)= 0$.\\

\ni
Finally, in order to find the black hole echo spectrum, we can simply study the evolution of an initial perturbation in the form of a Gaussian wave-packet consisting of all frequencies. Numerically, it can be easily achieved by discretizing the $(u,v)$-plane in the form of square grids of length $h << 1$ and express the evolution as a finite-difference equation,
%%%%%%%%%%%%%%%%%%%%%%%%%
\begin{align} \label{AQ_dis}
&\Psi(u+h,\, v+h)=  \Psi(u+h,\, v) + \Psi(u,\, v+h) - \Psi(u,v)& \nonumber \\ 
&- \frac{h^2}{8}\, \left[V(u+h,v)\, \Psi(u+h,v) + V(u,v+h)\, \Psi(u,v+h)\right]\, .
\end{align}
%%%%%%%%%%%%%%%%%%%%%%%%%
%%%%%%%%%%%%%%%%%%%%%%%%
\begin{figure}[!htp]
\centering
\includegraphics[scale=0.8]{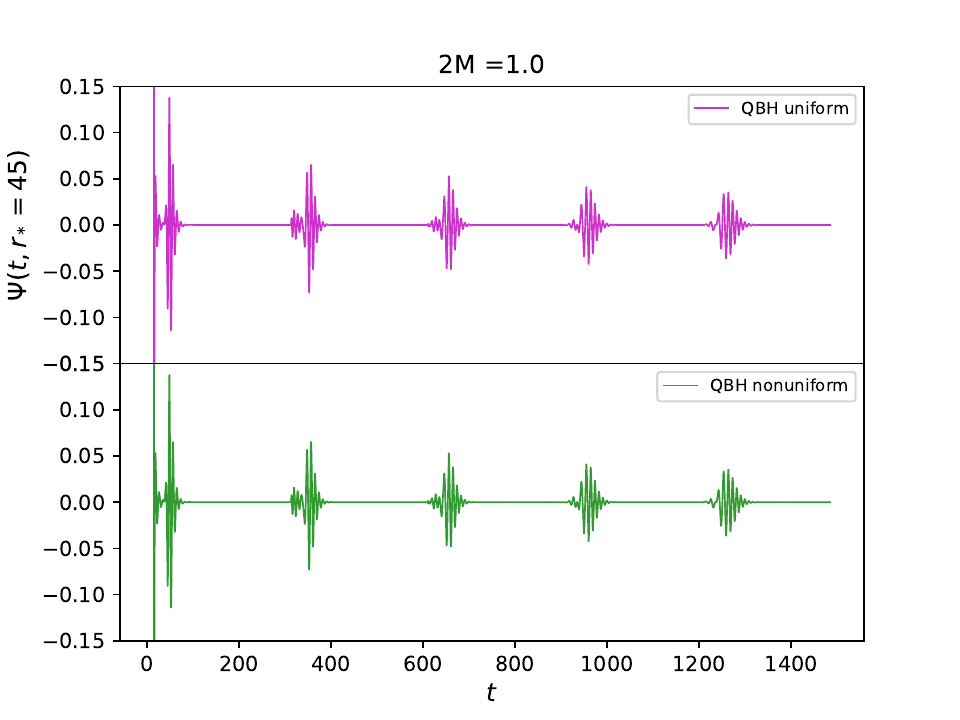}
\caption{\textbf{Echo spectrum for Model 1.} The parameters for the non-uniform quantization are $C = -3.6$, and $\nu = -1/90$. We have chosen $\epsilon = 10^{-59}$. Note that for both uniform and non-uniform area-quantization, echo-time remains the same.} 
\label{echo1}
\end{figure}
%%%%%%%%%%%%%%%%%%%%%%%%

%%%%%%%%%%%%%%%%%%%%%%%%
\begin{figure}[!htp]
\centering
\includegraphics[scale=0.8]{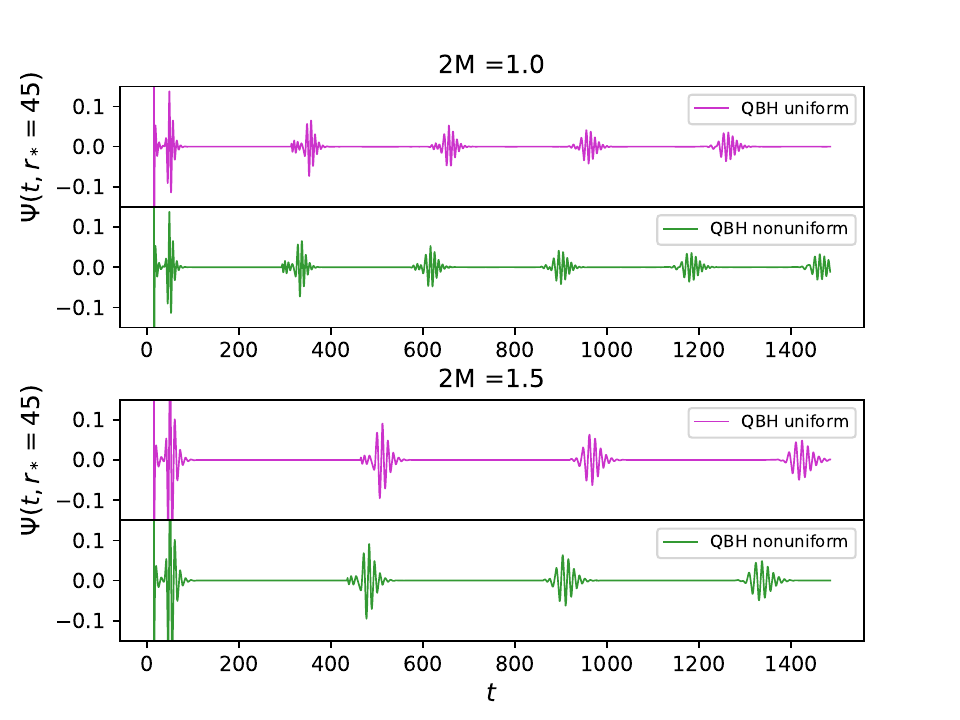}
\caption{\textbf{Echo spectrum for Model 2.} The quantization parameters are chosen to be $C = -3.6$, and $\nu = -1/90$. We have chosen $a = 10^{-70}$.} 
\label{echo2}
\end{figure}
%%%%%%%%%%%%%%%%%%%%%%%%

%%%%%%%%%%%%%%%%%%%%%%%%
\subsection{{\color{red!70!blue} The Echo Spectrum}}
%%%%%%%%%%%%%%%%%%%%%%%%
One of the important observables in echo signal is the so-called echo-time, which is the twice the light travel time between the rightmost barrier wall at $x_\epsilon$ and the LR at $r = 3M$. To the leading order in $\epsilon$, it is given by
%%%%%%%%%%%%%%%%%%%%%%%%%
\begin{equation} \label{AQ_echot}
M^{-1} \Delta t_{echo} \sim 2 \left[ 1 - 2\, \text{ln} 2 - 2\epsilon + 2\, \text{ln}(\epsilon^{-1})\right]\, .
\end{equation}
%%%%%%%%%%%%%%%%%%%%%%%%%
Thus, for Model 1, since $x_\epsilon$ is fixed and does not depend on the details of area-quantization, the quantity $M^{-1} \Delta t_{echo}$ is also independent of $(C, \nu)$ and also of the mass $M$. This is evident from \ref{echo1}. Also, note that for uniform area-quantization $C=0$, both the models will give the same echo spectrum.\\

\ni
In contrast, for Model 2, the surface of reflection $x_\epsilon$ varies with different choices of $(C, \nu)$, and so does its distance from the LR. As a result, the quantity $M^{-1} \Delta t_{echo}$ also depends the choice of quantization parameters $(C, \nu)$, see \ref{echo2}. Moreover, in this case, $M^{-1} \Delta t_{echo}$ also different for different mass $M$ of the QBHs unlike Model 1. \\
%%%%%%%%%%%%%%%%%%%%%%%%
\begin{figure}[!htp]
\centering
\includegraphics[scale=0.8]{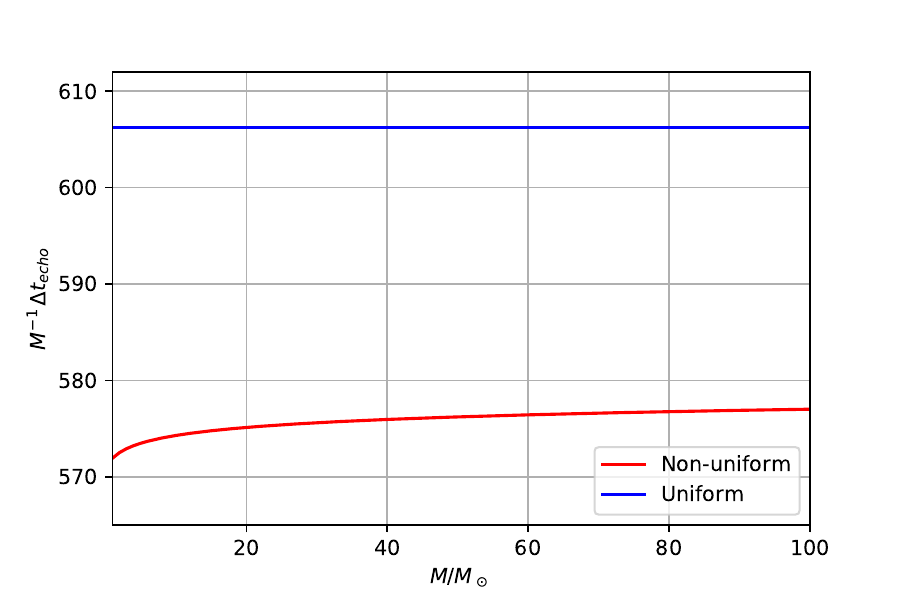}
\caption{\textbf{Variation of echo-time with mass of the BH for Model 2.} Note that the quantity $M^{-1} \Delta t_{echo}$ is dimensionless. The quantization parameters are chosen to be $C = -3.6$, $\nu = -1/90$, and $a=10^{-70}$.} 
\label{dt}
\end{figure}
%%%%%%%%%%%%%%%%%%%%%%%%

\ni
Hence, it immediately suggests a possible way to distinguish these two models. First, if the gravitational echo spectrum is observed by future detectors for more than one GW sources (having different masses), a significant variation of the quantity $ M^{-1} \, \Delta t_{\textrm{echo}}$ will be a strong evidence for non-uniform area-quantization in Model 2, see \ref{dt}. Moreover, performing a parameter estimation for $C$ using GW observations, we can be sure whether non-uniform area-quantization is preferred over the uniform one in case $C$ comes out to be different from zero.
%%%%%%%%%%%%%%%%%%%%%%%%
\section{{\color{blue!70!brown} Summary}}
%%%%%%%%%%%%%%%%%%%%%%%%
In this chapter, we have explored several observable imprints of BH area-quantization on the gravitational radiation emitted from a binary coalescence event. It is as if the astrophysical BHs magnify the near-horizon Planck scale physics to the domain of current GW observations. In particular, we have discussed the signatures of area discretization on both the inspiral and ringdown phase of a binary dynamics. In inspiral stage, these BHs leads to a modified TH phenomena due to their selective absorption at characteristic frequencies. Similarly, in the ringdown phase, area-quantized BHs give rise to repeated echo signals. These observables may contain valuable information about the quantum micro-structure of the underlying QBHs.\\

\ni
The entropy-area proportionality is a robust feature of BHs in GR. However, various putative modifications may lead to deviations from this fact. Using this general ground, we postulated a new model of BH area-quantization which, unlike the Bekenstein-Mukhanov's construction, is  non-uniform. Therefore, studying various signatures of area-quantization in current and future GW observations may give us a useful way to test the entropy-area proportionality. This, in our opinion, is a stepping stone towards understanding the quantum nature of gravity.\\

\ni
As suggested by our study, it is highly likely that with the advent of future GW detectors having better accuracy, sensitivity and signal-to-ratio, we will be able to put stringent bounds on various parameters of area-quantization. For this purpose, we have also sketched few methods that can potentially distinguish between uniform/non-uniform area discretization models and uplift other ambiguity too. Moreover, there are ample scopes to extend our work and look for more GW observables. For example, it would be interesting to study the near-merger features of BH area-quantization, for which one needs the powerful toolkit of numerical relativity. We leave such exercises for future attempts.

\chapter{{\color{red!60!black}Black Hole Perturbations and Quasi-Normal Modes} }\label{Chapter_7}
\large
\textbf{This Chapter is based on the work: Phys. Rev. D 108 (2023) 2, 024038 by R. Ghosh, N. Franchini, S. V\"{o}lkel, and  E. Barausse~\cite{Ghosh:2023etd}}.\\

\ni
In the previous chapter, we discussed how gravitational radiation from BH coalescence events can unravel hitherto unknown aspects of strong-field and highly relativistic regime of gravity. Our discussion will remain incomplete if we do not consider one of the most promising probes to the near-horizon BH physics, namely the emission of QNMs in the postmerger phase of these BH binaries~\cite{Detwiler:1980, Dreyer:2003bv, Berti:2005ys}. Such modes are emitted by perturbed BHs, akin to the dynamics of a damped-harmonic oscillator, as they try to attain their final state of equilibrium and therefore, contain valuable information of the underlying BH spacetime. In the framework of GR, QNM spectrum of BHs is uniquely determined by only two parameters, their mass $M$ and spin $a$~\cite{Regge:1957td, Zerilli:1970se, Brill:1972xj, 1974ApJ193443T, Leaver:1985ax}. Moreover, finding these modes are particularly simple as the governing perturbation equations (for scalar, vector and gravitational) in Schwarzschild/Kerr background decouple into radial and angular parts. However, in general, one may not have such luxury for BH solutions of modified theory. Then, the most straightforward way to progress is to compute the QNMs in a theory-by-theory basis. A few such notable efforts include theories like dynamical Chern-Simons gravity~\cite{Cardoso:2009pk,Molina:2010fb,Wagle:2021tam,Srivastava:2021imr}, Einstein-dilaton-Gauss-Bonnet gravity~\cite{Blazquez-Salcedo:2016enn,Blazquez-Salcedo:2017txk,Blazquez-Salcedo:2020rhf,Blazquez-Salcedo:2020caw,Pierini:2021jxd,Pierini:2022eim}, and Lorentz-violating gravity~\cite{Franchini:2021bpt}.\\

\ni
However, the scope of this theory-by-theory based approach is limited and one is forced to develop more general theory-independent methods for finding QNMs. In this general direction, a few works are worth mentioning. There has been a recent interest in generalizing the Teukolsky equation governing the gravitational perturbations beyond GR in such a way that is valid for arbitrary values of the BH spin parameter~\cite{Li:2022pcy,Hussain:2022ins}. Other theory-agnostic approaches include the parameterized QNM framework~\cite{Cardoso:2019mqo,McManus:2019ulj,Kimura:2020mrh,Volkel:2022aca,Franchini:2022axs}, the effective field theory of QNMs \cite{Franciolini:2018uyq,Hui:2021cpm}, and parameterized BH metrics \cite{Volkel:2020daa}. However, these methods are also limited to perturbations involving spherically symmetric or slowly rotating BHs. Thus for describing realistic astrophysical BHs, we need more general method that can tackle arbitrary spin values. In this front, there exist a few recent works that either study eikonal QNMs~\cite{Glampedakis:2017dvb,Glampedakis:2019dqh,Silva:2019scu,Bryant:2021xdh,Dey:2022pmv}, or add order-by-order corrections in powers of the spin parameter~\cite{Maselli:2019mjd,Carullo:2021dui}. Also, a technique was introduced in Ref.~\cite{Cano:2020cao} for finding QNMs of rotating BHs in higher-derivative gravity.\\

\ni
In this chapter, we present an alternative approach for efficiently calculating the QNMs of BHs in situations where the background geometries do not lead to separable perturbation equations. These geometries are, however, assumed to be perturbatively close to some background with known QNM modes, like Kerr/Schwarzschild BHs. To demonstrate its effectiveness, we delve into the technical intricacies of our proposed method and for concreteness, study scalar perturbations in Schwarzschild and Kerr BH endowed with an anomalous quadrupole moment. Our method also shows a universal structure of the eikonal QNMs for such deformed BHs.
%%%%%%%%%%%%%%%%%%%%%%%%
\section{{\color{blue!70!brown} Methodology}}
%%%%%%%%%%%%%%%%%%%%%%%%
Let us consider a stationary and axisymmetric spacetime that is perturbatively "close" to the Kerr metric $g_{\mu \nu}^{(0)}$ given by \ref{kerr},
%%%%%%%%%%%%%%%%%%%%%%%%
\begin{equation}\label{QNM_metric}
g_{\mu \nu}(r,\theta) = g_{\mu \nu}^{(0)}(r,\theta) + \epsilon\, g_{\mu \nu}^{(1)}(r,\theta)\ .   
\end{equation}
%%%%%%%%%%%%%%%%%%%%%%%%
Here, $g_{\mu \nu}^{(1)}$ represents the deviation from the Kerr background and $\epsilon$ is a small deviation parameter. We shall treat $\epsilon$ as an expansion parameter and neglect all higher order terms. For example, in the case of a slowly rotating Kerr BH, the background is the Schwarzschild metric (Kerr with spin $a = 0$) and the deviation metric is the linearized $g_{t \phi}$-coefficient of the Kerr metric. Then, in this particular scenario, the dimensionless spin $\epsilon = a/M$ serves as the expansion parameter. However, in general, we will not assume any additional restriction on the structure of $g_{\mu \nu}^{(1)}(r,\theta)$ except stationarity and axisymmetry.\\

\ni
Although the method we are about to present here is valid for any background metric $g_{\mu \nu}^{(0)}$ whose QNMs are known, we consider it to be the Kerr metric for there is a natural choice of angular basis in this spacetime, namely the spheroidal harmonics $S_{\ell m}(\theta)$ equipped with $e^{i m \phi}$. With this, we want to study the dynamics of a scalar perturbation $\Psi(x)$ in the BH spacetime given by \ref{QNM_metric}. The scalar wave (Klein-Gordon) equation in this spacetime takes the form $g^{\mu \nu}\nabla_{\mu}\nabla_{\nu} \Psi = 0$. Now, we use the usual decomposition of $\Psi(x)$ as~\cite{Teukolsky:1973ha, 
Brill:1972xj, Berti:2009kk},
%%%%%%%%%%%%%%%%%%%%%%%%
\begin{equation}\label{QNM_psi}
\Psi = \int d\omega\, \sum_{\ell, m} \frac{Z_{\ell m}(r)}{\sqrt{r^2+a^2}} \, S_{\ell m} (\theta)\, \ee^{-\ii\om t+ \ii m \cf}\ .  
\end{equation}
%%%%%%%%%%%%%%%%%%%%%%%%
In this context, one should note that the scalar spheroidal harmonics $S_{\ell m}(\th)$ form a complete angular basis~\cite{Teukolsky:1973ha, Brill:1972xj} and satisfy the bi-orthogonality relation~\cite{London:2020uva},
%%%%%%%%%%%%%%%%%%%%%%%%
\begin{equation}\label{eq:completeness}
    \int \dd\Om \,  S_{\ell m}(\th)\, \overline{S}_{\ell' m'}^*(\th) \ee^{\ii (m - m') \cf} = \de_{\ell \ell'}\, \de_{m m'}\,.
\end{equation}
%%%%%%%%%%%%%%%%%%%%%%%%
Then, for the Kerr background ($\ep = 0$), the Klein-Gordon equation decoupled into radial and angular parts:
%%%%%%%%%%%%%%%%%%%%%%%%
\begin{align}\label{QNM_psi0}
&\frac{\dd^2 Z_{\ell m}}{\dd r_*^2} + V_{\ell m}^{(0)} (r)\, Z_{\ell m} = 0 \, ; \\ 
&\frac{1}{\sin\th}\frac{\dd}{\dd\th} \left[\sin\th \frac{\dd S_{\ell m}(\th)}{\dd\th}\right] + \bigg[ a^2\om^2\cct + \la_{\ell m} - \frac{m^2}{\sst}\bigg] S_{\ell m}(\th) = 0 \,,\label{eq:spheroidal_scalar}
\end{align}
%%%%%%%%%%%%%%%%%%%%%%%%
where $r_*(r)$ is the Kerr tortoise coordinate defined by $dr/dr_* = h(r) = \De/(r^2+a^2)$, and $\lambda_{\ell m}$ is a separation constant. Interestingly, the second equation is automatically satisfied by the spheroidal harmonics discussed above. And, in case of a spherically symmetric background like Schwarzschild, we recover the spherical harmonics $Y_{\ell m}(\theta, \phi)$ as the angular basis.\\

\ni
On the other hand, the radial equation \ref{QNM_psi0} has to be solved with suitable boundary conditions to obtain the QNMs. It contains an effective potential $V_{\ell m}^{(0)}$~\cite{Berti:2009kk} given by,
%%%%%%%%%%%%%%%%%%%%%%%%
\begin{equation} \label{QNM_V0}
    V_{\ell m}^{(0)} (r) = \frac{K^2(r)-\lambda_{\ell m}\, \Delta(r)}{(r^2+a^2)^2} -\frac{d G(r)}{dr_*} -G^2(r)\, .
\end{equation}
%%%%%%%%%%%%%%%%%%%%%%%%
Here, $K(r)=(r^2+a^2)\omega - a\, m$, and $G(r) = r\, \Delta(r)/(r^2+a^2)^2$. In the limit $a \to 0$, we get back the corresponding results for Schwarzschild case.\\

\ni
Now, we want to study the most general scenario where $\ep \neq 0$ and the perturbation equation does not separate automatically like that of Kerr/Schwarzschild. In the linear order of the deviation parameter $\ep$, the Klein-Gordon equation boils down to the following expression,
%%%%%%%%%%%%%%%%%%%%%%%%
\begin{equation}\label{QNM_eomlm}
\begin{split}
\int d\omega \sum_{\ell, m} \ee^{-\ii\om t+ \ii m \cf} S_{\ell m} (\theta) \left[ \frac{d^2}{dr_*^2} + V_{\ell m}^{(0)} (r)\right] Z_{\ell m} = \epsilon\, \mathcal{J}[\Psi]\, , 
\end{split}
\end{equation}
%%%%%%%%%%%%%%%%%%%%%%%%
where using the notation $ \sqrt{-\det\, g} = g_{(0)} + \epsilon\, g_{(1)}$, the source term is given by,
%%%%%%%%%%%%%%%%%%%%%%%%
\begin{equation}\label{QNM_J}
\mathcal{J}[\Psi] = - \frac{h(r)\rho^2}{g_{(0)}\, \sqrt{r^2+a^2}}\, \partial_\mu \left[ g_{(0)}\, g^{\mu \nu}_{(1)}\, \partial_\nu \Psi + g_{(1)}\, g^{\mu \nu}_{(0)}\, \partial_\nu \Psi \right]\ ,
\end{equation}
%%%%%%%%%%%%%%%%%%%%%%%%
where the derivatives are taken in terms of $x^\mu = (t,r,\th,\cf)$, referring to the Boyer-Lindquist coordinates in which the metric \ref{QNM_metric} is written. Then, we may use the decomposition of $\Psi$ 
to rewrite the source term in a more suggestive form: $\mathcal{J} = \int \dd\om \sum_{\ell, m} \ee^{-\ii\om t+ \ii m \cf} J_{\ell m}  (r,\th)$, with its component having the following structure,
%%%%%%%%%%%%%%%%%%%%%%%%
\bea \label{eq:decomposition}
    J_{\ell m} (r,\th) = a(r, \th) Z_{\ell m}(r) S_{\ell m}'(\th) + b(r, \th) Z_{\ell m}'(r) S_{\ell m}(\th) + c_{\ell m}(r, \th) Z_{\ell m}(r) S_{\ell m}(\th)\,.
\eea
%%%%%%%%%%%%%%%%%%%%%%%%
The explicit expressions of $a, b,$ and $c_{\ell m}$ in terms of various metric components can be found in Appendix C~[\ref{app4}].\\

\ni
Now, using the completeness of spheroidal harmonics, we can decompose $J_{\ell m} (r,\th)$ as $J_{\ell m}(r, \th) = \sum_{\ell'} j_{\ell \ell' m}(r) S_{\ell' m}(\th)$ and then, the perturbation equation takes the form,
%%%%%%%%%%%%%%%%%%%%%%%%
\begin{equation}
    S_{\ell m}(\th) \left[ \frac{\dd^2}{\dd r^2_*} + V_{\ell m}^{(0)}(r) \right] Z_{\ell m}(r) = \ep \sum_{\ell'} j_{\ell \ell' m}(r) S_{\ell' m}(\th)\, .
\end{equation}
%%%%%%%%%%%%%%%%%%%%%%%%
We can completely decouple the above equation by employing the bi-orthogonality relation given by \ref{eq:completeness}. This will give us the longed for radial perturbation equation. However, our work is not finished because of the sum on RHS involves coupling between different $\ell'$-modes. To make it apparent, lets recall that $j_{\ell \ell'}(r)$ is a linear combination of just $Z$ and $Z'$, i.e., we have
%%%%%%%%%%%%%%%%%%%%%%%%
\begin{equation}\label{eq:sourcejl}
    j_{\ell' m} \equiv j_{\ell'\ell' m} =  \alpha_{\ell' m}(r)\, Z_{\ell' m} + \beta_{\ell' m}(r)\, \frac{\dd Z_{\ell' m}}{\dd r_*}\, ,
\end{equation}
%%%%%%%%%%%%%%%%%%%%%%%%
where the exact forms of $\alpha_{\ell' m}$ and $\beta_{\ell' m}$ depend on the exact form of $g_{\mu\nu}^{(1)}$ under consideration. With this structure at hand, the radial perturbation equation can be written as
%%%%%%%%%%%%%%%%%%%%%%%%
\begin{equation}\label{QNM_ceom}
\begin{split}
\frac{\dd^2 Z_{\ell m}}{\dd r_*^2} + \, V_{\ell m}^{(0)}(r)\, Z_{\ell m} = \epsilon\, j_{\ell m} + \epsilon\, \sum_{\ell'\neq\ell} j_{\ell' m} \, .
\end{split}
\end{equation}
%%%%%%%%%%%%%%%%%%%%%%%%
Note that this equation makes the coupling between $\ell' \neq \ell$ modes completely explicit. However, one can do further simplification by absorbing the $Z'_{\ell m}$ term coming from $j_{\ell m}$ into a field redefinition $Z_{\ell m} \rightarrow Z_{\ell m} \exp\left[-\ep/2\int \dd r \beta_{\ell m}(r)/h(r)\right]$ yielding \footnote{One must ensure that the boundary conditions at $r_* \to \pm\infty$ for the new functions $Z_{\ell m}$ are the same as for the old functions. This is achieved by requiring that $\beta$ falls off at least as $1/r_*^2$ as $r_* \to \pm\infty$. This is verified for all the examples we will consider later. In fact, in those examples, we will have $\beta(r)=0$ and thus, the transformation is not needed in the first place.}
%%%%%%%%%%%%%%%%%%%%%%%%
\begin{equation}\label{QNM_meom}
\frac{\dd^2 Z_{\ell m}}{\dd r_*^2} + \, V_{\ell m}(r)\, Z_{\ell m} = \epsilon\, \sum_{\ell'\neq\ell} j_{\ell' m} \, ,
\end{equation}
%%%%%%%%%%%%%%%%%%%%%%%%
where we have obtained the the master potential as
%%%%%%%%%%%%%%%%%%%%%%%%
\begin{equation}\label{QNM_pot}
V_{\ell m}(r) = V_{\ell m}^{(0)} (r) - \epsilon \left[ \alpha_{\ell m}(r) -\frac{1}{2}\, {\beta'}_{\ell m}(r)\, h(r) \right]\,.  
\end{equation}
%%%%%%%%%%%%%%%%%%%%%%%%
The above differential equation still represents a coupled system because of the coupling among $\ell' = \ell, \ell \pm 1, \ell \pm 2, \dots)$ modes present in the RHS of \ref{QNM_meom}. Thus, to decouple the system, it is suggestive to make another field redefinition: $Z_{\ell m}(r) = X_{\ell m}(r)+\epsilon\, U_{\ell m}(r)$, so that $U_{\ell m}(r)$ satisfies the following differential equation
%%%%%%%%%%%%%%%%%%%%%%%%
\begin{equation} \label{QNM_ulm}
\frac{d^2 U_{\ell m}}{dr_*^2} + V_{\ell m}^{(0)} (r)\, U_{\ell m} = \sum_{\ell'\neq\ell} j_{\ell' m}(r) := T_{\ell m}(r_*)\,.
\end{equation}
%%%%%%%%%%%%%%%%%%%%%%%%
Then, $X_{\ell m}$ obeys the QNM master equation, which is the main result of this chapter,
%%%%%%%%%%%%%%%%%%%%%%%%
\begin{equation} \label{QNM_master}
    \frac{d^2 X_{\ell m}}{dr_*^2} + V_{\ell m}(r)\, X_{\ell m} = 0\ ,
\end{equation}
%%%%%%%%%%%%%%%%%%%%%%%%
where the potential is given by \ref{QNM_pot}. Among these two equations, the first one involving $U_{\ell m}$ is an ordinary differential equation with known coefficients. Whereas, the later one involving $X_{\ell m}$ is an eigenvalue equation, which has to be solved to obtain the QNMs $\omega_{\ell m}$ by imposing ingoing boundary conditions at the horizon and outgoing boundary conditions at infinity. We shall demonstrate this by considering some explicit examples in the subsequent sections.\\

\ni
Though knowing the explicit form of $U_{\ell m}$ is not required to solve for the QNM spectrum from the master equation in \ref{QNM_master},
we must argue that such solution exist under the ingoing (outgoing) boundary conditions at the horizon (spatial infinity). We shall devote Appendix C~[\ref{app5}] to discuss this in great details. Moreover, we shall also demonstrate an explicit example on the construction of $U_{\ell m}$ in Appendix C~[\ref{app6}].
%%%%%%%%%%%%%%%%%%%%%%%%
\section{{\color{blue!70!red} Application and Results}}
%%%%%%%%%%%%%%%%%%%%%%%%
To illustrate the method described above, let us now consider a few concrete examples. In all the cases, the background metric is taken to be either Schwarzschild or Kerr.
%%%%%%%%%%%%%%%%%%%%%%%%
\subsection{{\color{red!70!blue} Slowly Rotating Kerr BH}}
%%%%%%%%%%%%%%%%%%%%%%%%
In this example, we want to study a slowly rotating Kerr spacetime. Thus, the background $g_{\mu \nu}^{(0)}$ is the Schwarzschild metric, and the deviation $g_{\mu \nu}^{(1)}$ has only two non-zero components, $ g_{t \phi (1)} = g_{\phi t (1)} = - 2 M^2\, \sin^2\th/r$ with the deviation parameter $\epsilon = a/M$. Therefore, $g_{(0)} = r^2\, \sin\theta$, $g_{(1)} = 0$, and $g^{\mu \nu}_{(1)}$ has only two non-zero components, $ g^{t \phi}_{(1)} = g^{\phi t}_{(1)} = - \frac{2 M^2}{r^2 (r-2M)}$. Also, due to spherical symmetry, we shall choose to work with the spherical harmonics $Y_{\ell m}(\theta, \phi)$ as the angular basis.\\

\noindent
Then, using \ref{QNM_J}, we get the source term as $\mathcal{J}[\Psi] = \frac{4\, M^2 \omega\, m}{r^2 (r-2 M)}\, \Psi$. Finally, the QNM master equation becomes
%%%%%%%%%%%%%%%%%%%
\begin{equation}\label{QNM_eomSR}
\frac{d^2 Z_{\ell m}}{dr_*^2} + \left[ V^\mathrm{Sch}_{\ell}(r) + \epsilon\, \frac{4\, M^2\, \omega\, m}{r^3} \right]\, Z_{\ell m} = 0\ , 
\end{equation}
%%%%%%%%%%%%%%%%%%%
where the Schwarzschild potential is denoted by
%%%%%%%%%%%%%%%%%%%
\bea \label{QNM_vschw}
V^\mathrm{Sch}_{\ell}(r) = \om^2 - f(r)\, \left[\frac{\ell(\ell+1)}{r^2} - \frac{2 M}{r^3}\right]
\eea
%%%%%%%%%%%%%%%%%%%
with $f(r) = 1- 2M/r$. The master equation, which can be solved for the QNM frequencies, matches exactly with that given in Refs.~\cite{Pani:2012bp,Pani:2013pma}.
%%%%%%%%%%%%%%%%%%%%%%%%
\subsection{{\color{red!70!blue} Schwarzschild Quadrupole BH}}
%%%%%%%%%%%%%%%%%%%%%%%%
As discussed in \ref{Chapter 1}, vacuum BH solutions of GR are characterized by only two parameters, namely mass $M$ and spin $a$. In other words, all higher multipole moments are uniquely fixed by these two parameters alone~\cite{Hansen:1974zz, Geroch:1970cd}. However, in the presence of putative modifications to GR, BHs may violate aforesaid novel results by growing extra hairs. Over the years, many extensive studies have been performed to understand the observational signatures of such non-Kerr BHs~\cite{Ryan:1995wh, Ryan:1997hg, Gair:2007kr, Bambi:2011jq, Bambi:2011vc, Bambi:2015ldr,Volkel:2020xlc,EventHorizonTelescope:2021dqv,Dey:2022pmv, Isi:2019aib}.\\

\ni
Motivated by these works, we now perform the QNM analysis of a Schwarzschild
BH with an anomalous (different from GR) quadrupole moment $\epsilon$. To achieve this task, let us construct a BH metric that has Schwarzschild-like structure near the horizon and boils down to the static Hartle-Thorne metric as $r \to \infty$~\cite{Glampedakis:2005cf, 1967ApJ1501005H, 1968ApJ153807H, Allahyari:2018cmg}. We are interested to study the effect of the quadrupole moment on QNMs, and neglect terms containing quadratic and higher powers in $\epsilon$. Up to the linear order in $\epsilon$, the non-zero covariant metric components are as follows: 
%%%%%%%%%%%%%%%%%%%%%%%%
\begin{align} \label{QNM_metricschq}
&g_{t t} = -f(r)\, \left[1+\epsilon\, f_1(r)\, P_{2}(\cos\theta)\right]\, , \nonumber \\
&g_{r r} = f(r)^{-1}\, \left[1-\epsilon\, f_1(r)\, P_{2}(\cos\theta)\right] \, , \\
&g_{\theta \theta} = (\sin\theta)^{-2}\, g_{\phi \phi} = r^2\, \left[1-\epsilon\, f_1(r)\, P_{2}(\cos\theta)\right]\, , \nonumber
\end{align}
%%%%%%%%%%%%%%%%%%%%%%%%
where $f_1(r) = 2\, f(r)\, (M/r)^3$, and $P_{2}(\cos\th)$ is the Legendre polynomial of second order. The choice of $f_1(r)$ ensures that the leading asymptotic fall-off proportional to the quadrupole moment is same as that of Hartle-Thorne metric. Moreover, since we have chosen the function $f_1(r)$ in such a way that it vanishes at the horizon $r = 2M$ (same as Schwarzschild BH), the corresponding near-horizon boundary condition remains the same as that of Schwarzschild BH. \\

\begin{figure}
\centering
\includegraphics[width=0.8\linewidth]{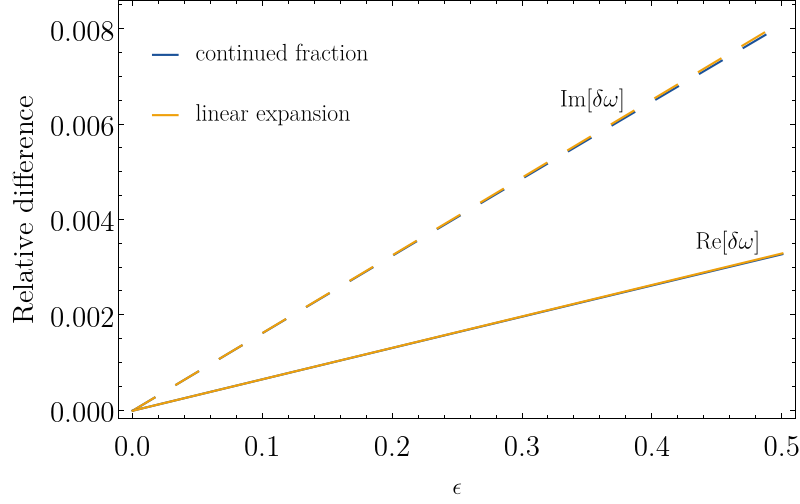}
\caption{\textbf{Absolute relative difference between the both real (solid) and imaginary (dashed) parts of the $\ell=m=2$ scalar QNM mode of a Schwarzschild BH with and without the quadrupolar correction.} QNMs calculated via the continued fraction method (blue line) and linear expansion in $\epsilon$ (orange line) matches quite well.}
\label{fig:qnm_nonrot}
\end{figure}

\ni
Now, we can calculate the source term as,
%%%%%%%%%%%%%%%%%%%%%%%%
\begin{equation} \label{QNM_Schj}
\mathcal{J}[\Psi] = \frac{4M^3\, \omega^2}{r^3}\, P_{2}(\cos\theta)\, \Psi\ .
\end{equation}
%%%%%%%%%%%%%%%%%%%%%%%%
The presence of the product $P_{2}(\cos\theta)\, Y_{\ell m}(\theta,\, \phi)$ in $\mathcal{J}$ will ultimately cause a coupling among various angular momentum components and the quadrupole. Then, to decouple the radial and angular parts of the perturbation equation, we shall use of the following relation~\cite{Pani:2013pma},
%%%%%%%%%%%%%%%%%%%%%%%%
\begin{equation}\label{QNM_cY}
(\cos\theta)\, Y_{\ell m} = Q_{\ell+1, m}\, Y_{\ell+1\,  m} + Q_{\ell m}\, Y_{\ell-1\, m}\, \, ,\, \text{with}\, \, \, Q_{\ell m} = \sqrt{\frac{\ell^2 - m^2}{4\ell^2-1}}\ .
\end{equation}
%%%%%%%%%%%%%%%%%%%%%%%%
An important observation that we shall use later is $Q_{\ell m}$ vanishes for $m = \pm \ell$. Moreover, two application of this recursion relation shows that the source term given by \ref{QNM_Schj} will lead to a coupling between $\{ \ell, \ell \pm 2\}$ modes in the radial perturbation equation, 
%%%%%%%%%%%%%%%%%%%%%%%%
\begin{equation} \label{QNM_Vq}
\frac{d^2 Z_{\ell m}}{dr_*^2} + V_{\ell m}(r) Z_{\ell m} =
3\, \epsilon\, \omega^2\, f_1(r) \left(B_{\ell+2 m}\, Z_{\ell+2 m} + B_{\ell m}\, Z_{\ell-2 m}\right)\, , 
\end{equation}
%%%%%%%%%%%%%%%%%%%%%%%%
where $\dd r_* = \dd r /f(r)$ is the Schwarzschild tortoise coordinate, $A_{\ell m} = Q_{\ell m}^2 + Q_{\ell+1\, m}^2$, and $  B_{\ell m} = Q_{\ell-1\, m} Q_{\ell m}$. Finally, following the methodology section (for more explicit calculation, see \ref{app6} in the Appendix), we can decouple this equation and recast it in the form of the master equation given by \ref{QNM_master} with potential 
%%%%%%%%%%%%%%%%%%%%%%%%
\begin{equation} \label{QNM_Schpot}
V_{\ell m}(r) = V_{\ell}^\mathrm{Sch}(r) - \epsilon\, \omega^2\, f_1(r)\, (3A_{\ell m}-1)\, .
\end{equation}
%%%%%%%%%%%%%%%%%%%%%%%%
With this setup, we shall now proceed to calculate the QNMs using two methods namely, the method of continued fraction~\cite{1985RSPSA402285L, 1986JMP271238L} and a linear expansion in $\ep$ as suggested in Refs.~\cite{Cardoso:2019mqo, Volkel:2022aca}. The results of these two methods
agree well, as shown in \ref{fig:qnm_nonrot}. 
%%%%%%%%%%%%%%%%%%%%%%
%%%%%%%%%%%%%%%%%%%%%%
As expected, the effect of $\ep$ on the QNMs is always small, reflecting the fact that we are working in the linear approximation regime. Moreover, to check the validity of our decoupling technique, we have compared these QNMs with that obtained from \ref{QNM_Vq} with the RHS coupling terms. For this, we truncate the coupling at different $\ell_{\textrm{max}}$ and use the numerical method given in Ref.~\cite{1986JMP271238L}. And, we found that as $\ell_{\textrm{max}}$ increases, say $\ell + 2$ to $\ell + 4$, the relative difference between these truncated QNMs and the previously obtained QNMs decreases at the linear order in $\ep$, suggesting the validity of our decoupling method.\\

\ni
Now, we proceed to study the eikonal limit ($\ell = m \gg 1$) of the QNM modes. For both Kerr and Schwarzschild BHs, there is a well-known correspondence between the eikonal QNMs and the motion of photons near the LR~\cite{Ferrari:1984zz, 1985ApJ291L33S, Iyer:1986np, Iyer:1986nq, Dolan:2010wr, Yang:2012he}. In particular, the real part of the eikonal QNMs is related to the orbital frequency of the photons' orbit at the LR, whereas the imaginary part correspond to the associated Lyapunov exponent. Thus, it is a natural question to ask whether such correspondence is valid also for the Schwarzschild quadrupole BH under consideration. This expectation is supported by the following reasoning. In the eikonal limit, the scalar perturbation takes the form $\Psi=A\, \exp(\ii \,S)$, where $S$ is a rapidly varying phase. Then, the Klein-Gordon equation reduces to the Hamilton-Jacobi equation $g^{\mu\nu}\partial_\mu S\, \partial_\nu S=0$, representing the geodesic motion of photons~\cite{deFelice:1990hu, Barausse:2019pri}. For this check, it is sufficient to consider only the equatorial motion of photons, as they remain confined in this plane if started with the initial conditions $(\theta = \pi/2,\, \dot{\theta} = 0)$. However, there is one subtle point that one should be careful about. The eikonal correspondence is only apparent when both the QNM equation and the null geodesic equation are expressed in terms of the same tortoise coordinate. In our case, however, the QNM equation contains the Schwarzschild tortoise coordinate $r_*$ which is different from the tortoise coordinate $\bar{r}_*$ of the Schwarzschild quadrupole metric. In fact, at the equatorial plane, they are related as: $d\bar{r}_* = dr_*\, (1+\epsilon\, f_1(r)/2)$. Therefore, one should first rewrite the master equation for $\ell = m \gg 1$ as, $d^2 X_{\ell m}/dr^2_*+ V_{\ell m}^{\textrm{eik}}(r) X_{\ell m}= 0$. Here, the potential is given by 
%%%%%%%%%%%%%%%%%%%%%%%%
\begin{equation} \label{QNM_eikold}
    V_{\ell m}^{\textrm{eik}}(r) \simeq \omega^2 \big[1+\epsilon\, f_1(r)\big] -\frac{\ell^2\, f(r)}{r^2}\, .
\end{equation}
%%%%%%%%%%%%%%%%%%%%%%%%
After some easy manipulation (for more details see Appendix C [\ref{app7}]), one gets the potential in $r_*$-coordinate as 
%%%%%%%%%%%%%%%%%%%%%%%%
\begin{equation} \label{QNM_eik}
    \widetilde{V}_{\ell m}^{\textrm{eik}} \simeq \big[1-\epsilon\, f_1(r)\big] V_{\ell m}^{\textrm{eik}}(r) \simeq \omega^2 - \frac{\ell^2\, f(r)}{r^2} \big[1-\epsilon\, f_1(r)\big]\, ,
\end{equation}
%%%%%%%%%%%%%%%%%%%%%%%%
where $V_{\ell m}^{\textrm{eik}}(r)$ is given by~\ref{QNM_eikold}.
On the other hand, the equatorial photon potential can be calculated from the geodesic equation as $V(r) = E^2 - L^2 f(r)\big[1-\epsilon\, f_1(r)\big]/r^2$, with $L$ and $E$ are respectively the angular momentum and energy of the photon. Thus, the eikonal QNM potential matches exactly with the the equatorial photon potential, provided we identify the energy $E$ and the angular momentum $L$ of the photon with the eikonal QNM frequency $\omega$ and angular momentum quantum number $\ell$.\\

\ni
For an explicit representation of the correspondence, we note that the location of the equatorial LR is given by $r_p = 3 M + \epsilon\, M/81$. Then, in the eikonal limit, the LR frequency and Lyapunov exponent can be computed to be $\Omega_p = \left(1 - \epsilon/81 \right)/\left(3\sqrt{3}\, M\right)$ and $|\lambda_p| = \left(243 - 4\, \epsilon \right)/\left(729\sqrt{3}\, M\right)$, respectively. See Refs. \cite{Glampedakis:2017dvb, Glampedakis:2019dqh} for more details. And, the eikonal QNM frequency is given by $\omega_{n \ell} \approx \ell\, \Omega_p - \ii (n+1/2)\, |\lambda_p|$, similar to the Schwarzschild case. Here, we have used the fact that the real part of the background Schwarzschild QNMs has the form $\omega_R^{\textrm{Sch}} \approx \ell\, \Omega_p^\textrm{Sch}$ and $A_{\ell m} \to 0$ in the eikonal limit. \\

\noindent
Interestingly, for a small perturbation $ |\epsilon| \ll 1$, the BHs remain stable as the imaginary part of $\omega_{n \ell}$ is always negative. In fact, it is expected since we are working in the linear expansion regime. However, there is a way to distinguish different signs of $\epsilon$. Since the relaxation time is inversely proportional to $|\lambda_p|$, positive (negative) values of $\epsilon$ leads to a slower (faster) decay of perturbations than Schwarzschild.
%%%%%%%%%%%%%%%%%%%%%%%%
\subsection{{\color{red!70!blue} Kerr Quadrupole BH}}
%%%%%%%%%%%%%%%%%%%%%%%%
In the same spirit of the previous Schwarzschild case, we may construct a rotating BH metric that has Kerr-like structure near the event horizon at $r_+ = M + \sqrt{M^2-a^2}$, and has same leading order asymptotic structure as that of the rotating Hartle-Thorne metric given by Eqs.(2.13--2.16) of Ref.\cite{Glampedakis:2005cf} at asymptotic infinity. The deviation parameter in this metric is $\epsilon = q-(a/M)^2$, where $q$ and $a/M$ are the dimensionless quadrupole moment and spin parameter of the background Kerr BH given by \ref{kerr}. Up to the linear order in $\epsilon$, he non-zero covariant metric components are as follows:
%%%%%%%%%%%%%%%%%%%%%%%%
\begin{equation}
\begin{split}
&g_{tt} = g_{tt}^\mathrm{Kerr} \left[1+\epsilon\, f_2(r)\, P_{2}(\cos\theta)\right] , \\
&g_{rr} = g_{rr}^\mathrm{Kerr} \left[1- \epsilon\, f_2(r)\, P_{2}(\cos\theta)\right], \\
& g_{\theta \theta} = g_{\theta \theta}^\mathrm{Kerr} \left[1- \epsilon\, f_2(r)\, P_{2}(\cos\theta)\right] , \\
& g_{\phi \phi} =  g_{\phi \phi}^\mathrm{Kerr} \left[1- \epsilon\, f_2(r)\, P_{2}(\cos\theta)\right], \\
&g_{t\phi} = g_{t\phi}^\mathrm{Kerr},
\end{split}
\end{equation}
%%%%%%%%%%%%%%%%%%%%%%%%
where $f_2(r) = 2\, F(r)\, (M/r)^3$ with $F(r)= \Delta(r)/r^2$ and $\Delta (r) = r^2 -2\, M\, r+a^2$. The above metric reduces to the Schwarzschild-quadrupole metric given by~\ref{QNM_metricschq} in the limit $a \to 0$. Let us now study the leading effect of the anomalous quadrupole moment on the QNM spectrum.\\

\begin{figure}
\centering
\includegraphics[width=0.8\linewidth]{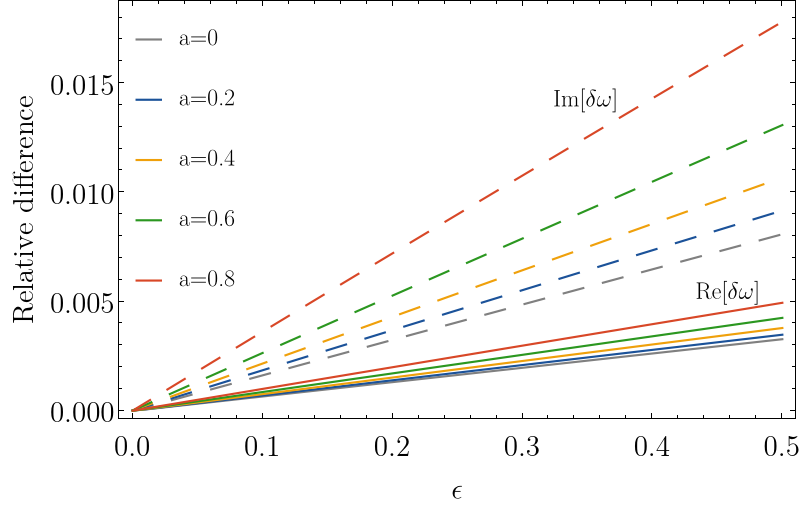}
\caption{\textbf{Absolute relative difference between the both real (solid) and imaginary (dashed) parts of the $\ell=m=2$ scalar QNM mode of a Kerr BH with and without the quadrupolar correction.} For various values of the spin parameter (in units of $M = 1$), the QNM modes are calculated using the continued fraction method.}
\label{fig:quadrupole_rot}
\end{figure}

\begin{figure}
\centering
\includegraphics[width=0.8\linewidth]{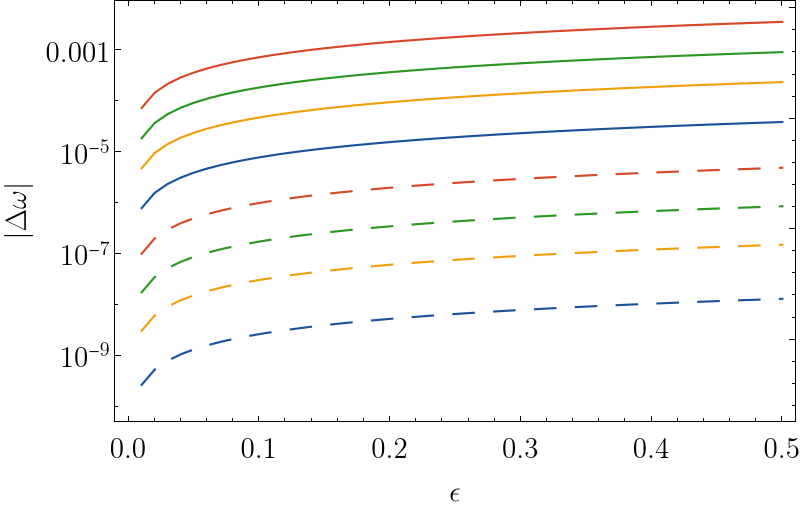}
\caption{\textbf{Comparison of the absolute difference for the $\ell = m = 2$ QNM modes at different order of expansion in $(a\om)^N$ of the potential in~\ref{QNM_KqV}.} Solid lines compare the modes between $N=1$ and $N=2$ truncation, whereas dashed lines show the difference between $N=2$ and $N=4$ approximations. The colors represent the same spin values as in ~\ref{fig:quadrupole_rot}.}
\label{fig:diff_quadrupole_rot}
\end{figure}

\ni
The first step is to find out the source term $\mathcal{J}[\Psi]$, which can be directly calculated using \ref{QNM_J}. Nevertheless, for simplicity, we make one more approximation and expand the source term in the powers of $a\om$, which is perfectly valid away from the extremal limit ($a \to M$). Under this assumption, we can express the spheroidal harmonics $S_{\ell m}(\th)$ as a linear combination of the generalized Legendre polynomials $P_{\ell m}(\th)$,
%%%%%%%%%%%%%%%%%%%%%%%%
\begin{equation}\label{QNM_SY}
    S_{\ell m} (\th) = \sum_{n=0}^{\infty} (a\om)^{2n} \sum_{k = -2n}^{N} K^{(N)}_{\ell \, k \, m} \, \, P_{\ell + k}^{m}(\th)
\end{equation}
%%%%%%%%%%%%%%%%%%%%%%%%
where the coefficients $K^{(N)}_{\ell \, k \, m}$ can be easily found using the Black Hole Perturbation Toolkit~\cite{BHPToolkit}. Finally, one gets the following perturbation equation,
%%%%%%%%%%%%%%%%%%%%%%%%
\begin{equation} \label{QNM_VKq}
\frac{d^2 Z_{\ell m}}{dr_*^2} + V_{\ell m} (r)\, Z_{\ell m}
= \epsilon\, \sum_{\ell'\neq\ell} j_{\ell' m} +\epsilon\, \mathcal{O}(a^{N}\om^{N})\ .
\end{equation}
%%%%%%%%%%%%%%%%%%%%%%%%
Then, following Appendix C [\ref{app7}], we can decouple the above equation in the form of~\ref{QNM_master}. And, the QNM potential for $N=2$ is given by,
%%%%%%%%%%%%%%%%%%%%%%%%
\begin{align} \label{QNM_Vqk}
V_{\ell m} = V_{\ell m}^{(0)}(r)  & +3\, \epsilon\, a^2\om^4\, f_1(r)\, \left(K^{(1)}_{\ell \, 2 \, m}\, B_{\ell+2}^{m} + K^{(1)}_{\ell \, -2 \, m}\, B_{\ell m} \right) \nonumber \\
&- \epsilon\, f_2(r)(a^2+r^2)^{-2}\, f_3(r)
+\epsilon\, \Ord(a^3\om^3)\, ,
\end{align}
%%%%%%%%%%%%%%%%%%%%%%%%
where $\left\{f_1,\, f_2,\, A_{\ell m},\, B_{\ell m},\, K_{\ell m}, V_{\ell m}^{(0)}\right\}$ are defined earlier, and 
%%%%%%%%%%%%%%%%%%%%%%%%
\begin{equation} \label{QNM_KqV}
\begin{aligned}
   f_3(r)=\, & r\Big[\om^2\, r^3-2 m a\, \om M+ a^2\om^2 (r+2\, M)\Big](3A_{\ell m}-1)\\
    &+a^2\om^2\, \Delta(r)\Big[ (3A_{\ell m} -1)A_{\ell m}+ 3 (B_{\ell+2}^{m})^2+ 3 (B_{\ell m})^2\Big]\, ,
\end{aligned}
\end{equation}
%%%%%%%%%%%%%%%%%%%%%%%%
Similar to the non-spinning case, we are now all set to plot the QNM modes and compare them with that of Kerr BHs. This is demonstrated in~\ref{fig:quadrupole_rot}, which plots the relative difference in the $\ell = m = 2$ scalar QNM modes between a Kerr quadrupole BH and a Kerr BH of equivalent mass and spin as a function of the anomalous quadrupole moments $\epsilon$ for various spin values $a$. It is noteworthy that as the values of $a$ and $\epsilon$ increase, both the real and imaginary parts of the QNM modes exhibit a rise in relative differences. Nevertheless, these differences consistently remain small, underscoring our reliance on the linear approximation.
%%%%%%%%%%%%%%%%%%%%%%
%%%%%%%%%%%%%%%%%%%%%%
Whereas in~\ref{fig:diff_quadrupole_rot}, we show that the QNM modes converges very quickly as we increase the order of expansion $(a\om)^N$ of the potential given by~\ref{QNM_Vqk}. It is apparent from this figure that the size $|\Delta \omega|$ of the relative difference in the $\ell = m = 2$ mode, when considering the quadratic and quartic expansions is smaller than that between the linear and quadratic expansions for any fixed values of $(a, \epsilon)$. This figure also clarifies that the spin expansion presented in our work differs from the pure slow-rotation expansion described in previous studies~\cite{Kojima:1992ie,Pani:2012bp,Pani:2012vp}.
%%%%%%%%%%%%%%%%%%%%%%
%%%%%%%%%%%%%%%%%%%%%%
Finally, we have also numerically checked that the QNMs in the eikonal limit is in very good agreement with the photon's orbit at the LR for large values of $\ell$ (say, $\ell = 10$). Moreover, even for moderate values of $\ell$ (say, $\ell = 4$)
such correspondence is quite apparent.
%%%%%%%%%%%%%%%%%%%%%%%%
\section{{\color{blue!70!red} Summary}}
%%%%%%%%%%%%%%%%%%%%%%%%
In this chapter, we have outlined a general methodology for computing the scalar QNMs of BHs in scenarios where the exact separability of perturbation equation is unattainable, but the departure of the background Schwarzschild/Kerr spacetime is small. Building on an underlying concept previously employed in the context of rotating BHs in higher-derivative gravity (as discussed in Ref.~\cite{Cano:2020cao}), our approach involves an approximate reformulation (keeping terms only up to the linear order in the deviation parameter) of the perturbation equation in the basis of spherical/spheroidal harmonics. Despite the coupling among radial functions with different quantum numbers, we prescribe a judicious redefinition of the radial function to diagonalize (or decouple) the system. Finally, the corrections to the QNM spectrum from that of Kerr/Schwarzschild BHs are achieved via usual methods (such as continued fraction method).\\

\ni
We have also checked the robustness and reliability of our approach in various ways. For example, when applied to a slowly rotating Kerr BH, our method yields the well-known analytical results for the scalar perturbation equation. Moreover, for the Kerr quadrupole case, the QNM frequencies calculated at different orders of expansion when transitioning from spheroidal to spherical harmonics exhibit a quick convergence. Finally, we have explored the eikonal limit, which relates the orbital frequency at the photon ring and its Lyapunov exponent to the QNMs. Our findings reveal a strong agreement, particularly for large values of $\ell$, and the correct trend for moderate values.\\

\ni
There are various ways to extend our method. It seems a generalization to higher-order deviation is possible, which may be useful for stability analysis of these deformed non-Kerr BHs. However, the most important extension of our method would be to incorporate gravitational perturbation. But, in this context, it should be noted that such a task is non-trivial and requires one to compute the background metric in a specific modified theory of
gravity and then derive the set of perturbation equations. Despite all these complications, the exercise is undoubtedly important as a potential tool to constraint the anomalous multipoles of BHs, which can be treated as a powerful probe for the no-hair properties of BHs. We leave this goal for an future attempt.

\chapter{{\color{red!60!black}Constraining the topological Gauss-Bonnet coupling from GW150914} }\label{Chapter_8}
\large
\textbf{This Chapter is based on the work: Phys. Rev. D 106 (2022) 4, L041503 (Letter) by K. Chakravarti, R. Ghosh, and S. Sarkar}.\\

\ni
Besides its crucial application in BH stability analysis discussed in the previous chapter, BH perturbation theory is especially useful to determine the characteristics of the remnant BH formed as an end state of binary mergers. In the framework of GR, the remnant BH achieve its final state of equilibrium as a Kerr BH by the emission of damped sinusoidal modes. These QNM modes bear the fingerprints of the mass and spin of the underlying central object, which can be inferred from an analysis of the post-merger ringdown signal of a binary event. In contrast, before the coalescence, the corresponding properties of the component Kerr BHs can be measured independently using the inspiral signal. Recently, these techniques have been used~\cite{Isi:2020tac} for the event GW150914 to test one of the most important properties of GR BHs, namely Hawking’s area theorem~\cite{Hawking:1971vc, Bardeen:1973gs}. In the context of BH mergers, the global version of this theorem dictates that the final BH's area must be larger than the total initial area of the component Kerr BHs in the inspiral phase. Moreover, in GR, due to entropy-area proportionality, an increase of BH area also implies an increase of BH entropy during the merger. Therefore, the aforesaid test of the area law can also be viewed as a test of the global version of the BH second law.\\

\ni
However, in the presence of putative higher-curvature modifications of GR, entropy-area proportionality is no longer valid and a verification of the area law does not generally entail the validity of the second law. For concreteness, let us consider the following theory, 
%%%%%%%%%%%%%%%%%%%%%%%%%%%%%
\bea 
{\cal A} = \frac{1}{16 \pi \, }\int \sqrt{-g}\,  \, d^4 x \left( R +  \alpha\, R^2 + \beta\, R_{\mu \nu} R^{\mu \nu} +\, \gamma\, R_{\mu \nu \rho \sigma} R^{\mu \nu \rho \sigma} \right)\ ,
\eea
%%%%%%%%%%%%%%%%%%%%%%%%%%%%%
which represents the most general $4$-dimensional theory of gravity containing up to quadratic order modification to GR. The dimensionful couplings $\alpha$, $\beta$ and $\gamma$ denote the length scales at which the corresponding higher curvature terms are important. Due to the presence of extra propagating modes, the GW signatures of this theory is starkly different from that GR. Thus, for our purpose, we shall consider a special combination (known as the Gauss-Bonnet term) of the quadratic curvature terms that do not modify the gravitational dynamics,
%%%%%%%%%%%%%%%%%%%%%%%%%%%%%
\bea \label{GB_ac}
{\cal A} = \frac{1}{16 \pi \, }\int \sqrt{-g}\,  \, d^4 x \left( R +  \gamma\, \mathcal{L}_{GB} \right)\ .
\eea
%%%%%%%%%%%%%%%%%%%%%%%%%%%%%
Here, as a consequence of celebrated Gauss-Bonnet theorem, the term $\mathcal{L}_{GB} = R^2 - 4\, R_{\mu \nu} R^{\mu \nu} + R_{\mu \nu \rho \sigma} R^{\mu \nu \rho \sigma}$ is topological in $D=4$. Therefore, the Kerr BH remains a solution of
this theory and the dynamics of the GWs propagating in this background is same as GR. Therefore, it may seem the coupling $\gamma$ has no consequence and thus, cannot be constrained via observations/experiments.\\

\ni
Nevertheless, the above claim is not entirely correct and there are some important consequences of this term. For example, it is known that $\gamma > 0$ can increase the instability of 4D de Sitter spacetime by allowing nucleation of BHs \cite{Parikh:2009js}. However, the most striking effect of this topological term is on the second law of BH mechanics. In fact, we shall see that unless the coupling $\gamma$ is constrained, the Gauss-Bonnet term may lead to a decrease of BH entropy during a merger process. Thus, demanding validity of the second law, we will put stringent bound on $\gamma$ using GW observation. And, as per our knowledge, this is the first
such bound on this coefficient.
%%%%%%%%%%%%%%%%%%%%%%%%
\section{{\color{blue!70!brown} Entropy Change in a Merger Process}}
\begin{figure}
\centering
\includegraphics[width=0.6\linewidth]{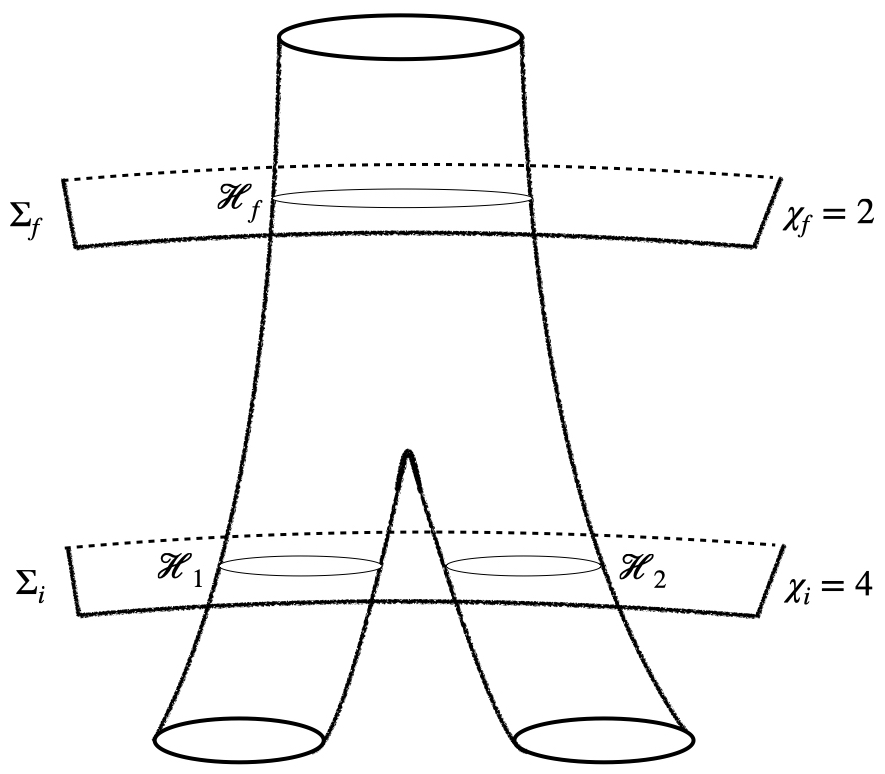}
\caption{\textbf{Formation of the final Kerr BH with horizon slice $\mathcal{H}_f$ by merging two initial Kerr BHs with horizon slices $\mathcal{H}_1$ and $\mathcal{H}_2$.} The Euler character changes discontinuously from the initial ($\Sigma_i$) to final ($\Sigma_f$) Cauchy slice, $\Delta \chi = 2-(2+2)=-2$.}
\label{merge}
\end{figure}

%%%%%%%%%%%%%%%%%%%%%%%%
We are interested in calculating the change of entropy in a BH-BH merger event. For this purpose, we may consider both the initial BHs (in the inspiral phase) and final one (remnant) are well approximated by Kerr metrics with masses and spins $(M_i, a_i)$ for $i = \{1,2\}$, and $(M_f, a_f)$, respectively. Then, using Wald's prescription for the theory in \ref{GB_ac}, the entropy of any one of these BHs receives a correction proportional to the Euler character $\chi = 2$ of its horizon cross sections (topological spheres)~\cite{Jacobson:1993xs, Wald:1993nt, Liko:2007vi},
%%%%%%%%%%%%%%%%%%%%%%%%
\bea \label{GB_S}
S = \frac{A}{4} + 2\, \pi\, \gamma\, \chi\ .
\eea
%%%%%%%%%%%%%%%%%%%%%%%%
It appears that the correction term remains constant and does not influence any dynamical change of the BH spacetime. Nevertheless, this conclusion is prematurely drawn and holds true only under the condition that the topology of the horizon cross sections remains unaltered. In particular, for the the merger of two Kerr BHs as depicted in \ref{merge}, the initial horizon slice $\Sigma_i$ exhibits a topology resembling a disjoint union two separate 2-spheres, while the final horizon slice $\Sigma_f$ adopts the topology of a single 2-sphere.
%%%%%%%%%%%%%%%%%%%%%%%%
%%%%%%%%%%%%%%%%%%%%%%%%
Hence, there is a transition between distinct topological configurations that takes place precisely at the moment of the merger. The specific location of this merger point may vary depending on the chosen spacetime foliation. Nevertheless, if the classical second law holds analogous to the area increase theorem in GR, then entropy should invariably increase regardless of the foliation. However, it seems that entropy change given by
%%%%%%%%%%%%%%%%%%%%%%%%
\bea \label{GB_entch}
\Delta S = \frac{\Delta A}{4} + 2\, \pi\, \gamma\, \Delta \chi = \frac{\Delta A}{4} - 4\, \pi\, \gamma\ ,
\eea
%%%%%%%%%%%%%%%%%%%%%%%%
could be negative ($\Delta S<0$) even for a positive area change $\Delta A>0$, since during the merger the surface area undergoes a continuous change, whereas the Euler number experiences an abrupt and instantaneous jump $\Delta \chi = -2$~\cite{Liko:2007vi, Sarkar:2010xp}. Therefore, a global
violation of the second law is unavoidable, unless the value of the coupling $\gamma$ is constrained. In the above argument, we have assumed $\gamma > 0$ motivated by the result from Ref.~\cite{Cheung:2016wjt}. In fact, for negative values of $\gamma$, the second law could be violated during the formation of BHs from collapse. It is because at the instant that the horizon first appears its area $A$ is arbitrarily small, and hence, \ref{GB_S} implies $S < 0$~\cite{Liko:2007vi, Sarkar:2010xp}. All these arguments justify the choice $\gamma > 0$, which we shall use in the subsequent discussions. \\

\ni
We can also examine the local violation of the second law. Recall that in the context of GR, the area of the event horizon increases continuously at each moment during a BH binary coalescence. However, the scenario changes drastically as we introduce the topological Gauss-Bonnet term. Then, although the area still undergoes a continuous increase, the Euler number experiences a sudden discontinuity during the merger. Consequently, there always exists a slice where $\Delta S$ is negative, resulting in an instantaneous entropy reduction, regardless of the value of the Gauss-Bonnet coefficient $\gamma > 0$. This reduction may even occur with a very small value of $\gamma$~\cite{Sarkar:2010xp}. However, it is worth considering that Wald's formula for BH entropy may not be applicable in a violent, non-stationary phase like BH merger. Nevertheless, for the case of a global violation of the second law we are interested in, the metrics at the initial and final time slices closely resemble stationary Kerr BHs and we can use Wald's prescription. Then, \ref{GB_entch} and all its consequences are non-negotiable.\\

\ni
Furthermore, in Ref.~\cite{Chatterjee:2013daa}, it is proposed that if we regard $4D$ EGB gravity as an effective theory, the violation of the second law would necessitate a regime in which the semi-classical approximation breaks down. While this argument holds some merit, it does not exclude the possibility of a violation beyond the scope of the effective theory assumption. It is also worth noting that we can employ any value for the parameter $\gamma$ without altering the classical equations of motion. Therefore, it becomes imperative to determine the constraints on this parameter to prevent any violation of the second law. Indeed, Ref.~\cite{Sarkar:2010xp} demonstrates that it is feasible to exploit entropy reduction in the context of the $4D$ topological Gauss-Bonnet term, creating conditions that can lead to entropy decrease even in higher-dimensional Lovelock gravity. This is made possible through a Kaluza-Klein compactification of the original $5D$ spacetime into a $4D$ one.
%%%%%%%%%%%%%%%%%%%%%%%%
\section{{\color{blue!70!brown} Bound on the Topological EGB Coupling}}
\begin{figure}
\centering
\includegraphics[width=0.8\linewidth]{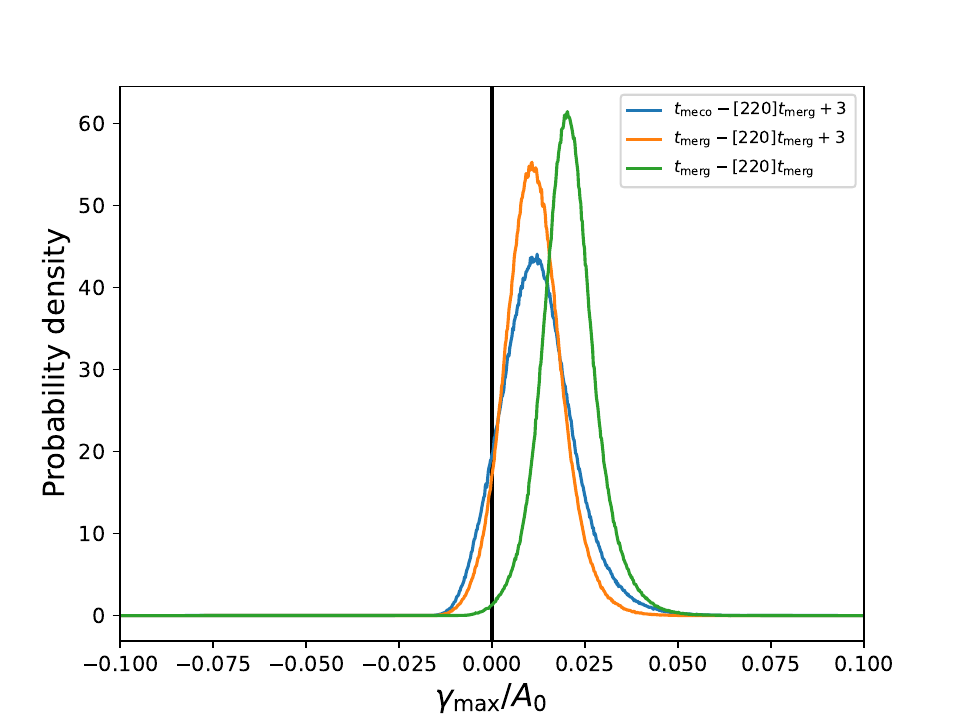}
\caption{\textbf{Plot of probability density of the dimensionless coupling $\left(\gamma_{max}/A_0\right)$.} The region to the left of the black vertical line violates the area law.}
\label{merger}
\end{figure}

%%%%%%%%%%%%%%%%%%%%%%%%
All the above discussions suggest that $\gamma$ must obey an upper bound so as to protect the validity of the BH second law, which using \ref{GB_entch} can be represented as
%%%%%%%%%%%%%%%%%%%%%%%%
\bea \label{GB_ineq}
\gamma < \frac{\Delta A}{16 \pi}  := \gamma_{max}\, .
\eea
%%%%%%%%%%%%%%%%%%%%%%%%
Therefore, a measurement of the 
area change during the merger of two Kerr BHs immediately
entails an upper bound on the coupling $\gamma$. For this purpose, we may use the analysis of Ref.~\cite{Isi:2020tac}, which obtains a posterior distribution of area change $\Delta A$ from the GW data of the event GW150914. However, since the corresponding data are not yet publicly available, one can only quote a rough point estimate, $\Delta A/ A_0 < 0.60$ from the $[220]$ mode~\cite{Isi:2020tac}. Consequently, \ref{GB_ineq} implies the result, $\left(\gamma_{max}/A_0\right) < 0.012$ with $A_0$ being the total area of the two initial Kerr BHs. However, to obtain a much better estimation for $\gamma_{max}$, we can utilize the probability distribution of $\left(\Delta A/A_0\right)$ provided in Ref.~\cite{Kastha:2021chr}. It can then be translated to the corresponding probability distribution for the dimensionless coupling $\left(\gamma_{max}/A_0\right)$. Our result is depicted in~\ref{merger}.\\
%%%%%%%%%%%%%%%%%%%%%%%%
%%%%%%%%%%%%%%%%%%%%%%%%

%%%%%%%%%%%%%%%%%%%%%%%%
\renewcommand{\arraystretch}{2} 
\begin{table}[h!]
\begin{center}
\begin{tabular}{|c|c|}
\hline
~\bf{Estimates}~ & ~\bf{Models}~ \\
\hline
$0.012^{+0.034}_{-0.005}$ & $(\text{NRSur7dq4 before}\, t_{\mathrm{meco}})$ \\ & Damped sinusoid $[220](\text{after}\,  t_{\mathrm{merg}}+3$ ms)\\ 
\hline
$0.011^{+0.027}_{-0.003}$ & $(\text{NRSur7dq4 before}\, t_{\mathrm{merg}})$ \\ & Damped sinusoid $[220](\text{after}\,  t_{\mathrm{merg}}+3$ ms)\\
\hline
$0.020^{+0.006}_{+0.037}$ & $(\text{NRSur7dq4 before}\,  t_{\mathrm{merg}})$ \\ & Damped sinusoid $[220](\text{after}\,  t_{\mathrm{merg}})$ \\
\hline
\end{tabular}
\caption{Upper bounds on $\left(\gamma_{max}/A_0\right)$ with $95\%$ credible interval from different inspiral-merger-ringdown models as suggested in Ref.\cite{Kastha:2021chr}.}
\label{table2}
\end{center}
\end{table}
%%%%%%%%%%%%%%%%%%%%%%%%

\ni
Also, for different choices of the model, Table~\ref{table2} summarizes various  upper bounds on the quantity $\left(\gamma_{max}/A_0\right)$ at $95 \%$ credibility $(2\sigma)$. We have used the abbreviations ``meco" and ``merg" as shorthands for the minimum energy circular orbit and merger, respectively. From the above table, we note a good agreement among the first two models and the point estimation presented earlier from Ref.~\cite{Isi:2020tac}. Whereas the last model peaks at a value of approximately $\left(\gamma_{max}/A_0\right) \approx 0.020$, exhibiting an approximately $80\%$ deviation from the other models. This anomaly can be attributed to inaccuracies stemming from the assumption that the damped sinusoidal QNMs initiate immediately after the merger at $t=t_{\textrm{merg}}$. Due this reason, we confine our analysis to the first two models. Finally, using the well-known estimates of masses and spins of the initial Kerr BHs for the event GW150914~\cite{Isi:2020tac}, we establish an approximate upper bound: $\gamma \lesssim \gamma_{max} \approx 2.804^{+7.946}_{-1.169} \times 10^{9}\, \textrm{m}^2$ at $95\%$ credibility. Interestingly, the peak value of this estimate aligns quite well with the point estimate from Ref.~\cite{Isi:2020tac}.
%%%%%%%%%%%%%%%%%%%%%%%%
\section{{\color{blue!70!brown} Summary}}
%%%%%%%%%%%%%%%%%%%%%%%%
In summary, we have used the recent test of area law reported in Refs.~\cite{Isi:2020tac, Kastha:2021chr} for a BH-BH merger event (GW150914) to put the first observational bound on the topological Gauss-Bonnet coupling $\gamma$, which is necessary to protect the validity of the global version of the BH second law. In future, with the observations of BH mergers generating GWs less efficiently (area change during merger is smaller), one can make this bound even stronger. Let us now contextualize our bound on $\gamma$ in perspective of the established constraints on analogous (quadratic) higher curvature couplings, such as $\alpha\, R^2$ and $\beta\, R_{\mu \nu} R^{\mu \nu}$.\\

\ni
The weak-field approximation of such theories shows Yukawa-type correction over the usual Newtonian potential~\cite{Capozziello:2004sm, Capozziello:2009ss, Berry:2011pb, PhysRevD.70.042004}, which has been verified via E\"{o}t-Wash experiment and stringent constraint as $\alpha < 2 \times 10^{-9}\, \textrm{m}^2$ has been reported~\cite{Berry:2011pb, PhysRevLett.98.021101}. This is perhaps the most stringent constraint on non-topological higher curvature theories attainable through local tabletop experiments. Turning our attention to astrophysical observations, the Gravity Probe B experiment has placed an upper bound on $\alpha$ of approximately $5 \times 10^{11} \, \textrm{m}^2$ \cite{Naf:2010zy}. Likewise, investigations into planetary precession rates have imposed an upper limit of $2.4 \times 10^{18} \, \textrm{m}^2$ on the coupling of $R^2$ gravity \cite{Berry:2011pb}. Analyzing the data from the Hulse-Taylor binary pulsar has yielded a constraint of $1.1 \times 10^{16} \, \textrm{m}^2$ on this coupling \cite{Vilhena:2021bsx}. Additionally, a relatively weaker bound on the coefficient of $R_{\mu \nu} R^{\mu \nu}$ has emerged from the analysis of the time delay between GW170817 and GRB 170817A, constraining $\beta \lesssim 10^{36} \, \textrm{m}^2$ \cite{Ghosh:2019twk}. Similar constraints on both $\alpha$ and $\beta$ have been derived from the study of gravitational waves generated by binary inspirals \cite{Kim:2019sqk}.\\

\ni
However, so far there was no such bound on the topological GB term, as it has no effect in the aforesaid gravitational phenomena.
Therefore, to the best of our knowledge, ours is the only
work that provides an observational constraint on such a
term. Furthermore, in magnitude, our constraint significantly surpasses the previous astrophysical bounds imposed on other higher curvature couplings.

\cleardoublepage

\chapter{{\color{red!60!black}Conclusion and Future Outlook} }\label{Chapter_9}
\large
Our current understanding of gravity is based on Einstein's theory of General Relativity, which describes gravitational interaction as a manifestation of spacetime curvature. This geometric description is extraordinarily successful in explaining numerous experiments and observations with a high degree of accuracy. Yet, unlike other fundamental forces of nature, there is still no consistent quantum theory of gravity and all attempts of reconciling GR with quantum mechanics have remained incomplete. In addition, as elucidated in the introduction section, GR also suffers from several other theoretical challenges, which dictate its failure to be a complete theory of gravity. Though the complete solution of these limitations is still unknown, various theoretical and observational efforts are constantly under inspection to obtain a systematic understanding of the nature of gravity, particularly in the strong field regime. In the same spirit, this thesis aims to provide a glimpse of the scientific excitement through a host of studies that explore various theoretical and observational consequences of modified gravity. On the theoretical side, we have considered BH thermodynamics, stability of compact objects, presence of BH hairs, and the issue of causality that may provide valuable inputs to discern the underpinnings of gravitational interactions. Whereas, on the observational side, we employ GW observations and BH perturbation theory to unravel unknown aspects of gravitational physics and put stringent bounds on the beyond-GR parameters. In this chapter, we summarize the main results stemming from these interesting studies and highlight their possible future generalizations.\\

\ni
After presenting a brief overview of GR, \ref{Chapter 1} showcases a few of its limitations and motivate the importance of considering the modified gravity framework. Subsequently, we move on to an in-depth discussion of various ways to modify GR, namely by (a) making dynamical changes to the GR action (e.g. higher curvature gravity), (b) performing
observation-oriented phenomenological changes in the solutions of GR (e.g. post-Kerr metric), and (c) invoking kinematical alteration in the spacetime structure (e.g. relaxing spacetime symmetry). Furthermore, we review some of the novel features of modified gravity, which provide a warm-up for the successive chapters of the thesis.\\

\ni
In Einstein’s theory, stationary BHs
possess some remarkable similarities
with ordinary thermodynamic systems. More precisely, the event horizon of such a spacetime can be endowed with various thermodynamic properties such as temperature and entropy, which follow the four laws of BH thermodynamics. Among them, the zeroth law dictates that the surface gravity (and hence, the temperature) of a stationary Killing horizon is constant. In GR, the validity of this important result follows as an immediate consequence of Einstein's field equations and the dominant energy conditions on matter. However, as is evident, the same proof does not apply to alternative theories. In addition, the complicated nature of modified gravity field equations makes the problem so non-trivial that there was no generalization of the zeroth law for a long time. In \ref{Chapter_2}, we demonstrate how this long-standing gap can be bridged in a special class of higher curvature theories known as the LL gravity, strengthening the idea that applicability of BH thermodynamics extends beyond GR. Interestingly, such an extension requires an extra assumption that various geometric quantities at the horizon have smooth limit to GR. Recently, our technique has been used to prove zeroth law in other modified theories, such as Horndeski and scalar-hairy Lovelock gravity. Finally, we conclude this chapter by mentioning other ways to further generalize our result. In particular, we prescribe a general form of the near-horizon field equations that may support the zeroth law. In future, it would be interesting to see which alternative gravity models obey such structure of field equations. If possible, this criteria could be used to classify consistent higher curvature theories based on their thermodynamic properties.\\

\ni
The landscape of modified gravity framework is full of surprises. For example, a seemingly healthy alternative theory obeying diffeomorphism invariance may still  give rise to pathological phenomena. Hence, it is absolutely essential to constrain the structure of higher curvature terms in the gravitational action using various consistency criteria, such as the causality constraint prescribed by CEMZ. They argued that theories in which the the Shapiro time shift experienced by a probe crossing a shock wave can take either sign (both positive and negative) are acausal, and such sick theories should be ruled out on the ground of causality. This criteria proves to be quite powerful in putting rather non-trivial constraints on the nature of modified gravity. For example, it has been shown that the presence of Gauss-Bonnet coupling in $D \geq 5$ dimensions is ruled out by CEMZ criterion. At this point, it is natural to ask whether all modified gravity theories suffer from such causality issue. In \ref{Chapter_3}, we tried to answer this question by considering QG as a toy model. Though this theory is generally condemned for the presence of a ghost mode, we argue that QG can still be considered a well-behaved low-energy effective theory as long as our probe energy scale lies below the ghost mass. Moreover, this theory appears naturally as the $1$-loop UV completed effective description of gravity, which obeys other desirable properties like the positive energy theorem. By performing a detail calculation of graviton dynamics in an exact shock wave background, we show that QG is free from any causality issue as the time shift is polarization independent and always positive. Our analysis is not just a duplication of the corresponding GR result in disguise, as the time delay of QG cannot be mapped to that of GR by using a field redefinition. Given the importance of this analysis, we have also proposed a general class of theories, of which GR and QG are two members, that might be free from the causality issue. In future, it will be interesting to study the causality constraints in theories involving cubic or higher order curvature terms. The result such obtained may provide us with invaluable input towards classifying consistent classical theories of gravity. \\

\ni
Compact objects, both with and without horizons, showcase a multitude of gravitational 
phenomena, which may reveal hitherto unknown aspects of strong gravity. Many of the probes, such as QNMs and shadows, hinge crucially upon the LRs (if exist) situated outside the central compact object. Given their observational relevance, there has been a recent interest to study the LR structure in a theory-independent way. For instance, it has been proven that any stationary, axisymmetric, asymptotically flat $4$-dimensional BH spacetime with a non-extremal and topologically spherical Killing horizon admits at least one LR outside the horizon for each rotation sense. However, the existence of LRs does not uniquely fix the nature of the central compact object. Several known horizonless objects, such
as boson stars and gravastars, also possess LRs. In fact, a similar theorem as in the case of BHs dictates that under the assumption of an initial LR, the spacetime of a horizonless compact object always supports an even number of them. And,
among these LRs, at least one must be stable. However, are such horizonless objects stable under perturbations? Because if they
prove to be unstable, we can exclude such objects on mere physical grounds. In this context, it has been argued that horizonless compact objects with stable LRs may trap perturbations without letting them decay, causing destruction of the central object due to nonlinear instability with moderate timescales. In \ref{Chapter_4}, we plug one subtle loophole in this argument. Note that the aforesaid reasoning only holds if there exists a stable LR, which is only guaranteed if the spacetime has an initial LR to begin with. However, in general, a horizonless compact object may not contain any such LR, and the instability argument fails. This is where our theorem comes into play, which dictates that at least for the particular case of stationary, axisymmetric and asymptotically flat spacetimes (both with and without horizons) having an ergoregion, the presence of one LR is always assured. Therefore, our work provides a strong support in favor of the BH hypothesis that claims objects with LRs are BHs. There are various ways to extend our result in future. For example, it would be interesting to see whether a similar result holds in higher dimensions and for asymptotically dS/AdS spacetimes. If true, these results will facilitate an unambiguous detection of BHs via observations, such as the shadow imaging by EHT.\\

\ni
We conclude our theoretical explorations by studying departures from the no-hair properties of BHs. In GR, it is widely believed that gravitational collapse washes
away all information about any additional parameters called hairs and the final BH can be specified only in terms of conserved quantities such as mass and angular momentum measured at asymptotic infinity. This claim is backed by a heuristic argument that any matter fields residing outside a BH would either be emitted away or absorbed by the BH itself unless those fields were associated with conserved charges at asymptotic infinity. However, this seemingly sound argument have been falsified time and again by finding several hairy BH solutions in GR. Furthermore, beyond the framework of GR, the presence of various putative higher curvature terms can also lead to the growth of BH hairs. However, from the observational point of view, the presence of extra hairs can only be captured via far-away observations, if these hairs must extend sufficiently outside the horizon. Thus, it is crucial to investigate whether BHs can support short hairs confined solely to the near-horizon region. In GR, assuming of the weak energy condition (WEC) and the non-positive trace condition on matter, it has been demonstrated that BHs cannot grow short hairs. In particular, the existing hairs must at least extend to the innermost LR outside the horizon. But, can we generalize this important result beyond GR? This is exactly what we have achieved in \ref{Chapter_5} by showing that all existing hairs of any static, spherically symmetric, and asymptotically flat $D$-dimensional BHs must extend at least to the innermost LR, regardless of the specific theory of gravity being considered. In addition, our analysis bears interesting implications on hairs of horizonless compact objects and the size of LRs as well. In future, one may generalize our work to scenarios where certain assumptions, like spherical symmetry, asymptotic flatness, or the WEC on matter, are relaxed. Most notably, investigating the short-hair properties of rotating BHs would be particularly important.\\

\ni
Next, we shift our attention to study various observational signatures of modified gravity. For this purpose nothing seems better suited than the GW observations by the LIGO-Virgo collaboration, which have ushered in a new era of testing GR in strong-gravity regimes. A notable avenue of exploration, thoroughly studied in literature, is to investigate potential signatures of quantum gravity near the BH event horizons. Such quantum effects may yield deviations from the all-absorbing nature of classical BHs.  One such model known as the BH area-quantization was proposed by Bekenstein and Mukhanov. According to this model, the area of BH event horizons is quantized in equidistant steps, resulting in selective absorption at discrete frequencies. Recent research efforts have been invested to study the manifestations of area-quantization in various contexts, including both the inspiral and ringdown phases of BH binaries. In \ref{Chapter_6}, we expand upon this model to encompass scenarios where BH entropy is no longer proportional to the horizon area and contains sub-leading correction terms. Consequently, uniform entropy quantization leads to non-uniform area quantization. After identifying the characteristic frequencies, we have discussed the possibility of overlap among nearby transition lines due to spontaneous Hawking radiation. Our exploration then extends to a comprehensive analysis of two important GW observables of area discretization, namely the dephasing phenomenon due to tidal heating in the inspiral stage and generation of late-time echo signals in the ringdown phase. In the later case, we argue that there is an ambiguity in the placement of the near-horizon quantum filter required to model the absorption profile of such BHs. This ambiguity ultimately leads to strikingly different echo signals, which can be used to distinguish between various models. Therefore, it seems that the astrophysical BHs magnify the near-horizon Planck scale physics in the realm of current GW observations. In future, as advancements in GW detectors bring about enhanced accuracy, sensitivity, and signal-to-noise ratios, we anticipate the ability to impose stringent constraints on various parameters related to area-quantization. Moreover, there is ample scope to extend our work and look for more GW observables. For instance, it would be interesting to study the near-merger features of BH area-quantization, which warrants the application of the powerful numerical relativity toolkit. \\

\ni
One of the most important probe to the near-horizon physics is QNMs, which are also crucial for BH stability analysis. In GR, finding these modes are particularly simple as the governing perturbation equation in Schwarzschild/Kerr background decouples into radial and angular parts. However, in general, one may not have such luxury for BH solutions of a modified gravity theory. Then, the most straightforward way to progress is to compute the QNMs in a
theory-by-theory basis. Due to its obvious limitations, one is
forced to develop more general theory-independent ways for finding QNMs. In \ref{Chapter_7}, we have devised an efficient method of computing scalar QNMs of BHs in situations where the background geometries are  perturbatively close to Schwarzschild/Kerr BHs. For concreteness, we considered various examples such as quadrupolar Schwarzschild and Kerr BHs. Our method also shows a universal structure of the eikonal QNMs for such deformed BHs. There are various ways to extend our method. For example, one can attempt to  incorporate gravitational perturbation in the scheme, which have crucial applicability in determining the properties of the remnant BH formed as an end state of binary mergers. \\

\ni
Recently, similar technique has been used for the event GW150914 to test one of the most important properties of BHs solutions of GR, namely Hawking’s area theorem. \ref{Chapter_8} uses this result along with the validity of the BH second law to put stringent constraint on the $4D$ Gauss-Bonnet coupling $\gamma$. The novelty of our analysis lies in the fact that such a coupling cannot be constraint (due to its topological nature) by other astrophysical observations. Future GW observations might help to make our bound on $\gamma$ even stronger.\\

\ni
Let us conclude with some final remarks. We live in an extraordinary time when we have diligently assembled most of the puzzle pieces needed to fathom the universe's grand tapestry. One of the remaining pieces is to understand the true nature of gravity, whose classical character is greatly explained by Einstein's theory of GR. Though GR is completely successful, there are ample reasons to believe it may still receive significant modifications. Nevertheless, in the absence of a fully consistent theory, we should consider all possible alternatives and subject them to scrutiny through observations. This quest for a complete theory is arguably as significant as eliminating potential substitutes. In the same spirit, this thesis presents some novel theoretical and observational studies with the aspiration that they might contribute valuable perspectives to the ongoing body of research.

\appendix
\chapter{{\color{red!60!black}Appendices for \ref{Chapter_3}} }\label{Chapter_10}

\section{Relation between Eikonal Scattering Amplitude and the Shapiro Time Shift}\label{app1}
This appendix derives the Shapiro time shift suffered by a massless probe scalar using tree-level scattering amplitude in the eikonal limit. The relationship between the Shapiro time shift and the phase-shift is given by,
%%%%%%%%%%%%%%%%%%%%%%%%%%%%%%%%%%
\begin{flalign}
\Delta v\, =\, -\, \frac{1}{p_{v}}\, \delta(\vec{b}, s)\ .
\end{flalign}
%%%%%%%%%%%%%%%%%%%%%%%%%%%%%%%%%%
The phase shift $\delta(\vec{b}, s)$ is given in terms of the tree-level amplitude in impact parameter representation, see \ref{CCscalar}. Using partial fraction decomposition of the $t$-channel eikonal amplitude given by \ref{CCeikonal}, one can rewrite the phase shift as follows,
\begin{flalign}
\delta(\vec{b}, s)\, =\, -\, 4\pi G s\, \int \frac{d^{D-2} \vec{q}}{(2 \pi)^{D-2}}\, \, e^{i \vec{b} \cdot \vec{q}}\, \left[\frac{1}{t}\, -\, \frac{1}{(t + \frac{1}{\beta})} \right]\ ,
\end{flalign}
where $t=-\lvert \vec{q} \rvert^2$, and  $b\, =\, (\vec{b} \cdot \vec{b})^{\frac{1}{2}}$. Then, assuming $\beta\, \leq\, 0$, the following integrals are easily obtained,
\begin{flalign}
\int \frac{d^{D-2} \vec{q}}{(2 \pi)^{D-2}}\, e^{i \vec{b} \cdot \vec{q}}\, \, \frac{1}{t}\, =\, -\, \frac{1}{4 \pi^{D/2-1}}\, \, \frac{\Gamma[D/2-2]}{b^{D-4}}\, ,
\end{flalign}
and,\\
\begin{flalign}
\int \frac{d^{D-2} \vec{q}}{(2 \pi)^{D-2}}\, e^{i \vec{b} \cdot \vec{q}}\, \, \frac{1}{t + \frac{1}{\beta}}\, =\, -\, \frac{(b \sqrt{-\beta})^{2-D/2}}{\left( 2 \pi \right)^{D/2-1}}\, \, K_{2-D/2} \left[\frac{b}{\sqrt{-\beta}}\right]\ .
\end{flalign}
Now, substituting $s\, =\, -4\, \vert P_{u}\vert\, p_{v}$, one can immediately verify \ref{CCscalar}. 

\section{Gravitational Perturbation in QG}\label{app2}
In this section, we outline the derivation of the general solution of the graviton EoM given by \ref{EoM}. We start by rewriting the equation in a suggestive form,
%%%%%%%%%%%%%%%%%%%%%%%%%%%%%
\bea \label{D}
\mathcal{D} \Delta h_{ij} = 0 ~;\ \ \text{where}\ \mathcal{D} = 1 + \beta \Delta ~.
\eea
%%%%%%%%%%%%%%%%%%%%%%%%%%%%%
Using the notation $\bar{h}_{ij} := \partial_v h_{ij}$, we can perform a Fourier transform in the $v$-direction,
%%%%%%%%%%%%%%%%%%%%%%%%%%%%%
\bea
\widetilde{h}_{ij}(u, p_v) = \int dv\ e^{-i p_v v}\ \bar{h}_{ij}(u,v) ~.
\eea
%%%%%%%%%%%%%%%%%%%%%%%%%%%%%
Then, the momentum space EoM is given by,
%%%%%%%%%%%%%%%%%%%%%%%%%%%%%
\bea \label{FT}
\left[ 1 - 4 i \beta p_v \left( \partial_u + i h_0 p_v \right) \right] \widetilde{K}_{ij}(u, p_v) = 0 ~,
\eea
%%%%%%%%%%%%%%%%%%%%%%%%%%%%%
where $\widetilde{K}_{ij}(u, p_v) = \left( \partial_u + i h_0 p_v \right) \widetilde{h}_{ij}(u, p_v)$. This is a first order differential equation in $u$, which can be easily solved to obtain
%%%%%%%%%%%%%%%%%%%%%%%%%%%%%
\begin{align} \label{K}
\widetilde{K}_{ij}(u, p_v, x_i) = \widetilde{K}_{ij}^{(0)}(p_v, x_i)\ \operatorname{exp}\left({\frac{u}{4 i \beta p_v}} \right)
\operatorname{exp} \left({-i p_v \int^{u} d u\ h_{0}\left(u, x_{i}\right)} \right) ~,
\end{align}
%%%%%%%%%%%%%%%%%%%%%%%%%%%%%
by reintroducing the $x_i$-dependence. Using the definition of $\widetilde{K}_{ij}(u, p_v, x_{i})$, we can now express $\widetilde{h}_{ij}(u, p_v, x_{i})$ in the form given by \ref{CCht}. Finally, we get the full solution for $h_{ij}$, see \ref{CCsol}, by integrating $\partial_v h_{ij} = \bar{h}_{ij}$.

\chapter{{\color{red!60!black}Appendix for \ref{Chapter_4}} }\label{Chapter_11}

\section{Derivation of the LR Equation}\label{app3}
Following Ref.~\cite{Cunha:2017qtt}, a LR can be defined as a circular null geodesic whose tangent filed is a linear combination of the two Killing vectors $\partial_t$ and $\partial_\phi$ alone. Thus, a photon with momentum $p_\mu$ orbiting along the LR must satisfy $p_\mu = \dot{p}_\mu = 0$, where $\mu = \{r, \theta\}$ and the dot represent derivative with respect to an affine parameter. Then, the vanishing of the proper interval $ds^2 = 0$ for the motion of photons along with $p_\mu = 0$ imply the condition, $V = g^{tt}\, E^2 - 2\, g^{t \phi}\, E\, \Phi+ g^{\phi \phi}\, \Phi^2 = 0$, at the location of the LRs. Here, $p_t = -E$ and $p_\phi = \Phi$ are two conserved quantities associated with the Killing isometries. In terms of the impact parameter $\sigma = E/\Phi$, one can re-express $V$ in a more suggestive form, $V = -\Phi^2\, g_{\phi \phi}\, (\sigma - H_+)\, (\sigma - H_-)/\Delta$, where the form of the potentials $H_{\pm}$ are given by 
%%%%%%%%%%%%%%%%%%%%%%%%%
\bea
H_{\pm}(r,\theta) = \frac{-g_{t \phi} \pm \sqrt{\Delta}}{g_{\phi  \phi}}\, ,
\eea
%%%%%%%%%%%%%%%%%%%%%%%%%
with $\Delta=g_{t \phi}^2-g_{tt}\, g_{\phi \phi} > 0$ outside the ergoregion. Using this equation, LR condition $V=0$ constraints the impact parameter $\sigma$ in terms of the potentials $H_{\pm}$. However, the location of the LRs are determined by the critical points of the equation $\partial_\mu H_{\pm} = 0$ as suggested by the remaining condition $\dot{p}_\mu = 0$.\\

\ni 
In fact, one can also infer the stability of the LRs by considering second derivatives of $H_{\pm}$ since~\cite{Cunha:2017qtt} 
%%%%%%%%%%%%%%%%%%%%%%%%%
\bea
\partial_\mu^2 V = \pm \frac{2\, \Phi^2}{\sqrt{\Delta}} \partial_\mu^2 H_{\pm} \, .
\eea
%%%%%%%%%%%%%%%%%%%%%%%%%
The two signs represent two opposite sense of rotations with respect to the central object. If the object is rotating in the "negative" sense, meaning $g_{t \phi} > 0$, the critical points of the potential $H_+$ ($H_-$) represent counter-rotating (co-rotating) LRs. Whereas for $g_{t \phi} > 0$, the roles of $H_{\pm}$ are swapped.

\chapter{{\color{red!60!black}Appendices for \ref{Chapter_7}} }\label{Chapter_12}

\section{The Source Term}\label{app4}
This appendix provides the explicit expressions of the coefficients $a$, $b$ and $c_{\ell m}$ appearing in \ref{eq:decomposition}. We have considered deviation metric $g_{\mu\nu}^{(1)}$ to be the most general stationary and axisymmetric configuration and no additional assumption such as circularity ~\cite{Wald:1984rg} are not assumed. Then, we have
\begin{align}
    & a(r,\th) = \dl_\th \left[ v_2 - u\, \frac{a^2\, r^2\, \sst}{\rho^2\, \De} \right] + 2\, f_{t\th}, \\
    & b(r,\th) = \De \,\dl_r \left[ v_1 - u\, \frac{a^2\, r^2\, \sst}{\rho^2\, \De} \right] + 2\, f_{tr}, \\
    & c_{\ell,m}(r,\th) = r \, \om \, w_{10} \left[ \frac{K}{\De} + r\, \om  \right] + a^2\, \om^2\,  w_{02}\, \cct \notag \\
    & \quad -\frac{4\, M\, a\, r\, u\, F}{\rho^2 \De} \left[ m \left(\rho^2 -r\right) + a\, r\, \om\, \sst \right]  \notag \\
    & \quad + \frac{m^2\,  w_{23}}{\sst} + w_{12}\, \la_{\ell m} \notag  + \dl_r f_{tr} \\
    & \quad + a\, m\, w_{13}\, \frac{2\, M\, r\, \om - a\, m}{\De} + \frac{1}{\sin\th}\dl_{\th}\left(\sin\th\, f_{t\th} \right),
\end{align}
where we have used the following notations:
\begin{align}
    u(r,\th) & = \frac{1}{2}\left( f_{tt} - 2 f_{t \cf} + f_{\cf\cf} \right), \\
    v_0(r,\th) & = \frac{1}{2} \left( - f_{tt} + f_{rr} + f_{\th\th} + f_{\cf\cf} \right), \\
    v_1(r,\th) & = \frac{1}{2} \left( f_{tt} - f_{rr} + f_{\th\th} + f_{\cf\cf} \right), \\
    v_2(r,\th) & = \frac{1}{2} \left( f_{tt} + f_{rr} - f_{\th\th} + f_{\cf\cf} \right), \\
    v_3(r,\th) & = \frac{1}{2} \left( f_{tt} + f_{rr} + f_{\th\th} - f_{\cf\cf} \right) ,
\end{align}
and $w_{ij} = v_i - v_j$. The functions are defined as,
\begin{align}
    & f_{tt}(r,\th) =g_{tt}^{(1)} / g_{tt}^{(0)}, \qquad f_{rr}(r,\th) = g_{rr}^{(1)} / g_{rr}^{(0)} , \\
    & f_{\th\th}(r,\th) = g_{\th\th}^{(1)} / g_{\th\th}^{(0)} , \qquad f_{\cf\cf}(r,\th) = g_{\cf\cf}^{(1)} / g_{\cf\cf}^{(0)} , \\
    & f_{t\cf}(r,\th) = g_{t\cf}^{(1)} / g_{t\cf}^{(0)}, \qquad f_{tr}(r,\th) = \De\, F\, g^{(1)}_{tr}, \\
    & f_{t\th}(r,\th) = F\, g^{(1)}_{t\th} , \qquad F = \ii \left( \om + \frac{r\, K}{\rho^2\, \De} \right).
\end{align}

\section{Existence of $U_{\ell m}$}\label{app5}
To study the existence of $U_{\ell m}(r)$ as a solution of~\ref{QNM_ulm}, we note that it is an ordinary differential equation with known coefficients and source. It is because they are functions of various background ($\ep = 0$) quantities, such as $Z_{\ell m}$  and frequency $\omega_{\ell m}$, already obtained by solving~\ref{QNM_psi0}.  
With this observation, we can employ the so-called ``variation of parameters''-method~\cite{nagle2012}, to  solve~\ref{QNM_ulm} by imposing that $U_{\ell m}$ satisfies ingoing boundary conditions for $r_*\to-\infty$ and
outgoing ones for $r_*\to\infty$. For this purpose, let us consider two linearly independent solutions $U_{\ell m}^{(1)}(r_*)$ and $U_{\ell m}^{(2)}(r_*)$ of the homogeneous part of~\ref{QNM_ulm} such that, 
\begin{align}
    & U_{\ell m}^{(1)}(r_*)\sim \ee^{\ii \omega_{\ell m} r_*} \mbox{  for  } r_*\to -\infty\,,\label{U1}\\
    & U_{\ell m}^{(1)}(r_*)\sim A\, \ee^{\ii \omega_{\ell m}  r_*}+B\, \ee^{-\ii \omega_{\ell m}  r_*}\mbox{  for  } r_*\to +\infty\,,\\
    & U_{\ell m}^{(2)}(r_*)\sim \ee^{-\ii \omega_{\ell m}  r_*} \mbox{  for  } r_*\to \infty\,,\\
    & U_{\ell m}^{(2)}(r_*)\sim C\, \ee^{\ii \omega_{\ell m}  r_*}+D\, \ee^{-\ii \omega_{\ell m}  r_*}\mbox{  for  } r_*\to -\infty\,,\label{U2}
\end{align}
where $A$, $B$, $C$ and $D$  are constant complex coefficients.
Since the Wronskian  of the differential equation~\ref{QNM_ulm} is a constant $W \neq 0$\footnote{Linearly independence of the two solutions make sure that $W \neq 0$. However, if one choose instead two solutions that satisfy ingoing/outgoing boundary conditions at the horizon/infinity, the Wronskian would vanish at QNM frequencies (by definition) obtained from~\ref{QNM_psi0}. This explains why it is more convenient to choose $U^{(1,2)}_{\ell m}$ satisfying~\ref{U1}--\ref{U2}.}, we can write a solution as
\begin{align}\label{particular}
    U_{\ell m}(r_*) = & U_{\ell m}^{(1)}(r_*) \int_{r_*}^{r_{*,2}} \frac{U_{\ell m}^{(2)}(x)}{W} \, T_{\ell m}(x)\, dx \nonumber \\
    & + U_{\ell m}^{(2)}(r_*) \int_{r_{*,1}}^{r_*} \frac{U_{\ell m}^{(1)}(x)}{W} \, T_{\ell m}(x)\, dx\,,
\end{align}
with $r_{*,1}$ and $r_{*,2}$ being constants. Note that from~\ref{eq:sourcejl}, the functions $\alpha_{\ell'm}$ and $\beta_{\ell'm}$ are finite as $r_*\to\pm \infty$, and the source $T_{\ell m}$ diverges at most as $\exp(\pm \ii \omega_{\ell'm} r_*)$, with $\ell'\neq \ell$, for $r_*\to\pm \infty$. However, these divergence is simply an artefact of working in the frequency domain. Once the time dependence $\exp(-\ii \omega_{\ell'm} t)$ is restored, it is apparent that $T_{\ell m}$ is finite at future null infinity and on the future event horizon. Thus, it completes our proof that $U_{\ell m}$ exists and in fact, can be obtained from \ref{particular}.\\

\ni
Moreover, using~\ref{U1}--\ref{U2} and integrating \ref{particular} by parts, it follows that for $r_*\to\pm\infty$ one has $U_{\ell m}\sim \exp(\pm \ii \omega_{\ell'm} r_*) {\cal A_\pm}$, where ${\cal A_\pm}$ depends linearly on the asymptotic values of $\alpha_{\ell'm}$ and $\beta_{\ell'm}$ as $r_*\to\pm\infty$. These behaviors are sensible as they correspond to outgoing/ingoing boundary conditions for the $(\ell'\neq \ell,m)$ modes, which also appear in~\ref{QNM_meom} due to the mode mixing term on the right-hand side.
Note, however, that the explicit form of $U_{\ell m}(r_*)$ is {\it not} needed to solve for the QNM spectrum from our master equation in~\ref{QNM_master}, but only if one wants to reconstruct the scalar eigenfunctions.

\section{Decoupling of the Wave Equation: An Example}\label{app6}
In this Appendix, we shall show an explicit method to decouple an QNM equations similar to the Schwarzschild quadrupole case, see~\ref{QNM_Vq},
%%%%%%%%%%%%%%%%%%%%%%%
\begin{align} \label{Zeq}
\frac{d^2 Z_{\ell m}}{dr_*^2} + V_{\ell m} (r)\, Z_{\ell m} = \epsilon\, \Big[ f_{\ell m}^{(1)}(r)\, Z_{\ell+2, m} + f_{\ell m}^{(2)}(r)\, Z_{\ell-2, m}\Big]\ .
\end{align}
%%%%%%%%%%%%%%%%%%%%%%%
We want to proceed in such a way that the result Ref.~\cite{Pani:2013pma} appears as a limiting case of our analysis. Note also that the ratio $\Big( f_{\ell m}^{(1)}/f_{\ell m}^{(2)} \Big)$ is $r$-independent for~\ref{QNM_Vq}. Now, we shall perform a field redefinition, $X_{\ell m}(r_*) = Z_{\ell m}(r_*) + \epsilon\, \tilde{Z}_{\ell m}(r_*)/n(r)+ \epsilon\, U_{\ell m}(r_*)$, where $r_*$ is the background tortoise coordinate defined by $dr/dr_* = h(r)$ and $ \tilde{Z}_{\ell m} = c_{\ell m}\, Z_{\ell+2,m} - d_{\ell m}\, Z_{\ell-2,m}$. We want to choose the $r$-independent coefficients $(c_{\ell m}, d_{\ell m})$ and the function $U_{\ell m}(r)$ in such a way that $X_{\ell m}$ satisfies the master equation given by~\ref{QNM_master}. One such choice is given by
\begin{align} \label{cdu}
&c_{\ell m} = \frac{f_{\ell m}^{(1)}(r)\, n(r)}{V_{\ell+2,m}^{(0)} - V_{\ell m}^{(0)}},\quad d_{\ell m} =  \frac{f_{\ell m}^{(2)}(r)\, n(r)}{V_{\ell m}^{(0)}-V_{\ell-2,m}^{(0)}}; \\
\nonumber \\ 
&\frac{d^2 U_{\ell m}}{dr_*^2} + V_{\ell m}^{(0)} (r)\, U_{\ell m} = \partial_{r_*} \Big( \frac{n'(r)\, h(r)}{n^2(r)}\Big)\, \tilde{Z}_{\ell m} + \frac{2n'(r)\, h(r)}{n^2(r)}\, \frac{d \tilde{Z}_{\ell m}}{dr_*}\ .
\end{align} 
where the wave function and the potential in the RHS of the above equation should be treated as the background (Schwarzschild) quantity. Then, it readily follows that the choice $n(r) =1$ and $U_{lm} = 0$ agrees with the slowly rotating Kerr case presented in~\cite{Pani:2013pma}. In contrast, for the Schwarzschild quadrupole BH in~\ref{QNM_Vq}, it is easy to see that $n(r) = r$, $d_{\ell m} = -3M^3 \omega^2 B_{\ell}^{m}/(2\ell-1)$, and $c_{\ell m} = d_{\ell+2}^{m}$ represent a valid solution. In any case, after replacing $\big(n(r), c_{\ell m}, d_{\ell m}\big)$, the field $U_{\ell m}$ satisfies an ordinary differential equation with known coefficients, which can be solved using the variation of parameters method~\cite{nagle2012}.\\

\ni
By a repeated application of the above method, we can decouple a general system with couplings among $\ell$ and $\{\ell-\bar{\ell}, \dots ,\ell-2, \ell-1, \ell+1, \ell+2, \dots, \ell+\bar{\ell}\}$ modes, see for example~\ref{QNM_VKq} in the Kerr quadrupole case. Such coupling are the consequences of higher order multipole deviations (such as quadrupole, octupole and so on) of the background Schwarzschild/Kerr metric. Therefore, from a phenomenological point of view, one may directly start with a potential $V_{\ell m} = V_{\ell m}^{(0)} + \sum_{i} \epsilon_i V_{\ell m}^{(i)}$ in~\ref{QNM_master}, where the background $V_{\ell m}^{(0)}$ has been modified by contributions $V_{\ell m}^{(i)}$ coming from various higher multipoles. This motivates the work presented in Ref.~\cite{Volkel:2022aca}. It is not hard to concoct a similar method for decoupling an equation with a source term containing derivatives of $Z_{\ell' m}$ as well.

\section{Wave Equation in Terms of $\bar{r}_*$}\label{app7}
This appendix addresses one more issue that may arise because the the tortoise coordinate $\bar{r}_*$ of the metric given by~\ref{QNM_metric} is different from the Kerr/Schwarzschild tortoise coordinate $r_*$. Then, except when the horizon location remains the same and the near-horizon (ingoing) boundary condition remains unaltered~\cite{Berti:2009kk, Pani:2013pma}, we have to rewrite the perturbation equation in terms of the new tortoise coordinate to properly incorporating the QNM boundary conditions.\\

\ni
To achieve this task, let us assume that the new and old tortoise coordinates are related by $d\bar{r}_* = dr_*\, [1+\epsilon\, g(r)]$, where the chosen radial coordinate is constant on the horizon (no angular dependence). Using this relation, we can express~\ref{QNM_master} as,
%%%%%%%%%%%%%%%%%%%%%%%%%
\begin{align} 
\frac{d^2 X_{\ell m}}{d\bar{r}_*^2} + \, \epsilon\, h(r)\, g'(r)\, \frac{d X_{\ell m}}{d\bar{r}_*}+ V_{\ell m}(r) \Big(1- 2\, \epsilon\, g(r)\Big) X_{\ell m}+ \mathcal{O}(\epsilon^2)= 0\ ,
\end{align}
%%%%%%%%%%%%%%%%%%%%%%%%%
where $dr/dr_* = h(r)$. The first-derivative term can be absorbed by a field redefinition, $X_{\ell m} \to \widetilde{X}_{\ell m}\, \exp\Big[-\epsilon/2\, \int d\bar{r}_*\, h(r)\, g'(r) \Big]$. And, the final master equation becomes
$d^2 \widetilde{X}_{\ell m}/d\bar{r}_*^2 + \widetilde{V}_{\ell m}(r)\, \widetilde{X}_{\ell m} = 0$, with the redefined potential as 
%%%%%%%%%%%%%%%%%%%%%%%%%
\begin{equation} \label{tilW}
 \widetilde{V}_{\ell m}(r) = V_{\ell m}(r)\Big[1-2\, \epsilon\, g(r)\Big] - \frac{1}{2}\, \epsilon\, h(r)\, \frac{d \big[h(r)\, g'(r) \big]}{dr}\, .
\end{equation}
%%%%%%%%%%%%%%%%%%%%%%%%%
We may demonstrate the above method by considering a Reissner-Nordstr\"{o}m (RN) BH with a small charge $|Q| \ll M$, and the location of the horizon is different from Schwarzschild. Here, $\epsilon = Q^2$ shows the deviation from the Schwarzschild spacetime, and $g(r) = -\big[ r^2\, f(r) \big]^{-1}$ and $h(r) = f(r)$. Then, using \ref{QNM_pot} and ~\ref{tilW}, the master QNM equation becomes
%%%%%%%%%%%%%%%%%%%%%%%%%
\begin{equation}
\frac{d^2 \widetilde{X}_{\ell m}}{d\bar{r}_*^2} + \left[ V_{\ell}^\mathrm{Sch}(r) - \epsilon\, \frac{6\, M + (\ell+2)(\ell-1)\, r }{r^5} \right]\, \widetilde{X}_{\ell m} = 0\ ,
\end{equation}
%%%%%%%%%%%%%%%%%%%%%%%%%
which can be verified by a direct calculation of QNM equation in RN spacetime.\\

\ni
As another interesting application, note that the potential given by~\ref{tilW} takes the following simplified form in the eikonal limit ($\ell = m \gg 1$),
%%%%%%%%%%%%%%%%%%%%%%%%%
\begin{equation} \label{tilVeik}
    \widetilde{V}_{\ell m}^{\textrm{eik}}(r) \simeq V_{\ell m}^{\textrm{eik}}(r)\Big[1-2\, \epsilon\, g(r)\Big]\, ,
\end{equation}
%%%%%%%%%%%%%%%%%%%%%%%%%
where $V_{\ell m}^{\textrm{eik}}(r)$ is the eikonal limit of the potential $V_{\ell m}(r)$ given in~\ref{QNM_pot}. We have used this equation to derive the eikonal QNM potential in the Schwarzschild quadrupole case, see \ref{QNM_eik}.

%\bibliography{References/Bibliography}
\providecommand{\href}[2]{#2}\begingroup\raggedright\endgroup

\bibliographystyle{References/utphys1}

\cleardoublepage
%%%%%%%%%%%%%%%%%%%%%%%%%%%%%%%%%%%%%%%%%%%%%%%%%%%%%%%%%%%%%%%%%%%%%%%%%%%%%%%%%%%%%%%%%%%%%%%%%%%
%%%%%%%%%%%%%%%%%%%%%%%%%%%%%%%%%%%%%%%%%%%%%%%%%%%%%%%%%%%%%%%%%%%%%%%%%%%%%%%%%%%%%%%%%%%%%%%%%%%
%%%%%%%%%%%%%%%%%%%%%%%%%%%%%%%%%%%%%%%%%%%%%%%%%%%%%%%%%%%%%%%%%%%%%%%%%%%%%%%%%%%%%%%%%%%%%%%%%%%
\end{document}